\documentclass[journal,table,xcdraw]{IEEEtran}

\usepackage{cite}
\usepackage{scalerel}
\usepackage{tikz}
\usepackage{amssymb}
\usepackage{xcolor}
\usepackage{graphicx}
\usepackage{caption}
\usepackage{supertabular}
\usepackage{float}
\usepackage{soul}
\usepackage{subfig}
\usetikzlibrary{svg.path}
\usepackage{svg}
\usepackage{comment}
\usepackage[11pt]{moresize}
\newsavebox{\measurebox}
\usepackage{longtable} 
\usepackage{pdflscape} 
\usepackage{pgf-pie} 
\newcommand*\rot{\rotatebox{-90}} 
\definecolor{orcidlogocol}{HTML}{A6CE39}
\tikzset{
    orcidlogo/.pic={
        \fill[orcidlogocol] svg{M256,128c0,70.7-57.3,128-128,128C57.3,256,0,198.7,0,128C0,57.3,57.3,0,128,0C198.7,0,256,57.3,256,128z};
        \fill[white] svg{M86.3,186.2H70.9V79.1h15.4v48.4V186.2z}
        svg{M108.9,79.1h41.6c39.6,0,57,28.3,57,53.6c0,27.5-21.5,53.6-56.8,53.6h-41.8V79.1z M124.3,172.4h24.5c34.9,0,42.9-26.5,42.9-39.7c0-21.5-13.7-39.7-43.7-39.7h-23.7V172.4z}
        svg{M88.7,56.8c0,5.5-4.5,10.1-10.1,10.1c-5.6,0-10.1-4.6-10.1-10.1c0-5.6,4.5-10.1,10.1-10.1C84.2,46.7,88.7,51.3,88.7,56.8z};
    }
}

\usepackage{calc}
\usepackage{ragged2e}
\newcolumntype{A}{>{\RaggedRight\arraybackslash}p{(\linewidth-7.215em-14\tabcolsep-(8\arrayrulewidth))/6}}

\newcommand\orcidicon[1]{\href{https://orcid.org/#1}{\mbox{\scalerel*{
                \begin{tikzpicture}[yscale=-1,transform shape]
                \pic{orcidlogo};
                \end{tikzpicture}
            }{|}}}}

\usepackage[colorlinks=false,urlcolor=black, hidelinks]{hyperref} 

\usepackage{amsmath}
\usepackage{algorithmic}
\usepackage{array}
\usepackage{balance}

\newcommand{\paragbf}[1]{\noindent\textbf{#1. }}
\newcommand{\paragit}[1]{\noindent\textit{#1. }}

\usepackage{multicol}

\begin{document}
%
\title{A Survey on Industrial Control System Testbeds and Datasets for Security Research}
%
%
%

\author{Mauro Conti$^{\textsuperscript{\orcidicon{0000-0002-3612-1934}}}$,~\IEEEmembership{Senior Member, IEEE}
        Denis Donadel$^{\textsuperscript{\orcidicon{0000-0002-7050-9369}}}$, 
        and~Federico Turrin$^{\textsuperscript{\orcidicon{0000-0001-5660-2447}}}$
\thanks{This work has been submitted to the IEEE for possible publication. Copyright may be transferred without notice, after which this version may no longer be accessible.}}

\maketitle

\begin{abstract}
The increasing digitization and interconnection of legacy Industrial Control Systems (ICSs) open new vulnerability surfaces, exposing such systems to malicious attackers. Furthermore, since ICSs are often employed in critical infrastructures (e.g., nuclear plants) and manufacturing companies (e.g., chemical industries), attacks can lead to devastating physical damages. In dealing with this security requirement, the research community focuses on developing new security mechanisms such as Intrusion Detection Systems (IDSs), facilitated by leveraging modern machine learning techniques. However, these algorithms require a testing platform and a considerable amount of data to be trained and tested accurately. To satisfy this prerequisite, Academia, Industry, and Government are increasingly proposing testbed (i.e., scaled-down versions of ICSs or simulations) to test the performances of the IDSs. Furthermore, to enable researchers to cross-validate security systems (e.g., security-by-design concepts or anomaly detectors), several datasets have been collected from testbeds and shared with the community.

In this paper, we provide a deep and comprehensive overview of ICSs, presenting the architecture design, the employed devices, and the security protocols implemented. We then collect, compare, and describe testbeds and datasets in the literature, highlighting key challenges and design guidelines to keep in mind in the design phases. Furthermore, we enrich our work by reporting the best performing IDS algorithms tested on every dataset to create a baseline in state of the art for this field. Finally, driven by knowledge accumulated during this survey's development, we report advice and good practices on the development, the choice, and the utilization of testbeds, datasets, and IDSs.
\end{abstract}


%
\IEEEpeerreviewmaketitle

\section{Introduction}

\IEEEPARstart{C}{ritical} infrastructures and the emerging Industry 4.0 are increasingly using more advanced technologies such as computers, electrical and mechanical devices to monitor the physical processes. 
The networks resulting from smart smart computing integration for the processes monitoring are called Industrial Control Systems (ICSs) or, sometimes, SCADA systems.

ICSs are composed of two macro areas. The Operational Technology (OT) network includes hardware and software used to monitor and manage industrial equipment, assets, processes, and events (e.g., Programmable Logic Controllers (PLCs), sensors, actuators). On the other side, the traditional Information Technology (IT) network contains workstations, databases, and other classical machines used to manipulate information. 
IT and OT networks were originally disconnected. However, due to the so-called IT/OT Convergence~\cite{alcaraz2019secure}, the two networks have been interconnected to facilitate the digitization of processes, opening new vulnerability surfaces.

For Cyber-Physical Systems (CPSs), which also contains ICSs, the classical CIA triad (Confidentiality, Integrity, Availability) is considered reversed, in order of importance, as Availability, Integrity, and Confidentiality~\cite{Stouffer2015nist}. 
In this context, reliability becomes the most critical request since, differently from IT systems where the main concerns are about the confidentiality of the data, for an ICS instead, the availability is fundamental since it can guarantee human safety and fault tolerance~\cite{sans2019}.
For instance, in a nuclear plant environment, data availability (e.g., the temperature of the core) is more important than its confidentiality~\cite{Neitzel2014}.

Since these systems control physical and sometimes dangerous processes, security is a fundamental need. However, in recent years several viruses attempting ICSs were identified.
One of the first cyberattack targeting SCADA systems dates back 1982~\cite{miller2012survey} when a trojan targeting the Trans-Siberian pipeline causes a massive explosion. In successive years, many incidents exposed the security weaknesses of ICSs. Stuxnet~\cite{Falliere2011stuxnet, StuxnetNYT} is probably the most famous malware discovered in this field. Stuxnet was a worm discovered in 2010 targeting Programmable Logic Controllers (PLCs) used in gas pipeline and power plants. It was able to cause self-destruction of 984 centrifuges in a uranium-enrichment plant in Iran. In 2014, the third version of a known trojan family, BlackEnergy~\cite{Shrivastava2016blackenergy}, was developed to target ICSs. 
In the following years, this trojan was spread mainly inside a Microsoft Word document that, once open, request to activate macros that hide the virus. 
Victims of these attacks are media and energy companies, mining industries, railways, and airports in Ukraine. On December 23, 2015, an attack employing BlackEnergy3 caused a three-hour disconnection of 30 substations in the Kyiv Power Distribution company, leading to several hours of blackouts in the area.
More recently, in 2017, another important malware, TRITON~\cite{Pinto2018triton}, was identified after an unscheduled shutdown of a Saudi Arabian petrochemical processing plant. 
TRITON reprograms some special PLCs used for safety purposes, causing them to enter a failed state.

According to a report of Kaspersky Lab~\cite{KasperskyLab2016}, in the second half of 2016 the 39.2\% of the industrial machines secured by Kaspersky's products have been attacked, a clear sign that threats to ICS are a growing problem nowadays. The vulnerabilities affecting these systems are also reported in recent studies on real ICS traffic over the Internet~\cite{barbieri2020sorry, nawrocki2020uncovering}, showing a dramatic lack of security features on the communication.


A successful attack on ICS implied a huge economic impact on the organization. These consequences include operational shutdowns, damage to the equipment, business waste, intellectual property fraud, and significant health and safety risks. Nozomi Network reports that known shutdown events of an ICS~\cite{web:nozomi_cost} due to an attack cost from 225K\$ up to 600M\$.
An increasing attack trend against ICS is Ransomware, which aims to obtain economic rescue~\cite{web:ibm_rans}.
According to Coveware~\cite{web:ransomware_coverware}, in Q4 of 2019, the average ransom payment increased by 104\% to 84,116\$, up from 41,198\$ in Q3 of 2019. One of the most recent Ransomware is EKANS, which was discovered targeting 64 ICS~\cite{web:ekans}.

To prevent such catastrophic events, it is fundamental to implement novel security-by-design approaches, and where it is impossible to apply them, prevention or mitigation techniques must be integrated.
However, to develop a new security-by-design concept, it is required a complete testing infrastructure.
Generally, researchers rely on scaled-down versions of a real ICS, created ad-hoc to reproduce real-world systems but in a controlled environment, called \textit{testbed}.
Testbeds could be based on physical devices to provide reliable data at the cost of being more expensive or virtual if the application does not require exact measures.
However, the development of a new testbed is not straightforward, instead is challenging from different points of view, ranging from implementation costs, sharing capability, and fidelity (Section~\ref{sec:testbed}).

To develop prevention and mitigation techniques, nowadays, researchers involve machine learning techniques that exploit big amounts of data to train classification algorithms to detect misbehavior or potential attacks.
The straightforward approach to collect data is to record and provide to researchers data from real ICSs. However, since these systems are generally critical and fundamental for society, this strategy can be challenging in many aspects. 
For instance, it is difficult, if not impossible, to deploy attacks in a real environment because they can damage the physical process or some devices. Moreover, privacy is a problem: private companies could be reluctant to share system data from their ICSs. In fact, disclosing this data can cause theft of intellectual property or reveal the infrastructure's vulnerabilities, attracting malicious attackers' attention. From a testbed, it is possible to generate data and share them with other researchers to compare and improve different detection algorithms' results. These captures are called datasets and can be composed of physical measures (i.e., data from OT sensors) and/or network traffic (i.e., data from network communications). 
Datasets are an excellent testing solution due to their simplicity and availability. However, they are also challenging from many points of view, for instance, in the generation process and lack modularity (Section~\ref{sec:dataset}).

\subsection{Contribution}\label{subsec:contribution}

In this paper, we present a comprehensive survey specifically targeting the security research platform in the ICS field. This work aims to collect all the information related to the testbed and dataset to support future research and studies on this sector. 
Furthermore, for each dataset identified, we report the best score achieved by an Intrusion Detection System (IDS) in terms of F1-Score, Accuracy, and Precision, which are the most common metrics.
We have accurately analyzed all the testbeds, datasets, and IDSs to provide the readers with an exhaustive overview of the current ICS state of the art. 
The paper aims to assist interested readers: (i) to discover the different testbeds and datasets which can be used for security research in ICS with a description of the design key points, (ii) to have a clear baseline when developing an IDS on a particular dataset, and (iii) to understand the challenges and the good practices to keep in mind when designing an ICS testbed or dataset. 

We summarize our main contributions as follows:
\begin{itemize}
    \item We provide a comprehensive background on ICS, which offers an overview on the reference architecture and the main components characterizing such systems;
    \item We present the most employed industrial communication protocols with a particular focus on the intrinsic security features and proposed security expansions of each protocol; 
    \item We provide an exhaustive overview of the current ICS state of the art by analyzing different testbeds, dataset, and IDS related to the ICS field available on literature to provide the reader with key points design concepts;
    \item We offer the reader an exhaustive survey of the different testbeds and datasets which can be used for security research in ICS;
    \item We describe the best performing IDS developed for the presented datasets. During the development of this work, we noted a lack of a defined methodology to test the detection frameworks (e.g., testing the IDS on the single attacks or the whole dataset) and a defined baseline to compare the developed IDS.
    We believe that this baseline can offer a starting point for future researchers to begin working on ICS security having a clear idea of the current state of the art direction and trend;
    \item Finally, we offer a review of the challenges and the good practices to keep in mind when designing an ICS testbed, dataset, or IDS, with some insight into the future directions useful to fill the field gaps.
\end{itemize}

To continue collecting the testbeds and datasets in the future and sharing them with the community, we also developed a website (Section~\ref{sec:testbed}) to support the resource sharing among the researchers in this field.

\subsection{Survey Organization}\label{subsec:organization}

The remainder of this paper is organized as follows. In Section~\ref{sec:related} we provide an overview of the previous survey on this field, highlighting how we differ from them. 
In Section~\ref{sec:ics_background} we provide background on Industrial Control Systems describing the reference architecture and the common devices employed. In Section~\ref{sec:protocols} we describe the most typical protocols for ICS communication, highlighting the main characteristics, security features and offering an analysis of their diffusion in the market. 
In Section~\ref{sec:ids_back} we briefly recall the concept of Intrusion Detection System, also describing conventional attacks implemented on ICSs. 
Then, in Section~\ref{sec:testbed} and Section~\ref{sec:dataset} we describe and analyze respectively the different testbeds and datasets present in the literature.
In Section~\ref{sec:good}, some advice and good practices are illustrated both for researchers that use these technologies both for institutions that want to create brand new datasets or testbeds. Finally Section~\ref{sec:conclusion} concludes the paper.

\section{Related ICS Surveys}\label{sec:related}


Literature includes different surveys comparing testbeds and datasets created for applications in the ICS field. However, to the best of our knowledge, no detailed analysis gathers and describes both the ICS datasets and testbeds, but also the main IDSs implemented on them. For every dataset, we also report the algorithm with the best performances and the most interesting and innovative detection approaches. We believe that this could be useful for future research in this field and set a baseline to compare the different detection results.

In 2015, Holm et al.~\cite{Holm2015survey} proposed a complete revision of several papers related to ICS testbeds. The authors then focused on the objective and the component's implementation of 30 different testbeds. Furthermore, the authors provided an analysis of each testbed's main requirements (i.e., fidelity, repeatability, measurement accuracy, and safe executions).
The paper's main scope was to provide an overview of the actual state of such systems' development without detailing each single testbed composition. In fact, except for a table that indicates each testbed's location, all the others show only aggregated information. 
Moreover, since the paper is not recent, some of the presented testbeds are quite old and not widely used nowadays, while others that were only designed have never been made (e.g., \cite{Christiansson2008europe}).

McLaughlin et al.~\cite{McLaughlin2016}, in 2016, presented a complete survey on state of the art in ICS security. The paper briefly introduces the ICS operation's key principles and the history of cyberattacks targeting ICSs. The authors then addressed the vulnerability assessment process, outlining the cybersecurity assessment strategy advised for ICS and providing a list of steps to study the security and the vulnerabilities of an industrial system. In the end, the authors focused on new attacks and mitigation techniques. Moreover, the paper briefly presents a small list of some testbeds which can be used for security research in this field. Differently, our work is less focused on providing a complete landscape on ICS security, while it offers a more in-depth analysis and review of all the most used testbeds and datasets for ICS security research.

In 2017, Cintuglu et al.~\cite{Cintuglu2017} presented a comprehensive survey focused on smart grid testbeds, providing a systematic study with a particular focus on their domains, research goals, test platforms, and communications infrastructures. There are some intersections between smart grid and ICS fields, such as some used components and protocols.
Nevertheless, some different concepts require a separate ICS analysis, like the specific applications and sensors used, combined with the complexity of smart grids. 
To classify smart grid testbeds, the authors provide different possible taxonomy. Some of them can be applied to the entire ICS field (e.g., platform type). Instead, some others are specific to Smart Grid (e.g., NIST grid domain). The employed testbed classification in~\cite{Cintuglu2017} is mainly based on the research area, which motivates the development of each system. In this work, instead, we classify testbed mainly based on the platform type, providing the reader an overview of the most suited ICS testbeds and datasets for his research.

Recently in 2019, Geng et al.~\cite{Geng2019testbeds} presented a survey on ICS testbeds based on the same four requirements of~\cite{Holm2015survey}.
Besides analyzing different ICS datasets, the authors also present the different techniques that can be employed to build a testbed, including application scenarios, the main challenges, and future development directions. However, the authors' only introduced an analysis of each testbed's structure without going into details or providing comparison tables.

In the same year, Choi et al.~\cite{Choi2019datasetSurv} gathered and analyzed datasets for ICS security research, providing different comparison tables to understand the most suitable dataset depending on the case study. The authors based the comparison on the attack vector strategy.
The paper includes 11 commonly used datasets. Some existing datasets not widely used or without attacks data are intentionally not considered. However, even if not suited for anomaly detection tasks, the latter could be useful to study the ICS environment's behavior.
Also, one of the presented dataset (i.e., DEFCON23) is no longer available.

In 2020, in~\cite{ani2020design} the authors present an exhaustive survey with guidelines and good practices to help the building of an ICS testbed, highlighting the main challenges and the results of a focus group involving security experts to identify relevant design factors and guidelines. In the same years, the same authors published~\cite{green2020ics}, with interesting guidelines for each ICS layer and a set of characteristics to consider when outlining testbed objectives, architecture, and evaluation process. While these works are interesting and give a comprehensive insight into the process of designing and evaluating a testbed, they do not consider datasets and IDSs, their requirements, and relationships.

Our survey aims to collect all the platforms (i.e., testbeds and datasets) useful for ICS security research. We base the existing literature to provide a detailed analysis of the current research issues, challenges, and future directions characterizing this field. We also report the best performance of the IDS on every dataset, which can be helpful for future IDS research baseline.

\section{Industrial Control Systems}\label{sec:ics_background}

In this section, we offer a background on ICSs, useful when start approaching this field and to understand the remainder of this paper. Firstly, in Section~\ref{subsec:ics_arc}, we focus on the architecture of such systems compared to the classical systems architecture. Then we present a summary of the widely used ICS components in Section~\ref{subsec:ics_comp}.

\subsection{ICS Architecture}\label{subsec:ics_arc}
Industrial Control Systems (ICSs) are composed of the interconnection of different computers, electrical and mechanical devices used to manage physical processes. These systems are usually very complex and include heterogeneous hardware and software components such as sensors, actuators, physical systems and processes being controlled or monitored, computational nodes, communication protocols, Supervisory Control And Data Acquisition (SCADA) systems, and controllers~\cite{Khorrami2016}. 
Control can be fully automated or may include a human in the loop that interacts via a Human Machine Interface (HMI). 
ICSs are widespread in modern industries (e.g., gas pipeline, water treatments) and critical infrastructures (e.g., power plant and railway).

Unlike classical IT systems, ICSs are composed of standard network traffic over TCP/IP stack and data from physical processes and low-level components. This interconnected and intertwined nature can open a wide space for new generation attacks exploiting new vulnerabilities surfaces. 
Several protocols are used in ICS, based on the specific purpose of each system.
Industrial protocols are specifically designed to deal with real-time constraints and legacy devices in an air-gap environment. Many protocols do not implement any encryption or authentication mechanism due to these constraints, opening several vulnerabilities surfaces. Moreover, sometimes, the industrial protocols are customized from the company opening, again, many documentation and vulnerability issues.

\begin{figure}[t]
    \centering
    \includegraphics[width=\columnwidth]{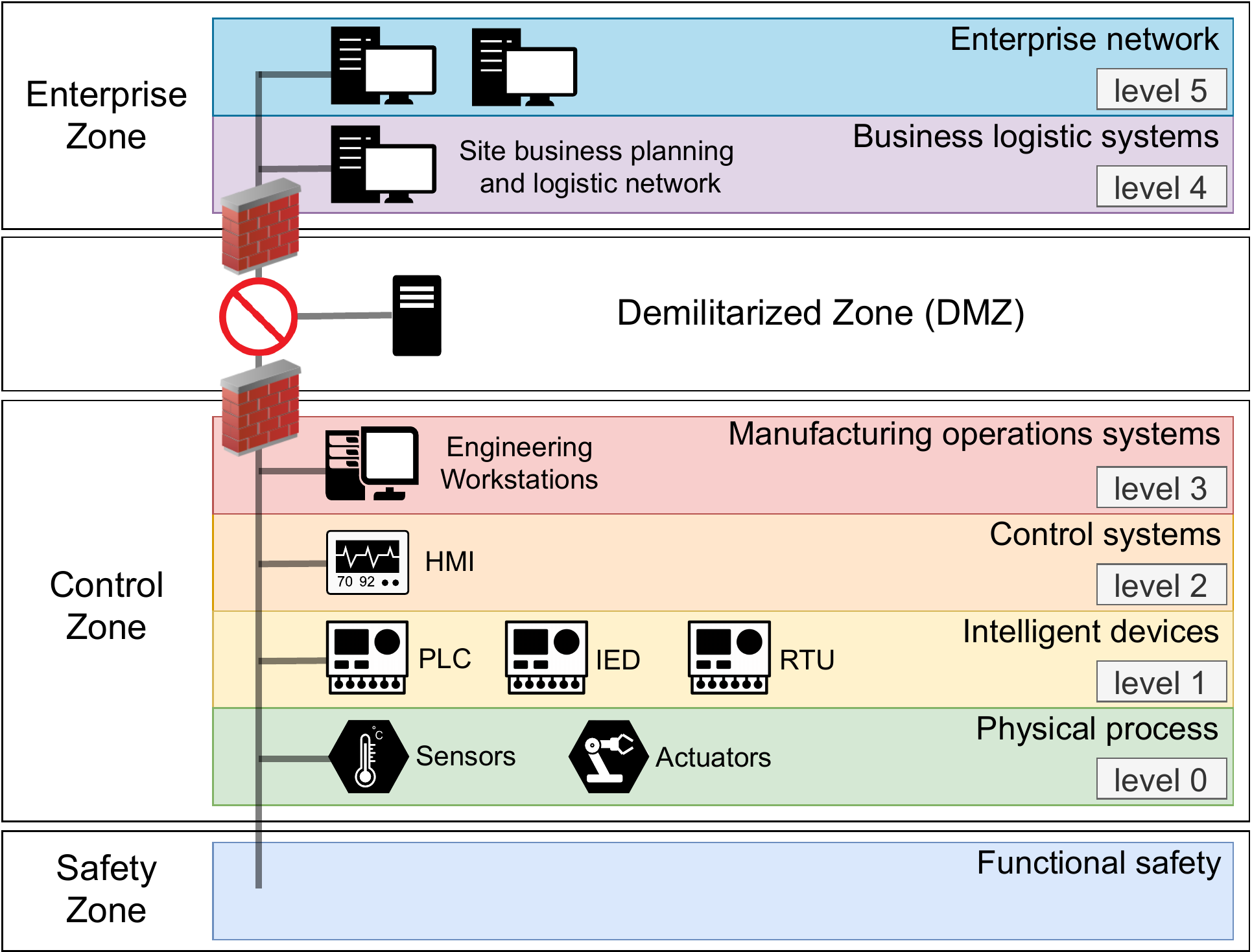}
    \caption{ICS Purdue Model architecture.}
    \label{fig:purdue}
\end{figure}

The reference architecture of the ICS is the Purdue Model~\cite{Geng2019testbeds, Obregon2014sans}. As depicted in Figure~\ref{fig:purdue}, the Purdue module divides an ICS network into logical segments with similar functions or similar requirements:

\begin{enumerate}

\item\textbf{Enterprise Zone}, or IT network, includes the traditional IT devices and systems such as the logistic business systems and the enterprise network.

\item\textbf{Demilitarized Zone (DMZ)} controls the exchange of data between the Control Zone and the Enterprise Zone, managing the connection between the IT and the OT networks in a secure way;

\item\textbf{Control Zone}, sometimes also referred to as OT network, includes systems and equipment for monitoring, controlling, and maintaining the automated operation of the logistic and physical processes. It is divided into four sub-levels:
\begin{itemize}
    \item \emph{Level 0} includes sensors and actuators that act directly on the physical process;
    \item \emph{Level 1} includes intelligent devices such as PLC, Intelligent Electronic Device (IED), and Remote Terminal Units (RTU);
    \item \emph{Level 2} includes control systems such as Human Machine Interfaces (HMI), alarms, and control room workstations;
    \item \emph{Level 3} includes manufacturing operation systems that are often responsible for managing control plant operations to produce the desired end product;
\end{itemize}
Level 2 and Level 3 devices can communicate with the Enterprise Zone through the DMZ.
\item\textbf{Safety Zone} includes devices and systems for managing ICS security by monitoring for anomalies and avoiding dangerous failures;
\end{enumerate}

The role of the DMZ is to filter the internal communication of the network. In fact, according to the Purdue model, all the traffic Exchanged between OT and IT networks must pass through the DMZ. However, this is rarely respected in the real-world, mainly due to the implementation difficulty or, more generally, the companies' insufficient attention to the industry building phase's security aspects. This condition exposes the critical part of the system (i.e., OT network) to potential attacks.

Compared with the classical IT environment, ICSs need a different risk handling strategy. The reliability is fundamental, and outrages are not tolerated due to the critical nature of the processes monitored, unlike IT, where occasional failures are acceptable. 
The risk impact is also different: in the IT environment, the principal risk is the compromising of privacy and confidentiality (e.g., loss or unauthorized alteration of data).
Instead, in the OT environment, a data compromise can cause a loss of production, equipment, and, in the worst case, a loss of lives or environmental damage. 

Another difference with respecting traditional IT systems relies on information handling performance: in an IT environment, the throughput must be high enough, while delays and jitter are accepted. On the other hand, in the industrial field, communication is defined with a regular polling time. Generally, this polling time is in second or millisecond orders, but delays are serious concerns. Finally, in IT systems, recovery can be made by rebooting. In contrast, in the OT system, fault tolerance is essential since a reboot would imply shutting down the entire industry and can lead to enormous economic losses~\cite{Christiansson2008europe}.

For all these reasons, and considering that nowadays most of the ICS are connected with the Enterprise zone, it is essential to protect them using new and precise technologies.

\subsection{ICS components}\label{subsec:ics_comp}

Industrial Control Systems are composed of a wide range of heterogeneous devices and components with a specif role in the system. In this section, we briefly introduce the most common devices in the ICS fields. These devices are generally installed or simulated in the testbed to replicate the ICS environment. 

\paragbf{Programmable Logic Controller (PLC)} 
PLC is a microprocessor-controlled electronic device that reads input signals from sensors, executes programmed instructions using these inputs and orders from supervisory controllers, and creates output signals that may change switch settings or move actuators. PLC is generally the boundary between the OT network and the physical process. It is often rugged to operate in critical environmental conditions such as very high or low temperature, vibration, or in the presence of big electromagnetic fields. As with most ICS components, PLCs are designed to last more than 10-15 years in continuous operations. The Real-Time Operations System (RTOS) installed in each PLC makes it suited for critical operations. The time to read all inputs, execute logic, and write outputs must last only a few milliseconds. Modern PLCs may use a UNIX-derived micro-kernel and present a built-in web interface that makes the management more simple but exposing the device to new vulnerabilities. A PLC has a power supply, central processing unit (CPU), communications interface, and analog/digital input/output (I/O) modules that can be connected to sensors (input) or actuators (outputs). These components are generally connected to the local network to communicate with supervisory processes. Based on the manufacturing company and the user requests, these communications can happen through different mediums (e.g., serial, fiber optic, wireless) using different protocols. 

\paragbf{Remote Terminal Unit (RTU)}
An RTU is a microprocessor-controlled electronic device. Like PLC, it is designed for harsh environments and is generally located far from the control center, for instance, in voltage switch-gear. There are two types of RTUs: station and field RTUs. Field RTU receives input signals from field devices and sensors and then executes programmed logic with these inputs. It gathers data by polling the field devices/sensors at a predefined interval. It is an interface between field devices/sensors and the station RTU, which receives data from field RTUs and orders from supervisory controllers. Then station RTU generates outputs used to control physical devices like actuators. Both field and station RTU has a power supply, CPU, and digital/analog I/O modules. For communication with the control center, RTU uses WAN technologies such as satellite, microwave, unlicensed radio, cellular backhaul, GPRS, ISDN, POTS, TETRA, or Internet-based links.

\paragbf{Intelligent Electronic Device (IED)}
An IED is a device containing one or more processors that can receive or send data from an external source. Examples of IEDs are electronic multi-function meters, digital relays, and controllers. Thanks to the higher complexity compared to a PLC and RTU, IED can perform more operations. An IED can be used for protection functions like detecting faults at a substation or for control functions such as local and remote control of switching objects and provide a visual display and operator controls on the device front panel. Other functions can be related to monitoring (for instance, a circuit breaker condition), metering (e.g., tracking three-phase currents), and communications with supervisory components.

\paragbf{Engineering Workstation}
The Engineering Workstation is generally a desktop computer or server running a standard operating system hosting the various software for controller and applications. Engineers use this platform to manage the controllers.

\begin{figure}[t]
    \centering
    \includegraphics[width=\columnwidth]{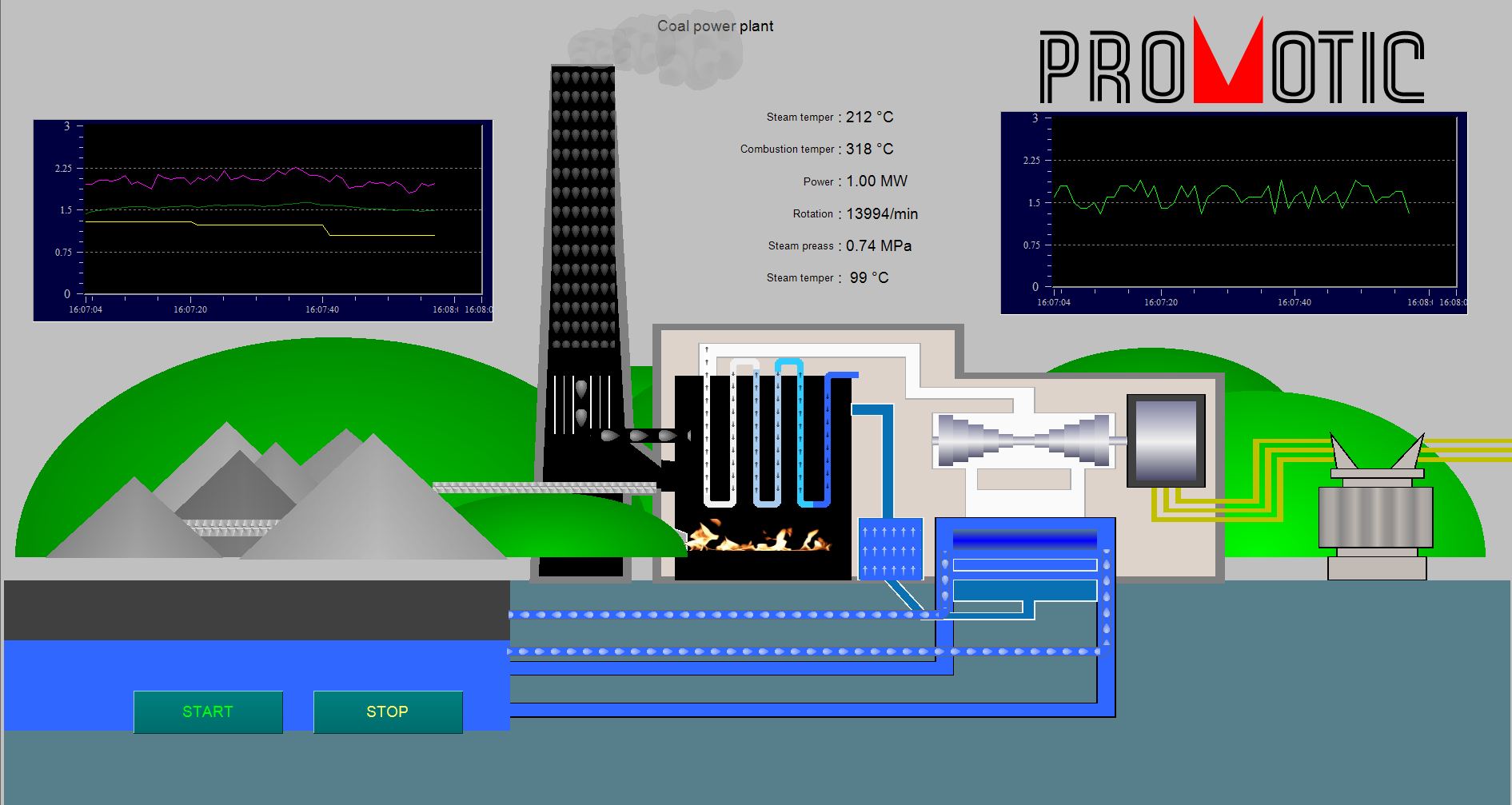}
    \caption{An example of HMI interface generated with Promotic Open-Source Tool~\cite{web_Promotic}.}
    \label{fig:hmi}
\end{figure}

\paragbf{Human Machine Interface (HMI)}
The HMI is a software installed on desktop computers, tablets, smartphones, or dedicated flat panel screens that permit operators to check and monitor the automation processes. 
As illustrated in Figure~\ref{fig:hmi}, 
the HMI shows the state of a plant operator, such as process values, alarms, and data trends. An HMI can monitor multiple process networks and several devices. An operator can use the HMI to send manual commands to controllers, for instance, to change some values in the production chain. Generally, the HMI shows a diagram or plant process model with status information to facilitate such a job.

\paragbf{Data Historian}
A Data Historian is a software application used to collect real-time data from the processes and aggregate them into a database for analysis. Data Historian mainly collects the same information shown in an HMI. The database and the hardware, generally a desktop workstation or a server, is designed for a very fast ingest of data without dropping data and uses industrial interface protocols.

\paragbf{Front End Processor (FEP)}
The FEP is a dedicated communications processor used to poll status information from multiple devices to give operators the possibility to monitor the system's overall status.

\paragbf{Communications Gateways}
A Communications Gateway is essential to make communications possible between devices from different manufacturers that use different protocols. Gateways can translate packets from a sending system to the receiver protocol.

\paragbf{Master Terminal Unit (MTU)}
MTUs manage the communication with the RTU or the PLC, gathers data from the PLCs, and process them. The communication between the MTU and the PLCs is bidirectional, but only the MTU can initiate the communication. Therefore, the MTU uses a master-slave communication where the MTU is the Master and PLCs are the slaves. Messages from the MTU to the PLCs can be triggered by an operator or be automatically triggered. These messages can either read memory parts representing current values like water flow, oil pressure, the temperature of a tank, or either write values in the memory and modify the configuration. 

\paragbf{Supervisory Control And Data Acquisition (SCADA)}
SCADA devices are placed on the higher level of the ICS hierarchy and are used to monitor and control centralized data acquired from different field sites. Furthermore, they manage the communication between the various devices and represent the remote connection point for the remote operators with the OT network. 
Over the year, SCADA systems protocols moved from proprietary standards towards open international standards, resulting in attackers knowing precisely the protocols. That is why there is a gain of interest in reinforcing industrial control systems security.

\begin{figure}[t]
\centering
\includegraphics[width=\columnwidth]{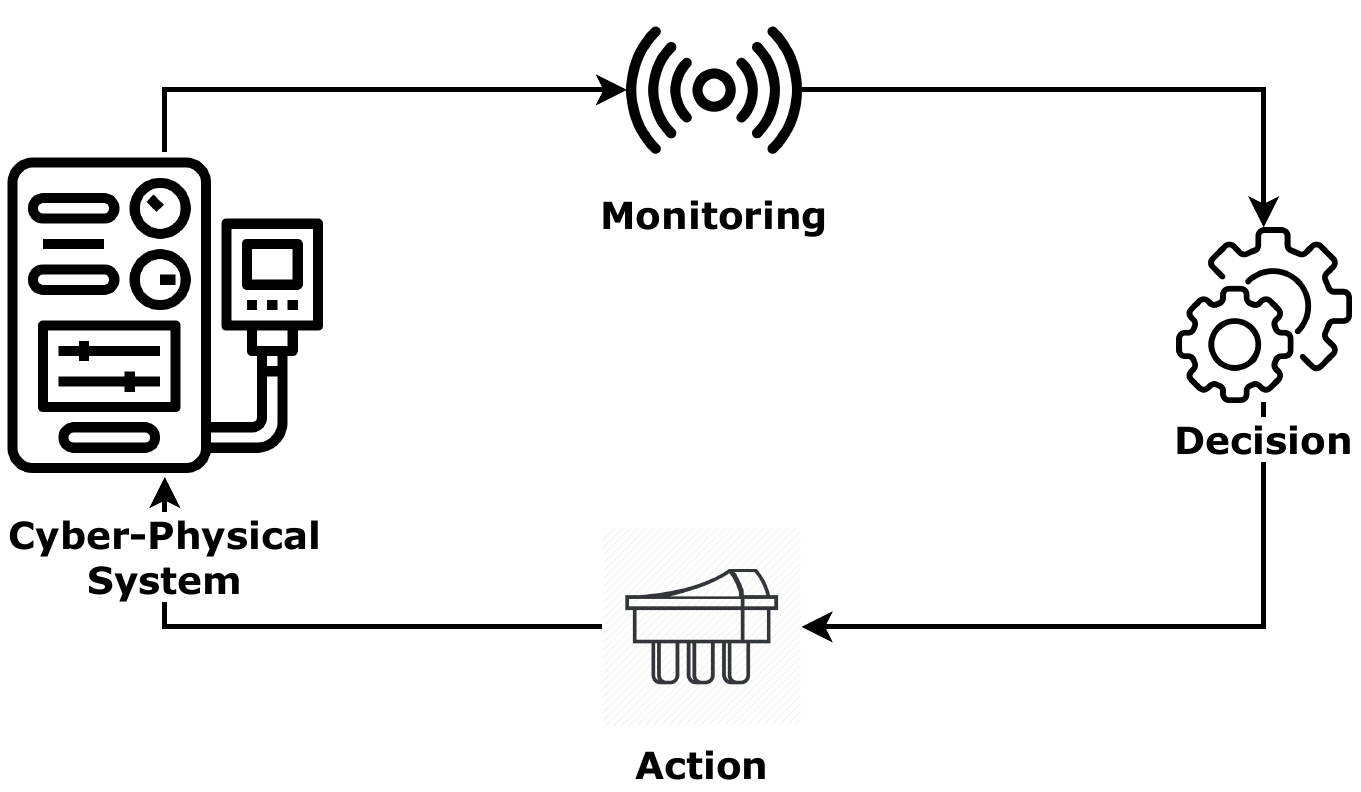}
\caption{The CPS close loop.}
\label{fig:close_loop}
\end{figure}

\paragbf{ICS Field Devices}
Field devices include all the components that are in direct contact with the physical process. The controllers can use them to get information regarding the physical process (e.g., the measure of temperature or pressure using sensors). Instead, actuators can interact with a physical process following commands from a controller (e.g., control motors, pumps, valves, turbines, agitators). The communication with the controllers is generally performed via I/O modules. 
Field Devices are implemented in the so-called CPS closed-loop to perform the three CPS main functions: monitoring using sensors, making decisions using PLCs, and applying actions using actuators. These three functions operate within a feedback loop covering, as shown in Figure~\ref{fig:close_loop}.

To sum up, controllers such as PLC, RTU, and IED are mainly used to interact with the Field Devices that can instead directly operate on the processes. HMI, Front End Processor, Engineering Workstation, and Data Historian are used to control and manage the system data. Instead, SCADA and Gateways Communications are used~\cite{Colbert2016} to set up all the connections between different components.

\section{Industrial Protocols Security}\label{sec:protocols}

With the growth of the ICSs, several new protocols have been developed to support the specific requirements of the OT environment, like fault tolerance and reliability. The majority of these protocols were designed to operate in an air-gapped environment. Therefore originally, less importance was given to the security aspects with respect to the real-time constraint. Some of them have no security features at all (e.g., Authentication, Encryption). However, after the IT and OT convergence, they have still been used in practice~\cite{barbieri2020sorry}.
This section reports the main industrial protocols focusing on the security proprieties initially and currently implemented.
Standards like PowerLink Ethernet, EtherCAT, Constrained Application Protocol (CoAP), Message Queue Telemetry Transport (MQTT), ZigBee PRO, WirelessHART, or ISA100.11.a are not used in the datasets and testbeds identified in this paper. Therefore we decided not to report them. However, some of these protocols are widely used in different fields, for instance, in Industrial Internet of Things (IIoT) scenarios.

Table~\ref{tab:sec_protocols} summarized the main information related to the protocols presented in this section. It includes:
\begin{itemize}
    \item Name of the \textbf{Manufacturer};
    \item Standard \textbf{Ports} according to IANA~\cite{web:IANA};
    \item Information related to the \textbf{Original} protocol:
    \begin{itemize}
        \item \textbf{Name} of the protocol;
        \item \textbf{Year} of release;
        \item A tick if \textbf{Encryption}, \textbf{Integrity}, or \textbf{Authentication} are available;
    \end{itemize}
    \item Information related to the \textbf{Enhancement} version with security measures offered by the manufacturer:
    \begin{itemize}
        \item Name of the new \textbf{Version} of the protocol;
        \item \textbf{Year} of release;
        \item A tick if \textbf{Encryption}, \textbf{Integrity}, or \textbf{Authentication} are available.
    \end{itemize} 
\end{itemize}

\subsection{Industrial Protocols}

\paragbf{Modbus}
Modbus is a serial communication protocol initially published by Modicon (now Schneider Electric) in 1979 for use with its programmable logic controllers (PLCs). Today Modbus~\cite{Modbus2006} is one of the most used and famous protocols in the ICSs. During the years, various versions of Modbus have been released. The first version was thought for serial communications, allowing to establish asynchronous serial communications on RS-232 and RS-485 interface. Modbus is also adapted to transmission means other than copper, such as optical fiber and radio links.
A typical communication via Modbus consists essentially of three stages: the formulation of a request from one device to another, the execution of the actions necessary to satisfy the request, and the resulting information's return to the initial device.  
This approach's main advantage lies in the interaction mode between the various network nodes: being a client-server type, each server device can exchange data simultaneously with more than one client.
The Modbus/TCP variant is substantially identical to the original serial version, but with the addition of a TCP/IP encapsulation module.


\paragit{Security} Implementations of serial Modbus use both RS232 and RS485, which are physical layer communication protocols. It makes no sense to speak of security on this layer, as these are functionalities developed on higher layers. 
Modbus was designed to be used in environments isolated from the Internet regarding the application layer's security. Therefore it does not include any security mechanism on this layer. 
These deficiencies are magnified by the fact that Modbus is a protocol designed for legacy programming control elements like remote terminal units (RTUs) or PLCs making the injection of 

{\renewcommand{\arraystretch}{1.3}
\onecolumn
\begin{center}
\begin{longtable}{|l|l|l|l|c|c|c|l|l|c|c|c|}
\caption{Summary of the main protocol's characteristics and them security measures implemented in the enhancement extensions. In particular, the table shows \textbf{E}: Encryption; \textbf{I}: Integrity; and \textbf{A}: Authentication. The Enhancement part refers to the version proposed by the manufacturer.}\label{tab:sec_protocols} \\
\hline
\rowcolor[HTML]{EFEFEF} 
                      &                & \multicolumn{5}{c|}{\cellcolor[HTML]{EFEFEF}\textbf{Original}}                                                              & \multicolumn{5}{c|}{\cellcolor[HTML]{EFEFEF}\textbf{Enhancement}}                                                                 \\ \cline{3-12} 
\rowcolor[HTML]{EFEFEF} 
\textbf{Manufacturer} & \textbf{Ports} & \textbf{Name} & \multicolumn{1}{l|}{\cellcolor[HTML]{EFEFEF}\textbf{Year}} & \textbf{E} & \textbf{I} & \textbf{A} & \textbf{Version}    & \multicolumn{1}{l|}{\cellcolor[HTML]{EFEFEF}\textbf{Year}} & \textbf{E} & \textbf{I} & \textbf{A} \\ \hline

\endfirsthead

Schneider Electric    & 502/802        & Modbus        & 1979                                                       &               &               &                & Modbus/TCP Security & 2008                                                       & \checkmark    & \checkmark    & \checkmark     \\ \hline
GE Harris             & 20000          & DNP3          & 1990                                                       &               &               & \checkmark     & DNP3-SA             & 2020                                                       & \checkmark    & \checkmark    & \checkmark     \\ \hline
Siemens               & 102            & S7Comm        & 1994                                                       &               &               &                & S7CommPlus          & 2014                                                       & \checkmark    &               & \checkmark     \\ \hline
PROFINET Int.                    & -              & PROFINET      & 2003                                                       &               &               &                & PROFINET Security Classes     & 2019                                                       & \checkmark    & \checkmark    & \checkmark     \\ \hline
ODVA                  & 44818          & EtherNet/IP   & 2000                                                       &               &               &                & CIP Security        & 2015                                                       & \checkmark    & \checkmark    & \checkmark     \\ \hline
OPC Foundation        & 4841           & OPC           & 1996                                                       &               &               &                & OPC-UA              & 2006                                                       & \checkmark    & \checkmark    & \checkmark     \\
\hline

\end{longtable}

\end{center}

{\renewcommand{\arraystretch}{1.3}
\begin{center}
\begin{longtable}{|>{\columncolor[HTML]{EFEFEF}}l|c|c|c|c|c|c|l|}
\caption{Protocol used in the presented testbeds and datasets. $\bullet$ indicates that the protocol is supported/available. In some work it is not indicated the version of Modbus (TCP, RTU, or ASCII) adopted. In such cases, a $\bullet$ is used to indicate a general version of the protocol.}\label{tab:protocol}\\

\hline
\textbf{Name}                               & \cellcolor[HTML]{EFEFEF}\textbf{Modbus} & \cellcolor[HTML]{EFEFEF}\textbf{S7Comm} & \cellcolor[HTML]{EFEFEF}\textbf{EtherNet/IP} & \cellcolor[HTML]{EFEFEF}\textbf{DNP3} & \cellcolor[HTML]{EFEFEF}\textbf{Logs} & \cellcolor[HTML]{EFEFEF}\textbf{Phy.} & \cellcolor[HTML]{EFEFEF}\textbf{Others} \\ \hline
\endfirsthead

    \multicolumn{8}{c}%
    {{\bfseries \tablename~\thetable{} -- continued from previous page}} \\
    \hline
\textbf{Name}                               & \cellcolor[HTML]{EFEFEF}\textbf{Modbus} & \cellcolor[HTML]{EFEFEF}\textbf{S7Comm} & \cellcolor[HTML]{EFEFEF}\textbf{EtherNet/IP} & \cellcolor[HTML]{EFEFEF}\textbf{DNP3} & \cellcolor[HTML]{EFEFEF}\textbf{Logs} & \cellcolor[HTML]{EFEFEF}\textbf{Phy.} & \cellcolor[HTML]{EFEFEF}\textbf{Others} \\ \hline
\endhead

\textbf{4SICS}                   & TCP                                     & $\bullet$                               & $\bullet$                                    & $\bullet$                             &                                       &                                           &                                         \\ \hline
\textbf{Aghamolki et al.}        & $\bullet$                               &                                         &                                              & $\bullet$                             &                                       &                                           & IEEE-C37.118                            \\ \hline
\textbf{Ahmed et al.}            & $\bullet$                               &                                         & $\bullet$                                    &                                       &                                       &                                           & Profinet                                \\ \hline
\textbf{Alves et al.}            & TCP                                     &                                         &                                              &                                       &                                       &                                           &                                         \\ \hline
\textbf{BATADAL}                 &                                         &                                         &                                              &                                       &                                       & $\bullet$                                 &                                         \\
\hline

\end{longtable}

\end{center}
} 

\begin{figure}[b]
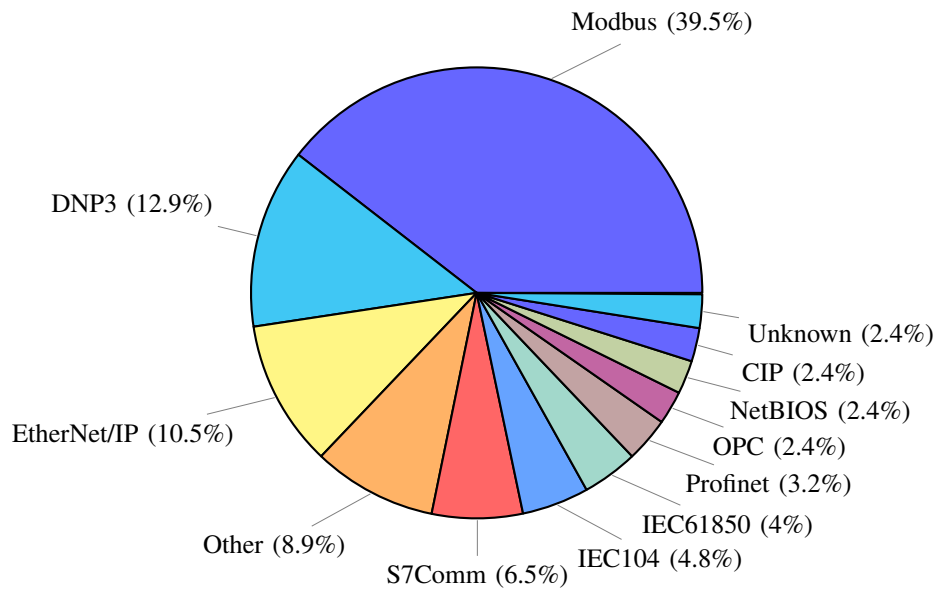

    \centering
    \caption{Percentage distribution of different protocols in the datasets and testbeds analyzed.}
    %
\end{figure}
\twocolumn
} 

\noindent malicious code into these elements easier. Modbus Security~\cite{ModbusSecurity2018} offers a Modbus/TCP version enhancement focused on using the port 802. This new version enables TLS to provide confidentiality, integrity, and authentication using x.509v3 certificates. Moreover, it specified certificate-based authorization using role information transferred via certificate expansions.
Researchers have also proposed different modifications to introduce confidentiality~\cite{shahzad2015real} or authentication~\cite{bernieri2020tambus} via covert-channel on Modbus.

\paragbf{DNP3}
Westronic, Inc. (now GE Harris) designed DNP in 1990. In January 1995, the \textit{DNP Users Group Technical Committee} was formed to review enhancements and recommend them for approval to the general Users Group. One of the most important tasks of this body was to publish the ``DNP Subset Definitions'' document, which establishes standards for scaled-up or scaled-down implementations of DNP3~\cite{DNP32002}. DNP3 is an open, intelligent, robust, and efficient SCADA protocol organized into four layers: physical, data link, pseudo-transport, and application. In serial implementations, commands are issued broadcast. DNP3 contains significant features that make it more robust, efficient, and interoperable than older protocols such as Modbus, at the cost of higher complexity. The protocol's primary goal is to maximize system availability by putting less care into confidentiality and data integrity factors.
DNP3 organizes data into data types such as binary inputs/outputs, analog inputs/outputs, counters, time and date, file transfer objects.

\paragit{Security} As previously mentioned, DNP3 is a protocol designed to maximize system availability by putting less care into confidentiality and data integrity factors.
Data link level includes the detection of transmission errors through Cyclical Redundancy Check (CRC) calculation. However, CRC is not a proper security measure since if an attacker can modify a packet, he/she can also change the CRC. 
At the application level, some efforts have been made to provide a safe authentication standard in DNP3. While in the beginning, pre-shared keys were used to authenticate, according to the standard IEEE 1815-2010 (deprecated), the latest versions implement Public Key Infrastructure (PKI) with remote key changes (standard IEE 1815-2012).
Recently, in 2020, GE Harris presents DNP3 version 6, introducing DNP3-SA~\cite{DNP3Seucirty2020}, a separate protocol layer that supports Message Authentication Codec (MACs) to provide secure communication sessions, including authentication and integrity. Moreover, this version supports encryption to offer data confidentiality by using the AES-256 algorithm.
Some other solutions have been proposed in the literature to implement cryptography protections, such as end-to-end encryption~\cite{bagaria2011flexi} and VPN for IP networks~\cite{majdalawieh2007dnpsec}.

\paragbf{S7Comm}
Introduced in 1995, S7comm (S7 Communication)~\cite{Kleinmann2014s7} is a Siemens proprietary protocol that runs between standard PLCs of the Siemens S7-200/300/400 family and new generation PLCs like S7-1200/1500. 
It is a proprietary and closed standard without significant literature related to it.
Siemens has a proprietary HMI software for the SIMATIC products and an Ethernet driver that provides connectivity to devices via the Siemens TCP/IP Ethernet protocol. In addition to this driver, there are also 3rd-party communication suites for interfacing and exchanging data with Siemens S7 PLCs.



\paragit{Security} S7Comm is a closed protocol, so there is no related documentation. However, as various works underline, the base version of S7Comm does not include security features, and it is vulnerable to replay attacks~\cite{lei2017spear}. However, in 2010 Stuxnet exploited the security vulnerabilities of S7Comm to compromise a Nuclear Plant in Iran. As a result of this incident, Siemens has developed a new version of the protocol, called S7CommPlus, with replay-attack protection. It has been proven that this version is also vulnerable to reverse debugging attacks~\cite{lei2017spear}.

\paragbf{PROFINET}
Developed by PROFIBUS \& PROFINET International (PI), PROFINET~\cite{Profinet2014} is an open standard for Industrial Ethernet standardized in IEC 61158 and IEC 61784. Introduced in 2003, it is an evolution of the PROFIBUS standard, whose lines can be integrated into the PROFINET system via an IO-Proxy. 
This protocol follows the provider-consumer model for data exchange in a cascading real-time concept. It is compatible with Ethernet thanks to its flexible line, ring, star structures, and copper and fiber-optic cable solutions. It is also compatible with radio communications such as WLAN and Bluetooth. Thanks to Ethernet-based communication, it provides a direct interface to the IT level.
The primary functions include a cyclic exchange of I/O data with real-time properties, acyclic data communication for reading and writing of demand-oriented data, including the identification and maintenance function, and a flexible alarm model for error signaling with three alarm levels.

\paragit{Security} PROFINET is a protocol operating in the application, link, and physical layers. The link layer in this protocol uses FDL (Fieldbus Data Link) to manage access to the medium. FDL operates with a hybrid access method that combines master-slave technology with the passing of a token, indicating who can initiate communication and occupy the bus. These measures ensure that devices do not communicate simultaneously. However, FDL constitutes any safety mechanism and may be susceptible to attacks involving traffic injection or Denial Of Service (DoS). 
In 2019, PI introduced three Security Classes to offer a way to select security measures based on the consumer needs~\cite{Profibus2019sec}. Class 1 improves robustness through a digital signing of General Station Description (GSD) files using a PKI infrastructure, an extended Simple Network Management Protocol (SNMP) configuration, and a DCP in read-only mode. Class 2 expands the previous class by offering integrity and authenticity via cryptographic functions and confidentiality only of the configuration data. Instead, Class 3 offers all the previous characteristics and the confidentiality of all the data.
Furthermore, it is worth mention that PROFIBUS offers some services that can use TCP/IP as a transport protocol, but only during an initial phase for device assignment. It is possible to add some of the classical TCP/IP cryptography and authentication security elements in these services.

\paragbf{ODVA’s networks}
Founded in 1995, ODVA~\cite{ODVA2016} is a global association whose members comprise the world’s leading automation companies with the mission of developing advance open and interoperable communication technologies for industrial automation. The primary interest is developing the Common Industrial Protocol (CIP), supporting the various network adoptions such as DeviceNet, CompoNet, ControlNet, and the widely used EtherNet/IP. CIP encompasses a comprehensive suite of messages and services to collect industrial automation applications such as control, safety, energy, synchronization, motion, information, and network management. This protocol allows users to integrate these applications with the IT Ethernet networks and the Internet. 
The protocol follows a model for objects: each one is made up of attributes (data), services (commands), connections, and behavior (the relationships between data and services). 
CIP also defines device types, with each device type having a device profile. The device profiles indicate which CIP objects must be implemented, what configuration options are possible, and the formats of I/O data.


EtherNet/IP is an adaption of CIP to the Ethernet TCP/IP stack, while DeviceNet provides a way to use CIP over the CAN technology. ControlNet uses CIP over a Concurrent Time Division Multiple Access (CTDMA) data link layer, and CompoNet implements CIP on a Time Division Multiple Access (TDMA) data link layer.

\paragit{Security} Recently, in 2015, ODVA introduced the CIP Security framework~\cite{CIPSecurityWeb} to provide security measures to CIP protocol. Since different systems might need different security levels, CIP Security provides different security specifications profiles to help users configuring inter-operable devices.
On EtherNet/IP, it enables TLS and DTLS to secure the TCP and UDP transport layer protocols. TLS and DTLS provide authentication of the endpoints using X.509 certificates or pre-shared keys, message integrity and authentication employing TLS message authentication code (HMAC), and optional message encryption.

\paragbf{Open Platform Communications (OPC)}
The classic OPC~\cite{Gonzalez2019}, developed in 1996, was designed to provide a communication protocol for personal computer-based software applications and automation hardware. It was based on Microsoft's distributed component object model, making them platform-dependent and not suitable for cross-domain scenarios and the Internet. 
Nowadays, the classic OPC is no anymore developed. In 2006 a new version, OPC United Architecture (OPC-UA), was released as an operational framework for communications in process control systems. It provides greater interoperability, eliminating MS-Windows dependency, but maintaining retro compatibility with its predecessor. This specification is built around Service-Oriented Architecture (SOA) and is based on web services, making it easier to implement OPC connections over the Internet.
The general layout of the communication is simple: the hardware devices (e.g., PLC, Controller) act as data sources, and the software applications (e.g., SCADA, HMI) play the role of data consumers, whereas the OPC interface acts as connectivity middleware, enabling the data flow. 
Using the OPC, the client applications access and manage the field information without knowing the physical nature of data sources. With OPC-UA improvements, the protocol is widely used in critical and industrial fields such as energy automation, virtualized environment, and building automation.

\paragit{Security} The use of Distributed Component Object Model (DCOM) and Remote Procedure Call (RCP) make OPC very susceptible to different attacks~\cite{rolston2006security}.
Since it is inherently difficult to apply patches to industrial control systems, many discovered vulnerabilities with available patches continue to be potentially exploitable industrial control networks.
Instead, OPC-UA implements a security model and five security classes, bringing greater security to the architecture at the cost of slightly higher complexity~\cite{Renjie2010ocpuaSec}. It is also possible to implement only a fraction of the security measures by using one of the five security classes provided. The security model allows generating a secure channel that provides encryption, signatures, and certificates at the communication layer. Furthermore, a session in the application layer is used to manage user authentication and user authorization. Thanks to these security measures, it is advisable to deploy OPC-UA rather than the classic version of OPC and to update the already deployed versions wherever possible.

\paragbf{IEC 60870-5-104 (IEC 104)}
Released in 2000, IEC 60870-5-104 (IEC 104) protocol~\cite{Clarke2004} is an extension of the IEC 101 protocol with the changes in transport, network, link, and physical layer services to suit the complete network access. 
The standard uses an open TCP/IP interface to network to connect to the LAN (Local Area Network), and routers with different facilities can be used to connect to the WAN (Wide Area Network). 
There are two different methods of transporting messages. The first provides bit-serial communications over low-bandwidth communications channels. In the second, introduced with IEC 104, the protocol's lower levels have been completely replaced by the TCP/IP transport and network protocols. 
Thanks to the IEC 104 simple structure in terms of its data types and data addressing options, it is possible to quickly achieve interoperability with other protocols.

\paragit{Security} IEC 104, has been proven to be vulnerable to different types of attacks, such as man-in-the-middle and replay attacks~\cite{maynard2014towards}. A more recent and secure standard of the IEC family is IEC 62351. This version implements end-to-end encryption to prevent attacks such as replay, man-in-the-middle, and packet injection. However, due to the higher complexity, industries rarely upgrade IEC 104 to IEC 62351.

\paragbf{IEC 61850}
Like IEC 104, IEC 61850~\cite{Baigent2013} was originally designed to enable communications inside substations automation systems. In recent versions, an extension of IEC 61850 allows substation-to-substation communication and provides tools for translation with other protocols such as IEC 60870-5, DNP3, and Modbus.
The protocol is devised using an object-oriented design suited for communication between devices of different vendors.
To provide long-term stability, IEC 61850 divides the information model and communication protocols. For not time-critical applications, the protocol uses MMS via TCP/IP as the communication protocol. Instead, GOOSE can be employed over Ethernet if the time constraint is critical. In the case of voltage and current sample information transportation, SV over Ethernet is generally used. However, recent versions of the standard provide GOOSE/SV mapping over TCP/IP using UDP packets at the transport layer for inter-substation information exchange.

\paragit{Security} IEC 62351 standard provides various security measures, offering guidelines and developing a secure operation framework. Since in time-critical application encryption is not suitable due to the 3ms delivery overhead, the standard recommends using digital signature generated by SHA256 and RSA public key algorithms.
For MMS communications instead, TLS is recommended, with optional end-to-end encryption of all the packets exchanged~\cite{Hussain2020iecSec}.
Furthermore, the employment of IEC 61850 in heterogeneous networks exposes the system to protocol mapping vulnerabilities. It is possible to prevent these vulnerabilities by developing ad-hoc security by design architectures~\cite{Yoo2016}.

\paragbf{Other protocols} In addition to previously presented protocols, some datasets contain few packets related to other generic protocols used in a wide range of applications. Published in 1995, BACnet is a data communication protocol for building automation and control networks supported by some HVAC components but not widely used~\cite{BACnetWeb}.
Distributed Computing Environment/Remote Procedure Calls (DCE/RPC) is a remote procedure call system that allows programmers to write distributed software as if it were on the same computer. One of the datasets presented in this paper includes DCE/RPC, together with NetBIOS, a networking protocol allowing applications on separate computers to communicate over a LAN. Other generic packets are visible in some datasets like Address Resolution Protocol (ARP) and Domain Name System (DNS) requests but are generally not related to the industrial field.

\subsection{Industrial Protocols Employment}\label{subsec:proto_empl}

In Table~\ref{tab:protocol}, we provide the complete list of the datasets and testbeds analyzed in this work, together with the protocol used in the specific platform. In detail, the table associates to each testbed the protocols supported and to each dataset the protocols available. Moreover, it indicates if data logs and physical measures are provided in the datasets.
As previously described, there are several different protocols employed in the ICS field. In Figure~\ref{fig:protocols_pie} we reported the percentage of usage of each protocol in the testbeds and datasets investigated in this survey. Modbus, and its different versions (i.e., TCP, RTU, ASCII), are the most used protocols, while EtherNet/IP, DNP3, and S7Comm follow with a lower but significant employments.


Since the testbed and dataset employed should represent an approximation of real-world scenarios, it is interesting to compare if the distribution of the protocols implemented in the different datasets is similar to the protocol's distribution in the real industrial system. Verifying this claim is a challenging task due to the various privacy concerns of companies in disclosing information. 
Various works tried to deal with this problem by measuring the industrial traffic present on the Internet. Although with limitations, the traffic measurement can represent a reasonable estimate of the most popular industrial protocols currently used.
By leveraging Censys search engine, Xu et al.~\cite{xu2018landscape} scanned the Internet for about two years (from 2015 to 2017), examining for industrial devices exposed. In particular, they focused on five protocols: Modbus, S7Comm, DNP3, BACnet, and Tridium Fox. 
Results show a significant prevalence of Modbus and Tridium Fox devices (with more than 20K devices found), a middle spread of BACnet (about 11K devices) and S7Comm (about 4.5K devices), and a lower number of devices using DNP3 (less than 1K). Furthermore, the authors noticed an increasing number of Modbus and S7Comm devices during the two years of recording, while the number of DNP3 devices decreased.
A similar study, presented by Barbieri et al.~\cite{barbieri2020sorry}, leveraged Shodan and an Internet Exchange Point (IXP) in Italy to measure ICS host exposure. They discover many devices using Modbus, MQTT, and Niagara Fox. Furthermore, the authors also identified EtherNet/IP, S7Comm, and BACNet devices but with significantly lower samples.

In addition to measurement works, we can also rely on market analysis. According to an HMS report~\cite{web:hms2020report}, the overall market share of Industrial Ethernet protocols increased in 2020. In particular, EtherNet/IP and Profinet obtained first place with 17\% of market share, while in third place there is EtherCat with a share of 7\%. On the other side, Fieldbus protocols such as Profibus and DeviceNet showed a decrease of 5\% on the market share with respect to the previous year. Interestingly that Modbus protocol (TCP and RTU variants), despite being heavily employed in testing, results in a 10\% of market share (5\% RTU, 5\% TCP).

Based on these findings, Modbus/TCP results as the most employed protocol in testbeds, datasets, and Internet measurements. Nevertheless, it obtains a low market share in the last year (i.e., 2020), outranked by EtherNet/IP, which is also employed in a significant part of the testbeds and datasets presented in this survey, and Profinet, which instead is used in only the 3.2\% of the testing system analyzed in this work. Therefore, Profinet can be an interesting protocol to introduce in future testbeds and datasets to follow the market trend. Other protocols that an increasing market share are BACnet, TridiumFox, and NiagaraFox, which are not present in the testing platforms, except for one dataset that contains BACnet packets. Finally, another protocol that could be interesting to include in testing platforms is EtherCAT, which has a 7\% of market share. However, no testing system currently supported it.

\section{ICS Attack and Defence}\label{sec:ids_back} 

In this section, we offer an overview of the various attacks and defense mechanisms in ICSs. 
In particular, in Section~\ref{subsec:attacks} we present an overview of the typical attacks which can target ICSs and are implemented in the different testbeds and datasets. Instead, in Section~\ref{subsec:IDS} we proposed a brief overview of the techniques which can be employed to detect and mitigate cyberattacks in this field.

\subsection{Typical Attacks}\label{subsec:attacks}

ICSs are extremely complex systems, which connecting IT components with sensors, actuators, and other OT devices. Such an interconnected scenario with a wide variety of various components may hide attack surfaces caused, for instance, by device-specific vulnerabilities or misconfigurations.

Having a clear idea of the different attack typologies is essential in building and testing defenses. Based on that, testbeds should be capable of simulating verisimilar attacks, while datasets should include not only normal operation data but also attack data. Reproducing attacks is a challenging task because the simulation should precisely emulate a realistic abnormal operating condition. However, it is impossible to replicate every type of attack due to the devices' potential damages. Some attacks could also shift the testbed's operating behavior in a dangerous state and seriously damage the machines. Furthermore, the limited class of attacks implemented could raise a generalization problem of the detection strategy, not transferable to novel and unknown attacks.

In a CPS scenario, by definition, there are two possible attacks vector surfaces on the system. \textit{Network-based} attacks, targeting the networking part of the network such as packets, protocols or routing policies, and \textit{Physical-based} attacks, aimed at corrupting the physical process of the devices. Sometimes, these two attack categories' goals may also converge or combine to reach a specific target.

\paragbf{Network Attacks}
The most common attack models include the Control Zone network access by the attacker to compromise an ICS. An attacker can obtain the network control through a phishing attack to the site operators~\cite{Pinto2018triton} or by exploiting the security lack of the legacy devices connected to the Internet~\cite{barbieri2020sorry}. There are different actions that malicious actors can perform, but it is possible to categorize the main ones into five different classes~\cite{Gomez2019electra, Rodofile2013} of network attack. These attacks are also implemented in the testbeds to generate abnormal operating conditions.

\begin{itemize}
    \item\textbf{Reconnaissance Attack} aims at the identification of potential victims within a network. Usually, this class of attack is used to plan other moves, such as identify other vulnerable devices. These attacks can be passive (e.g., port mirror) or active (e.g., nmap).
    
    \item\textbf{Man-in-the-Middle (MitM) Attack} allows an attacker to sit in the middle of communicating parties. The attacker is then able to read or modify the communications, inject commands, or drop packets. The final aims can range from the control of some devices to the disruption of the ICS's normal state to damage the system's owner or the system itself. 
    
    \item\textbf{Injection Attack} aims at supplying untrusted and malicious inputs to a system. Typically, in an ICS, an attacker can inject data such as false measures from sensors or actuators (Data Injection Attack) or command (Command Injection Attack). Often a compromised node launch this type of attack, but, in some cases, the injected data can originate from other sources (e.g., a new entry point for the network). 

    \item\textbf{Replay Attack} is based on the retransmission of a valid message that has been previously seen in the network. This attack is difficult to be detected, and it can lead to malfunctions of the system. For example, in a nuclear plant context, the attacker could retransmit a message with a low temperature of the reactor instead of rising, inhibiting the activation of safety measures.  
    
    \item\textbf{Denial of Service (DoS) Attack} is used to make devices unavailable by overloading the system resources to disrupt the communication between machines in the system. Usually, a common technique is packet flooding and, if packets are generated from many different sources, it is called Distributed Denial of Service (DDoS). This attack can stop some devices, making them unavailable and lead to unpredicted behaviors in the ICS.

\end{itemize}

\paragbf{Physical Process Attacks}
This class of attacks aims to alter the physical process and the complex relations of the system to manage it. Cyber-Physical Systems enable such attack surfaces due to the field device (i.e., sensors and actuators), sometimes in remote places. To achieve these attacks, the attacker could have previously obtained access to the system with one or more of the network attacks previously described. Generally, physical process attacks represent the final attack chain goal, which started with the network as an entry point.

\begin{itemize}
    \item\textbf{Stealth Attack} generates small perturbations in the system process to create long term damages (e.g., loss in production terms or the devices' degradation). The stealth attack can use a static perturbation, by introducing a constant error in the physical measure (e.g., increasing or decreasing the production), or dynamic, by rapidly oscillating between upper and lower measurement bounds (e.g., causing turbulence in the flows). This class of attacks is generally difficult to detect since it maintains the process inside its limits.
    
    \item\textbf{Device Manumission} is achieved by physically tampering with the field device to comprise the data recorded. This attack aims to induce wrong measurements in the system exploiting the distributed and, therefore, less monitored nature of these systems.
    
    \item\textbf{Direct Damage Attacks} aims to disrupt and damage the entire process or physical equipment by introducing significant process variations that bring the system into an unsafe state. This attack may also have severe consequences on the population or the environment around the site.
\end{itemize}

\subsection{ICS Defence Techniques}\label{subsec:IDS}

It is possible to enforce ICS security by implementing security-by-design network architectures. For instance, it is possible to use DMZ as specified in the Purdue Model (Figure~\ref{fig:purdue}), enforcing network separation and segregation. Furthermore, boundary protections and firewalls with ICS-specific rules help protect an ICS from external attacks. The National Institute of Standard and Technology (NIST) proposed a complete guide explaining how to set up a secure network to protect an ICS~\cite{Stouffer2015nist}. However, security-by-design can be challenging to consider in ICSs, due to implementation constraints. Sometimes, it could also happen that companies consider the security aspects after the construction phase. Detection mechanisms can solve this limitation and be integrated into the system, even if not always easy, after the construction phase, for instance, in central nodes or with network tap.



It is possible to deploy process-aware techniques to detect attacks that cannot be identified by code execution monitoring or other traditional methods used in IT environments. There are several techniques to do so, and the idea behind them is to exploit the massive amount of data collected from the sensors and predict an ICS's operations. 
Moreover, they represent a cost-effective solution since they can be installed without changing the system topology or substituting every network device.

In the following, we briefly report the two main categories of IDS, which employ two different approaches to detect attacks or domain drifts.

\emph{Knowledge-based} intrusion detection (also called \emph{misuse-based}) focus on looking for runtime features, such as physical values of sensors and actuators or network traffic, that match a specific pattern of misbehavior. This method aims to exploit the stationary of ICSs, which, unlike IT systems, are characterized by control loop operation regulated by a constant polling time communication.
However, these detection systems only react to known dangerous behavior, so there is no protection against zero-day vulnerabilities. For this reason, the research community is focusing on developing a dynamic mechanism that can identify domain shifts without the need for signatures~\cite{Mitchell2014ids}.

The current research trend focuses on the \emph{anomaly-based} intrusion detection, which looks for runtime features that differ from normal behavior. The normal behavior pattern can be defined using unsupervised approaches training the model with live data or semi-supervised utilizing a set of truth data. This approach is called \emph{behavior specification-based} intrusion detection. It represents a suitable ICS solution since it aims to dynamically learn the regular behavior model of network traffic and physical models.

This last method is promising thanks to modern machine learning and deep learning techniques that can be used for anomaly detection classification. A common requirement of these algorithms is the need for a considerable quantity of data: generally, the more data you provide to the training phase, the more precise your detection will be.

IDS are also classified according to the data source. \emph{Network-based} IDS uses network adapters to collect and analyze packets in real-time. On the contrary, \emph{host-based} IDS monitors the documents, processes, and other information specific to a particular device to identify. The disadvantage is that monitoring regard only one node in the network, while with the former approach, all the network is under control. On the other hand, \emph{host-based} can detect also threat coming from sources other than the network (e.g., USB sticks)~\cite{Hu2018ids}.
A novel detection design concept that exploits the correlation of multiple ICS points was proposed by Bernieri et al.~\cite{10.1145/3362743.3362961}. In this work, the authors proposed a distributed detection approach to consider the different information characterizing ICSs to identify more complex vulnerabilities.




\section{ICS Testbeds}\label{sec:testbed} 

In this section, we present a comprehensive analysis of the various ICS testbeds available in the literature. Firstly, in Section~\ref{subsec:testbed_classification} we introduce the classification method we employ in this work. Instead, in Section~\ref{subsec:testbed_requiremts} and Section~\ref{subsec:challenges} we recall, respectively, the requirements for an effective testbed and the main challenges in developing an ICS testbed. 
Then, we propose a detailed description of a set of interesting testbeds we choose testbed, dividing them into the three categories that we design. In particular, Section~\ref{subsec:testbed_physical} contains physical testbeds, Section~\ref{subsec:testbed_virtual} presents virtualized testbed, and Section~\ref{subsec:testbed_hybrid} describes hybrid testbeds which are a conjunction point of the other two categories.

\subsection{Testbeds Classification}\label{subsec:testbed_classification}

\begin{figure}[t]
    \centering
    \includegraphics[width=\columnwidth]{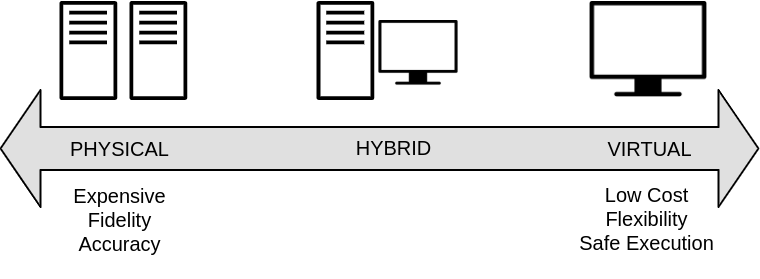}
    \caption{A summary of differences between testbed types.}
    \label{fig:testbedsChoice}
\end{figure}

There are different possible classifications of a testbed, basing on its sector, construction methodology, or the process involved. In this survey, and particularly in this section, we consider the functional elements involved in the testbed, classifying them as \emph{Physical}, \emph{Virtual}, or \emph{Hybrid} testbed. The different testbed categories are illustrated in Figure~\ref{fig:testbedsChoice}, even if sometimes the difference between can be minimal. For instance, many of the virtual testbeds presented can be interconnected with physical devices or wholly virtualized. Instead, Hybrid systems were designed with some real components and, without them, they could not work correctly.

Physical testbeds use real hardware and software to configure both the network and physical layers. They are a suitable approach when researchers need a solution to collect realistic measurement variation and latencies. Furthermore, it is possible to exploit the vulnerabilities of a specific device.
On the other hand, physical testbeds are expensive both in construction and maintenance. They generally have a long building time, and they may not provide a safe execution of dangerous physical processes (e.g., nuclear sector).

On the contrary, virtual testbeds leverage software simulations and emulations with single or multiple programs to reproduce the entire network and all the different components.
A virtual testbed represents a low-cost solution, but it is not easy to simulate high fidelity physical processes due to the virtualized environment. Despite this lack of precision, dangerous and risky processes (e.g., Nuclear sector) can be, in this way, simulated in a laboratory. Matlab, Modelica, Ptolemy, and PowerWorld are software used in the process simulation phase. 
Other tools are used to model control center communication networks (e.g., DETER, Emulab, CORE, ns3) and other devices used in the system such as PLCs (e.g., STEP7, RSEmulate, Modbus Rsim, Soft-PLC). Despite not generating data with perfect fidelity, these approaches are easy to update and upgrade, which gives them good flexibility and extendibility.

A widely diffused approach is developing testbeds composed of both physical devices and software simulations. This approach represents a good trade-off between physical and virtual solutions and is called a hybrid testbed. The main difference between the complete physical testbeds is that part of the components is simulated using specialized software. This solution can reduce the system's fidelity, but on the other hand, it permits to contain the cost and development time. 
However, as stated before, the separations between Virtual and Hybrid testbed is not always well defined. Sometimes virtual testbeds can be modified to work as a hybrid testbed by supporting physical devices. For example, VTET~\cite{Xie2018vtet} can be deployed using physical PLCs to replace the simulated ones. In this work, we consider as Hybrid a virtualized testbed composed of at least one real industrial device (e.g., PLC, IED, actuator, sensor).

\begin{figure*}
    \centering
    \includegraphics[trim={0 50 0 55},clip,width=0.9\textwidth]{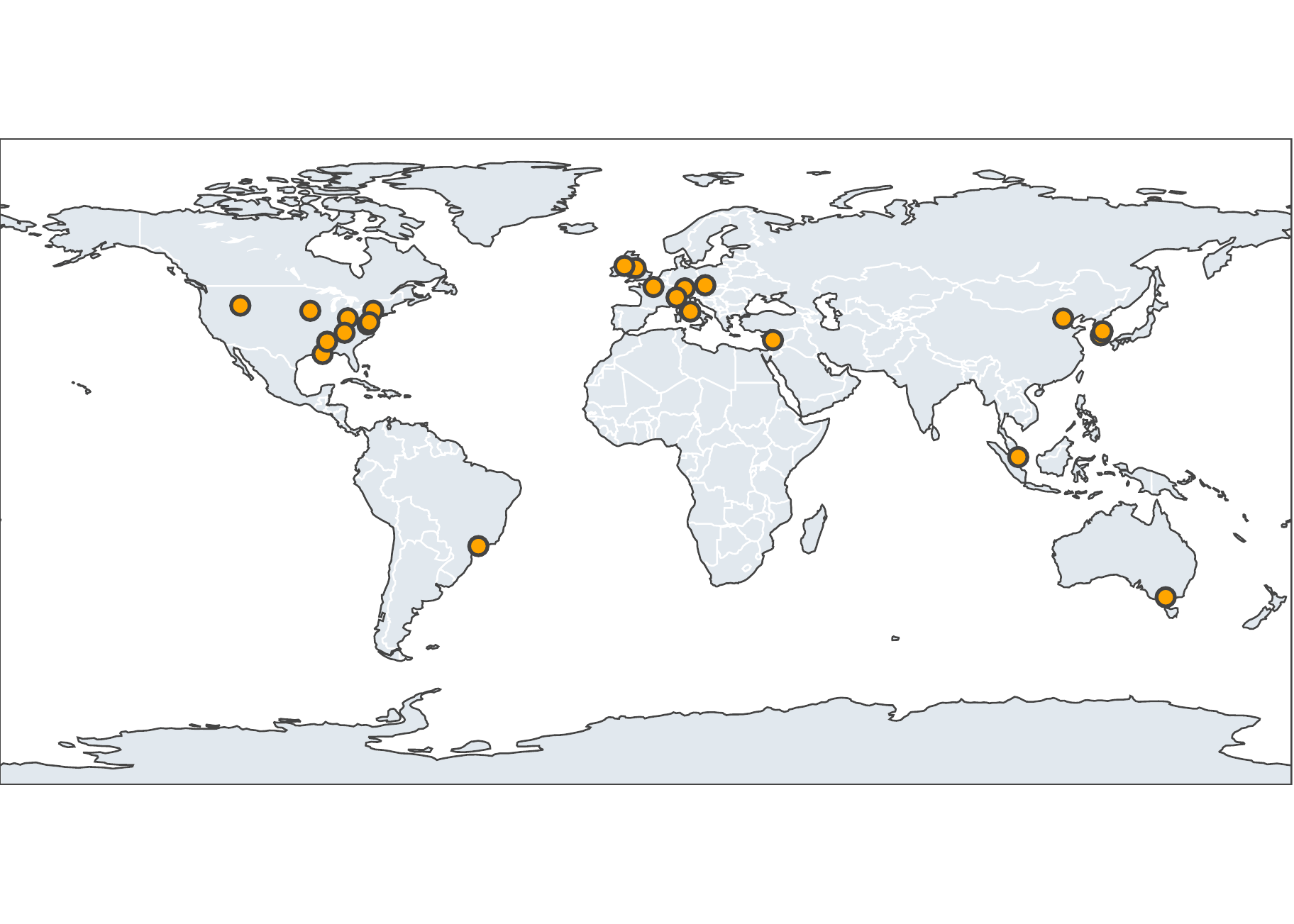}
    \caption{Physical and hybrid with physical process testbeds distribution around the World.}
    \label{fig:testbed_world}
\end{figure*}

\begin{figure*}
     \centering
     \subfloat[Legend]{\includegraphics[width=0.5\textwidth]{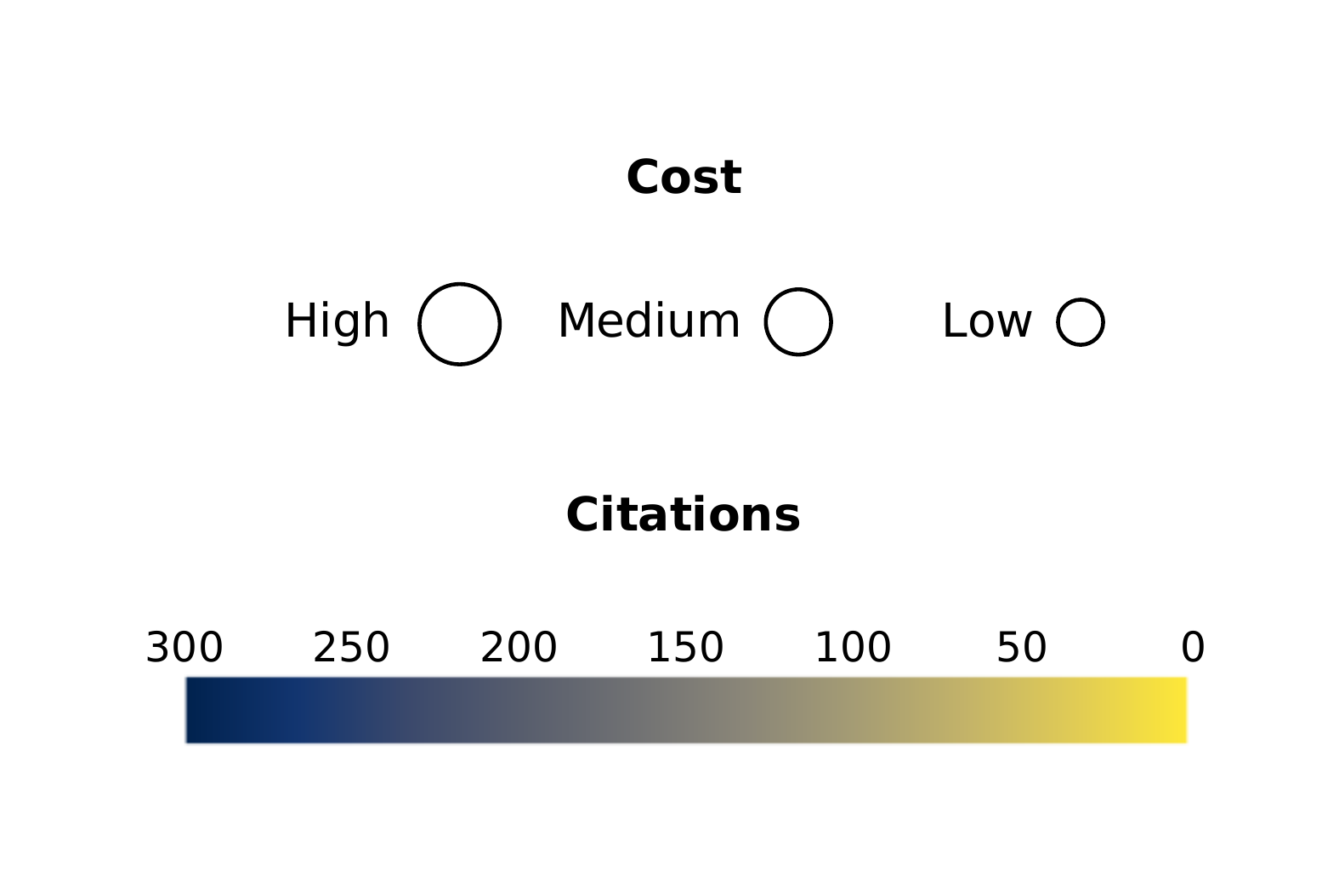}\label{subfig:testbed_legend}}
     \subfloat[North America]{\includegraphics[width=0.5\textwidth]{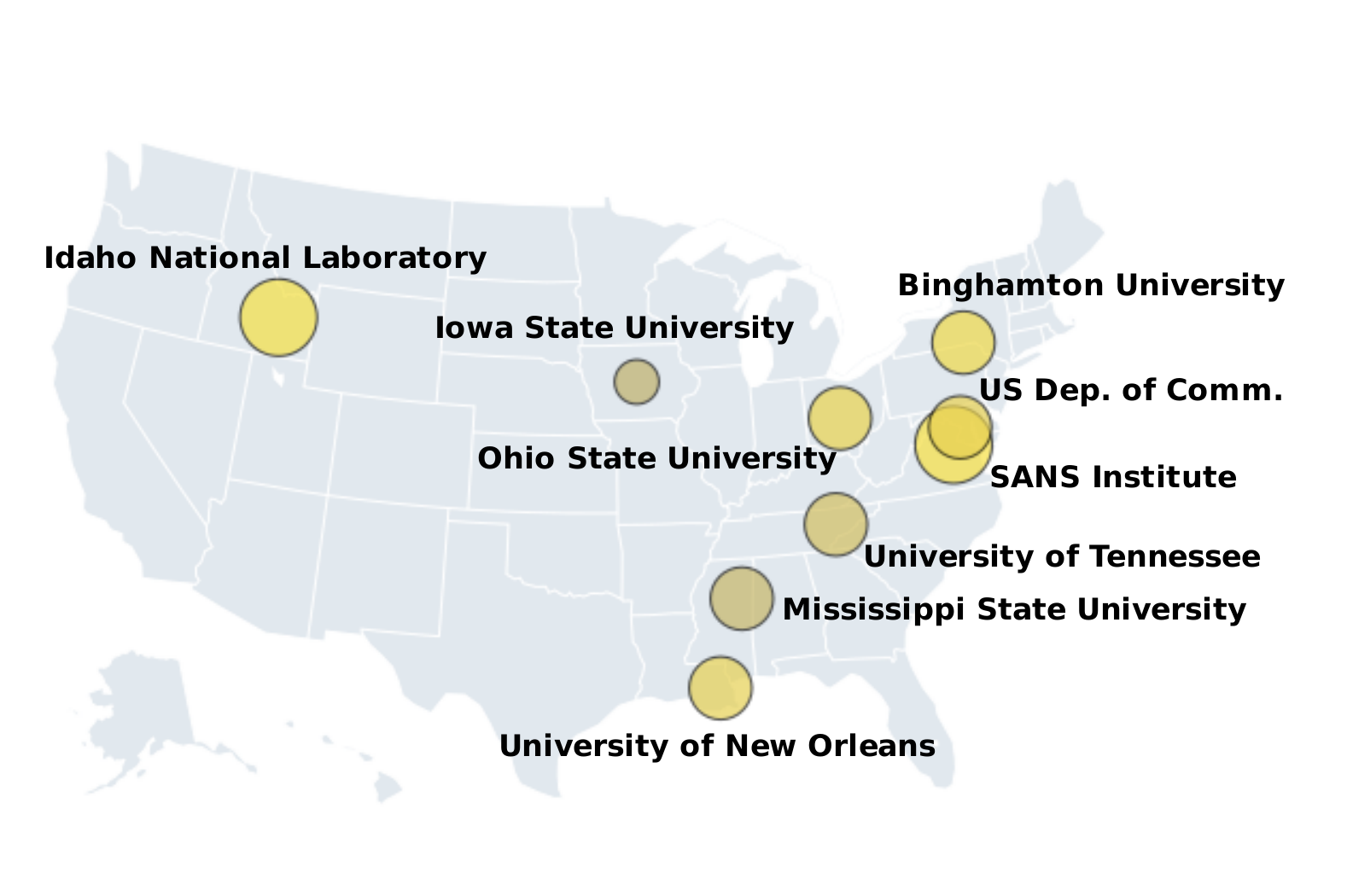}\label{subfig:testbed_us}}\\
     \subfloat[Europe]{\includegraphics[width=0.5\textwidth]{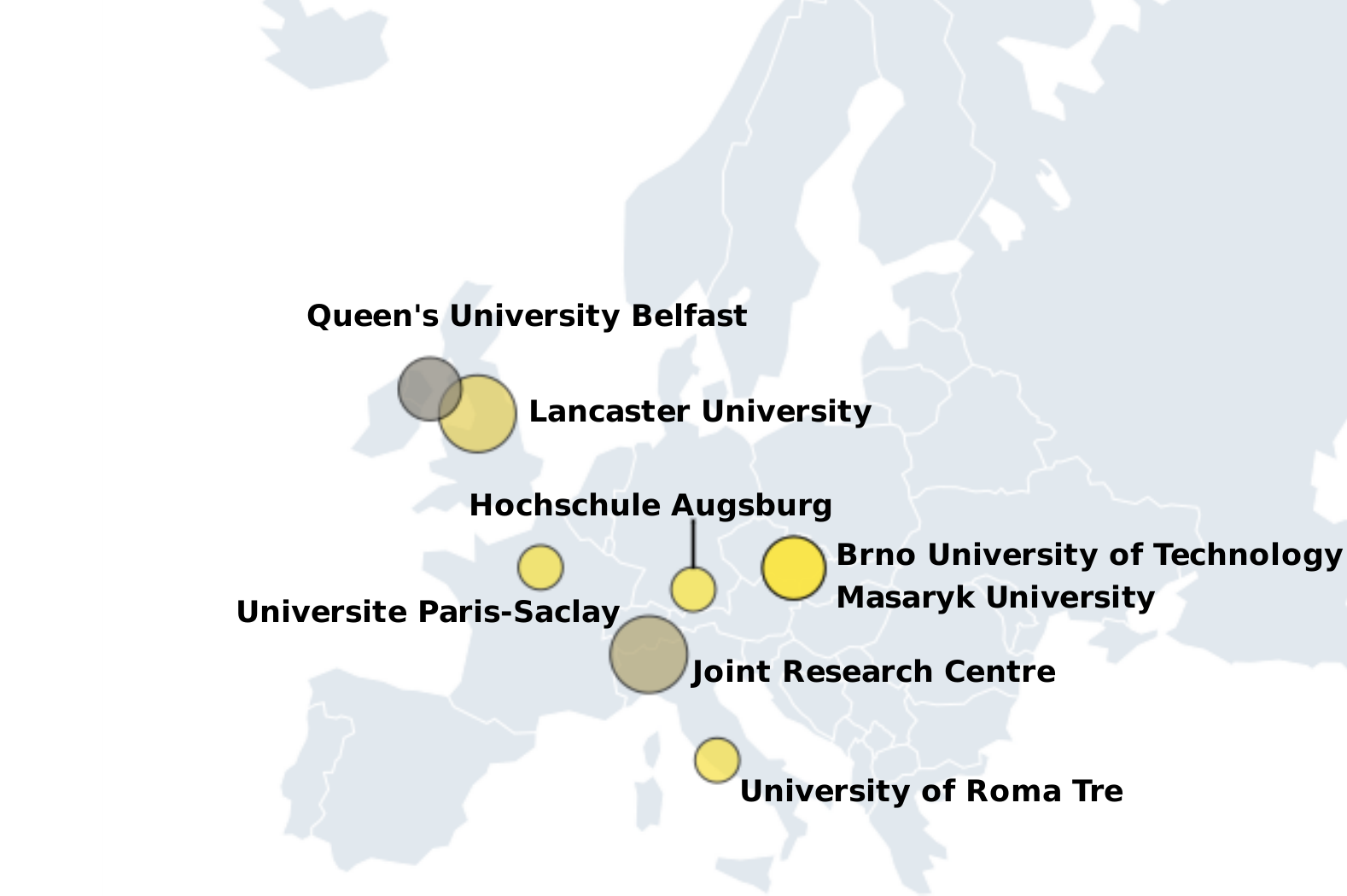}\label{subfig:testbed_eu}}
     \subfloat[Asia]{\includegraphics[width=0.5\textwidth]{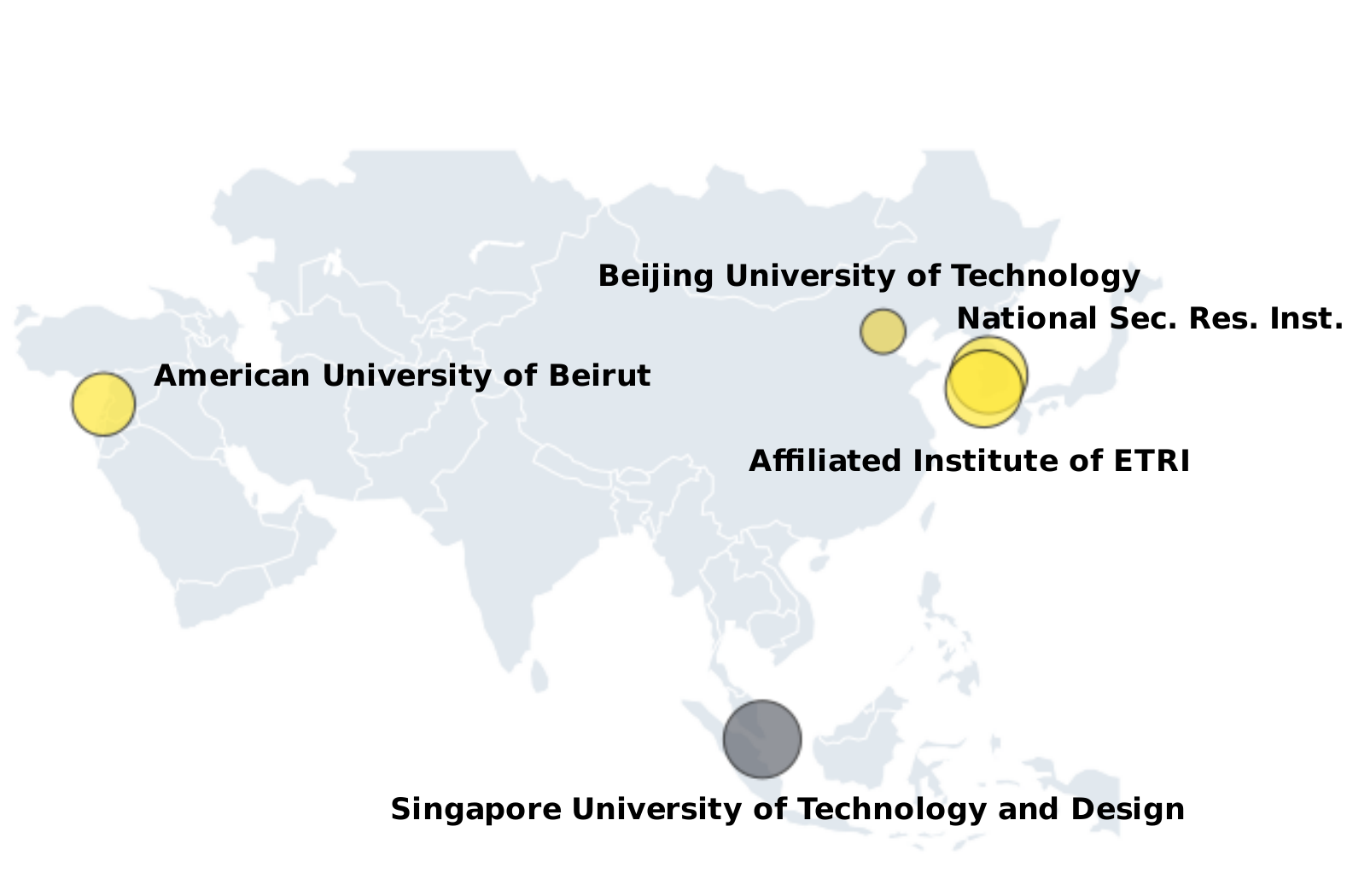}\label{subfig:testbed_asia}}
     \caption{Physical and hybrid with physical process testbeds distribution on the continents with more than one testbed: North America, Europe, and Asia. If there is more than one dataset in a place (e.g., Singapore SUTD), we aggregated the information.}
     \label{fig:testbed_fig}
\end{figure*}

In Figure~\ref{fig:testbed_fig}, we reported the geographic distribution of the Physical and Hybrid around the world. We think that this representation could help the reader see the current research trend in this sector in the world. 
In particular, Figure~\ref{fig:testbed_world} provide an high level view of where the testbeds are placed in the world, while Figures~\ref{subfig:testbed_asia}, \ref{subfig:testbed_eu}, and \ref{subfig:testbed_us} show close-up on the countries with more than one testbeds. In these figures, the marker size represents the estimated cost of the testbed. Simultaneously, the color indicates the Citations of the associated reference according to Google Scholar at the writing time.
Furthermore, we developed a website with an interactive map to collect and also provide information about future ICS testbeds and datasets\footnote{\url{https://spritz.math.unipd.it/projects/ics_survey/}}. Our goal is to continue to update this collection in the future.
Moreover, in Table~\ref{tab:testbed} we reported a brief comparison between the testbed presented in this paper, highlighting their main information and features. In particular for every testbed we reported the following information.
\begin{itemize}
    \item \textbf{Name} of the testbed (or of the authors if a name is not provided);
    \item \textbf{Institution} in which the testbed has been developed;
    \item \textbf{Country} The country on which is based the testbed or the institution of the first author;
    \item \textbf{Sector} indicates the field of the represented process;
    \item \textbf{Category} of the testbed. It can be \emph{Physical}, \emph{Virtual}, or \emph{Hybrid};
    \item \textbf{Physical Process} indicates how is implemented the physical level. It can be \emph{Simulated} with a software or \emph{Real} if consists of a physical implementation;
    \item \textbf{License} of the testbed. It can be:
    \begin{itemize}
        \item \emph{Open-source} if the source code is freely available;
        \item \emph{Open description} if, despite the source code is not provided, the description is sufficiently detailed to allow a reader developing a similar copy;
        \item \emph{Education} if it is maintained by an university and open to collaborator;
        \item \emph{Collaborations} if it is maintained by an institution which can accept collaborations;
        \item \emph{Not available} if it is owed by a private company and so not accessible or not available online.
    \end{itemize}
    \item \textbf{Scope} indicates the applications of the testbed. It can be:
    \begin{itemize}
        \item \emph{Security} if the main scope is related to cyber security research;
        \item \emph{Forensic} if the target scope is to provide a way to perform forensics research;
        \item \emph{Pedagogy} if the main scope is to provide education to students;
        \item \emph{General} if a precise scope is not specified.
    \end{itemize}
    \item \textbf{Cost} the estimated testbed implementation cost. It can be:
    \begin{itemize}
        \item \emph{Low} for a cost estimated $<500~\$$
        \item \emph{Medium} for a cost estimated between $500~\$$ and $10k~\$$
        \item \emph{High} for a cost estimated $>10k~\$$
    \end{itemize}
    \item \textbf{Reference} includes a reference to a description of the testbed.
    \item \textbf{Resource}, if available, indicates a resource for the download.
\end{itemize}
This information was not always available or easy to retrieve; therefore, the degree of detail may vary according to the specific dataset.

\subsection{Testbeds Requirements}\label{subsec:testbed_requiremts}

When researchers need to work with a real-world ICS environment, the proper solution is to build a testbed for conducting rigorous, transparent, and replicable testing of new technologies. The different testbeds vary in dimension, complexity, or sector.
According to~\cite{Geng2019testbeds}, an effective testbed needs to satisfy four main requirements: i) \textit{Fidelity}, ii) \textit{Repeatability}, iii) \textit{Measurement Accuracy}, and iv) \textit{Safe execution}. Sometimes, it could be challenging to satisfy all these requirements together; therefore, it is important to determine an optimal trade-off based on the research needs during the design phase.

A testbed should be developed to achieve a good \textit{fidelity} by accurately replicate the devices and processes from a real-world ICS. This is an expensive and space-consuming task, making it difficult for other researchers without much funding to replicate the same environment to verify the results. In these cases, mathematical models can be employed to virtualize physical processes in a cheap but less accurate way.

\textit{Repeatability} is an essential property for a testbed: it allows other researchers to reproduce the findings and compare other solutions on the same system. This property can be easily achievable for completely simulated testbeds, while it can be extremely challenging for ICSs that employ physical components or processes.

A testbed should monitor a physical process and take \textit{accurate measurements} without interfering with it. Sensors must be placed smartly, and if different points of measures are available, they must be carefully synchronized to provide accurate and reliable data.

Often ICSs are used to manage critical physical processes (e.g., chemical reactions, nuclear plants). If under attack, these kinds of \textit{processes can be dangerous} and can cause physical damage to the system itself. Since researchers need to study countermeasures' effect and effectiveness to attacks, testbeds must be provided with safe execute risky processes. This design challenge can be mitigated by employing simulations at the cost of a loss of accuracy. In other cases, processes are instead less critical. However, they can have an expensive or time-consuming recovery after an attack (e.g., after an attack completely empties a container into a water treatment system, it will take time to refill it again). In these scenarios, a virtual approach can be an excellent alternative to the physical replication~\cite{Xie2018vtet}.

\clearpage
\onecolumn

{\renewcommand{\arraystretch}{1.3}
\begin{center}
\footnotesize

\begin{longtable}[l]{|l|l|l|l|l|l|l|l|l|l|l|}
\caption{Summary of testbeds presented in the literature. We denote \textbf{Category} as H: Hybrid, P: Physical, and V: Virtual. To specify the ICS process \textbf{Physics} we use S: Simulated, R: Real, M: Mixed, and No: if there is not a physical process. To denote the \textbf{License}, we use OD: Open Description, E: Education, OS: Open-Source, C: Collaboration, and NA: Not Available. To denote the \textbf{Scope}, we use S: Security, G: General, P: Pedagogy, and F: Forensic. To denote the \textbf{Cost} we use L: Low, M: Medium, and H: High. The entries in Yellow indicates that the testbed is hybrid, the entries in blue mean that the testbed is physical, while entries in green correspond to virtual testbeds.}\label{tab:testbed}\\
\hline
\rowcolor[HTML]{EFEFEF} 
\rotatebox{-25}{\textbf{Name or authors}}                & \rot{\textbf{Category}} & \rot{\textbf{Institution\space}}    & \rot{\textbf{Country}} & \rot{\textbf{Sector}} & \rot{\textbf{Physics}} & \rot{\textbf{License}} & \rot{\textbf{Scope}} & \rot{\textbf{Cost}} & \rot{\textbf{Reference}}            & \rot{\textbf{Resource}} \\     \hline
\endfirsthead

\multicolumn{11}{c}%
{{\bfseries \tablename~\thetable{} -- continued from previous page}} \\
\hline
\rowcolor[HTML]{EFEFEF} 
\rotatebox{-25}{\textbf{Name or authors}}                & \rot{\textbf{Category}} & \rot{\textbf{Institution\space}}    & \rot{\textbf{Country}} & \rot{\textbf{Sector}} & \rot{\textbf{Physics}} & \rot{\textbf{License}} & \rot{\textbf{Scope}} & \rot{\textbf{Cost}} & \rot{\textbf{Reference}}            & \rot{\textbf{Resource}}\\ \hline
\endhead

\rowcolor[HTML]{FFFFC7} 
\textbf{Aghamolki et al.}    & H                       & USF                     & Florida, US      & Power Grid            & S                      & OD                     & S                    & M                   & \cite{7335093}                 & -                                \\ \hline
\rowcolor[HTML]{FFFFC7} 
\textbf{Alves et al.}        & H                       & UAH                     & Alabama, US      & Gas Pipeline          & S                      & OD                     & S                    & L                   & \cite{alves2016virtualization} & -                                \\ \hline
\rowcolor[HTML]{FFFFC7} 
\textbf{CockpitCI}           & H                       & University of Coimbra   & Portugal         & Power Grid            & S                      & OD                     & S                    & M                   & \cite{7542169}                 & -                                \\ \hline
\rowcolor[HTML]{FFFFC7} 
\textbf{CyberCity}           & H                       & SANS Institute          & -                & City                  & M                      & E                      & S, P                 & H                   & \cite{BorgesHink2016}          & -                                \\ \hline
\rowcolor[HTML]{FFFFC7} 
\textbf{EPIC (IPSC)}         & H                       & JRC Ispra               & Italy            & General CPS           & S                      & OS                     & S                    & L                   & \cite{Siaterlis2013epic}       & \cite{EpicWeb}                   \\ \hline
\rowcolor[HTML]{FFFFC7} 
\textbf{EPS-ICS}             & H                       & BIT                     & China            & Generic ICS           & R                      & NA                     & G                    & L                   & \cite{Gao2013eps}              & -                                \\
\hline
\end{longtable}
\end{center}

}

\begin{multicols}{2}

\subsection{Challenges in Developing a Testbed}\label{subsec:challenges}

The development of an industrial testbed is challenging from several points of view. Different works analyze the challenges in developing a well-designed ICS testbed~\cite{tippenhauer2019design, Almgren2018}. Based on the existing literature, in the following, we present the main problems related to the development of such a testbed.

\begin{itemize}
    \item \textbf{Design Guidelines}: When a research group decides to venture into building a testbed, it is fundamental to have a clear idea of the architecture. A clear and defined architecture can be useful in the development phases and to project further expansions. Moreover, it can guide other groups in building their own testbeds and therefore enabling the experiment repeatability. However, it is difficult to identify clear guidelines that help design a testbed from the engineering perspective.

    \item \textbf{Real Word Representation}: An industrial system must represent a real-world industrial scenario, including all the physical processes related to the environment. Furthermore, the Industrial testbed must include the most common industrial devices installed in the real world ICSs and supporting the most used protocols. Also, it is crucial to consider different versions of devices, knowing their different security features~\cite{Edgar2011, Gardiner2019oops}. The testbed should also include the different vulnerabilities that could, however, lead to a bias in the attack strategy vector.
    
    \item \textbf{Replication in Safety}: The physical processes controlled by ICSs are wide different, ranging from manufacturing processes to critical nuclear plants. The most delicate processes cannot always be replicated in a scaled-down version inside a laboratory. Furthermore, during attacks targeting the process's stability, even the less critical operation can express important safety issues~\cite{Gardiner2019oops}.
    
    \item \textbf{Complexity}: Industrial systems devices can be hard to configure and maintain due to their specificity and because they are designed to perform a precise and unique task. It is also challenging to find IT experts who have the needed knowledge to manage and maintain a complex ICS containing several OT specifications. The maintenance requirements must be considered from the early design stages since the increasing complexity can become even more expensive and difficult to manage~\cite{Gardiner2019oops}.
    
    \item \textbf{Cost}: To build physical industrial testbeds, research groups have to deal with building and maintenance costs. Expenses are one of the main reasons why there are not many testbeds available for research, and the ones that exist are generally not easily accessible by everyone. To overcome this problem, virtualized and emulated solutions are relatively diffuse in the field, even if they cannot provide the same fidelity and replication accuracy.
    
    \item \textbf{Lack of Documentation}: Another challenge in ICS research is the lack of documentation of the existing systems. Companies do not share internal information related to their system's architecture, the devices implemented, or the devices' software version. This is primarily due to the companies' privacy concerns, protection of intellectual proprieties, and security reasons. In fact, if a company discloses the presence of legacy devices with well-known vulnerabilities, it can attract several malicious actors' attention. This absence of documentation made the implementation of effective real-word testbeds difficult.
Furthermore, the lack of documentation can be problematic for the in-laboratory testbed. If poor documentation is provided, new researchers who start to work on a testbed might spend much time understanding the system's behavior and components and have a concrete idea. 
    To provide exhaustive documentation, it is essential to write it step-by-step during the testbed building process, avoiding writing it after the testbed is entirely built, which can be difficult and not cost-effective~\cite{Green2017pains}.
    
    \item \textbf{Reproducibility}: Due to the complexity of an ICS, it could be challenging to reproduce the experimental conditions of another research to replicate the results or test other solutions. The differences between the original conditions and the reproduced one can be minimal but, in some cases, can be sufficient to lead to different results. To facilitate the deployment, experiment-management systems can help researchers with the setup and the management of a testbed (e.g.,~\cite{eide2007experimentation}) by using template or code generation. Moreover, scripts for auto-configuration of an emulated testbed can be offered by developers (e.g.,~\cite{Antonioli2015mini}) to simplify the sharing process.
    However, suppose the testbed is composed of physical processes and components. In that case, it could be difficult to perfectly replicate them since many external variables can influence the system behavior (e.g., the temperature, the pressure)~\cite{Eide2010}.
    
    \item \textbf{Scalability}: If expanding a simulated or emulated testbed is generally straightforward, doing it with a physical testbed can be challenging. 
    Real devices are expensive, and researchers are not always able to afford them. Alternatives to expand physical processes are Hardware-In-the-Loop (HIL), i.e., mathematical representations of physical processes inserted in the chain. HILs offer great scalability of the system even if generally they are not advisable due to the lack of accurate mathematical models. A cheap way to add new devices is to employ software simulations. Software simulations are cost-effective solutions with the drawback of less precise and reliable physical representation. To provide system scalability and intelligent reconfiguration of all the physical devices implemented, virtualization and VLANs can be an excellent solution to be implemented in ICS without any substantial disadvantages~\cite{Gardiner2019oops}.
    
    \item \textbf{Data Collection}: A not trivial aspect of building a testbed is the data access and recording. It is generally a manual process, but it is vital to develop strategies to automate the collection precisely, providing reliability and synchronization between the different data collection points, for example, by introducing a central historian server.
    
\end{itemize}

\subsection{Physical Testbed}\label{subsec:testbed_physical}  

\textbf{Ahmed et al.}~\cite{Ahmed2016} presented a physical testbed built at the University of New Orleans, which models three industrial processes on a small scale but by employing real-world equipment such as transformers and PLCs.
A small gas pipeline that transports compressed air was built using a pipe fed with an air compressor. A valve regulates the other end of the pipe. Instead, the second system is a power transmission and distribution that carries electricity from power generation sources to individual consumers. This system is composed of a power station and four substations. Finally, the third system developed is a wastewater treatment system composed of sedimentation, aeration, and clarification processes. All the systems are installed at the top of a trolley, making the testbed easily transportable and particularly suitable for pedagogy and research. Each system is controlled by one PLC connected through a switch to a historian and an HMI. This last device makes it possible to visualize and control the systems. The industrial protocols employed are Modbus, EtherNet/IP, and PROFINET.

\begin{figure}[t]
    \centering
    \includegraphics[width=0.9\columnwidth]{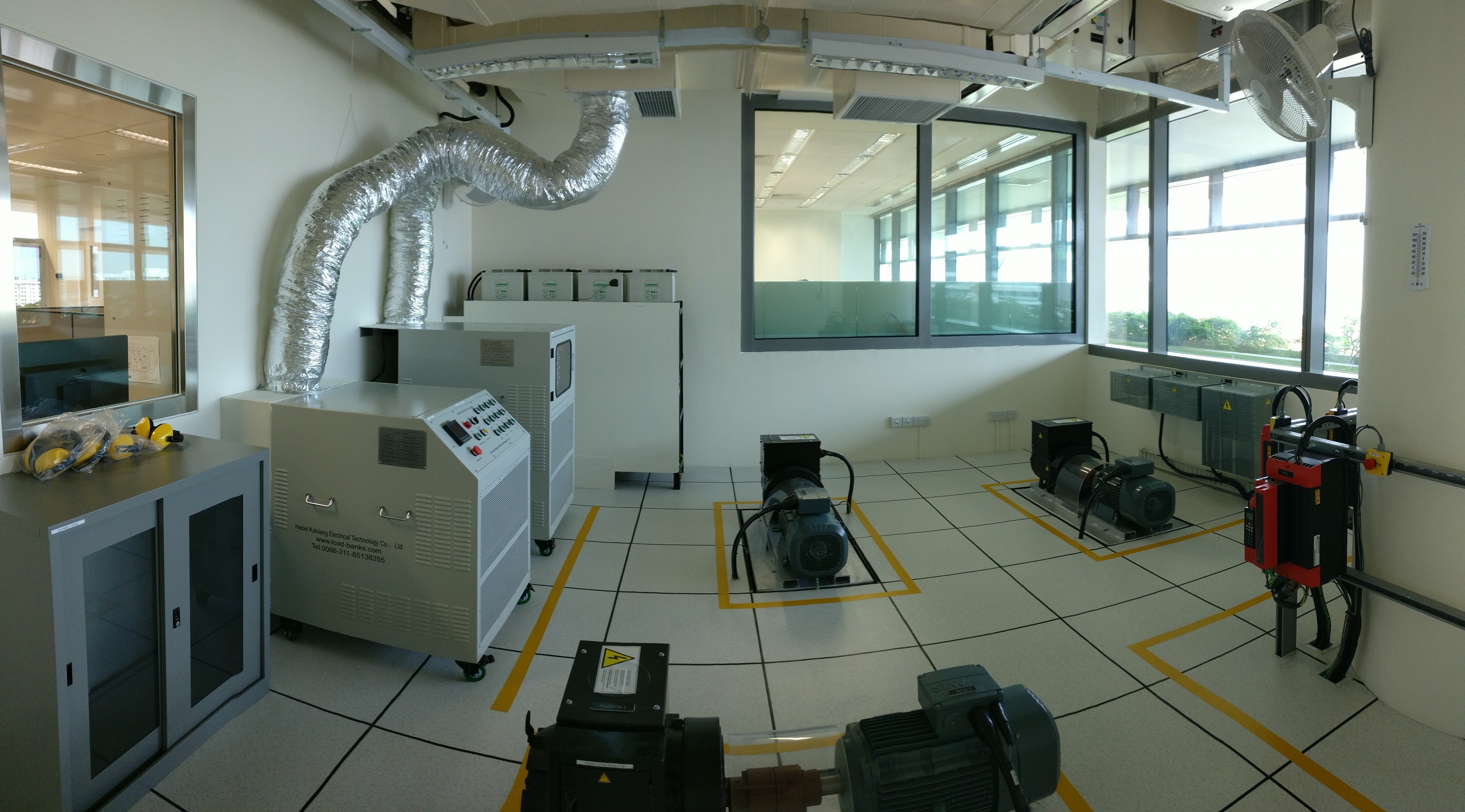}
    \caption{EPIC Testbed by iTrust in Singapore.}
    \label{fig:epic}
\end{figure}

\textbf{Electrical Power and Intelligent Control (EPIC)}~\cite{Adepu2018epic, iTrustEPICWeb} is a high-cost 72kVA electric power testbed that mimics a real-world power system in small scale smart-grid, and it is available for rent. The testbed is shown in Figure~\ref{fig:epic} and it is composed of four stages, namely: Generation, Transmission, Micro-grid, and Smart Home. Each stage is controlled by PLCs connected to a master PLC using switches and then to a SCADA gateway. 
The physical process is entrusted to two motor-driven generators, photovoltaic panels, and a battery.
Communications occur using the IEC 61850 standard protocol for the electrical substation and automation system that runs over TCP/IP stack. 
The authors also present false data injection attacks, malware attacks, power supply interruption attacks, and physical damage attacks, together with possible mitigation techniques. The testbed resides at the Singapore University of Technology and Design (SUTD), and it is used to supply power to two other testbeds inside the same institution (i.e., SWaT~\cite{Mathur2016swat}, and WADI~\cite{Ahmed2017wadi}) to create also the possibility for research related to a cascade-connected ICSs. The authors also shared a related dataset, which will be analyzed in Section~\ref{sec:dataset}.

\textbf{HAI Tesbted} (HIL-based Augmented ICS)~\cite{Shin2019, Choi2020} is an extensive and expensive interconnection of three independent real ICSs coordinated by a real-time Hardware-in-The-Loop (HIL) developed at The Affiliated Institute of ETRI, Republic of Korea.
Emerson's boiler control system, GE's turbine control system, and FESTO's water treatment control system are built-in small-sized by employing components used in industrial environments.
The HIL is used to simulate the power plant to combine the three control systems and form an integrated power generation system.
The interconnection employs Ethernet at Level 2, while different proprietary Fieldbus versions are used to communicate with the field devices~\cite{HAIWeb}.
The authors' developed a tool to schedule HMI tasks for long periods without human intervention. This tool also helps to schedule attacks (e.g., MitM attacks) only when a particular ICS state occurs. In~\cite{Shin2019} the authors present various physical attacks targeting the pump and the pressure of the boiler system. An expansion of the testbed~\cite{Choi2020} was built to make it possible to launch also network attacks using tools like Nessus or Acunetix.

\textbf{BU-Testbed}~\cite{korkmaz2016industrial} is a physical reproduction of two power generation systems developed at Binghamton University. 
The first one is composed of an AC motor directly coupled to a permanent magnet DC motor, generating up to 400V. The other one instead contains an AC motor used to drive a 12-volt DC blower motor used to generate electricity. The testbed also includes two types of Alley Bradley PLCs and a private computer with an LCD monitor used as HMI. The communication uses the EtherNet/IP protocol.
Furthermore, the authors explain some cyber-physical attacks which are practicable on the testbed. These attacks regard different categories: 1) attacks on networks (i.e., MitM, DNS poisoning); 2) network congestions and delay (i.e., DoS); 3) attacks on controllers, sensors, and drivers (i.e., malicious software injection and firmware modification); and 4) attacks on HMI and programmable stations (malware injection).
In another work~\cite{korkmaz2016attack}, Korkmaz et al. presented a similar testbed in which the vulnerability to time delay attacks has been evaluated. Results show the feasibility of such attacks, which can stop the power generation \end{multicols}\twocolumn\noindent 
process and shut down the testbed.

\textbf{Lancaster's testbed} by Green et al.~\cite{Green2017pains} at the Lancaster University is a big physical scaled-version of a generic industrial ICS (the testbed does not explicitly the physical processes involved). It is composed of six Manufacturing Zones, a DMZ, and an Enterprise Zone. Each core zone is split at the network level using VLANs. The legacy serial-based communications have been upgraded to IP to reduce the complexity and allow communications with a vast number of ICS devices. The connections are almost all physical, apart from two manufacturing zones connected using 3G, 4G, and satellite communications. 
To account for a changing landscape and to add flexibility, all the desktop and server-based software applications run inside a VMWare vSphere server as virtual machines.
The authors are continuously improving the testbed to make it more usable and more complete.
Students and researchers of the university use the testbed, but the authors also plan to make it more available for external researchers.

Morris et al.~\cite{Morris2011testbeds} at the Mississippi State University built seven different small physical testbeds for security research and pedagogy purposes. Five of them have communications based on Modbus/ASCII, Modbus/RTU, and DNP3 (henceforth called \textbf{Mississipi Serial}) and represent respectively: 1) a gas pipeline used to move petroleum products to market; 2) a storage tank used in the petrochemical industry; 3) a raised water tower used to provide pressure in the water distribution system; 4) a factory conveyor belt control system, and 5) an industrial blower used to force air through an exhaust system. These five systems are controlled by the same HMI but on different screens. It enables the control of all the systems from the same point and simulates a more extensive system by making them operate simultaneously.
The remaining two testbeds are connected through an Ethernet network (and then are called \textbf{Mississipi Ethernet}) and include: 1) a steel rolling operation; and 2) a smart grid transmission system. 
The authors also use the testbeds to generate datasets that are freely available online~\cite{MorrisWeb}.


\textbf{Smart Grid Test Bed (SGTB)}~\cite{moreno2017development, INL2017} deployed by Idaho National Laboratory is the world's first full-scale replication of a smart grid, and it is part of the United States National SCADA Test Bed Program. It is a 61-mile transmission massive testbed connecting twelve facilities with power distribution networks that can selectively operate at various voltages (12.47kV, 24.9kV, and 34.5kV). 
Portions of the power loop can be isolated and reconfigured for independent, specialized testing.
As planned in 2017, the authors obtain more funds to expand SBTB with a SCADA testbed to be installed in the command and control shelter to allow operators to observe, manage, and manipulate test line configurations and record testbed operating parameters. However, to the best of our knowledge, the authors never release updates about the project.
This testbed is not an ordinary scaled-down version of real systems. Instead, SGTB is a full-size plant. Even if it represents an impressive and valuable work, unfortunately, students and researchers have limited access to such a facility~\cite{Hahn2010}.

\begin{figure}[t]
    \centering
    \includegraphics[width=0.8\columnwidth]{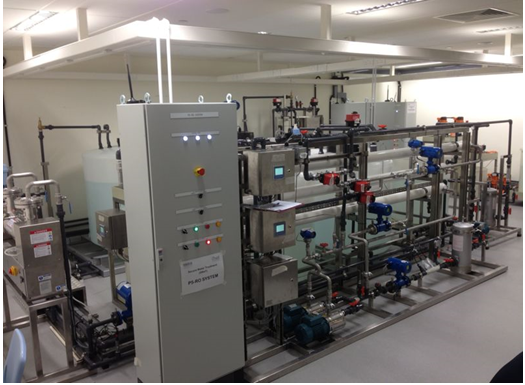}
    \caption{SWaT Testbed by iTrust of Singapore.}
    \label{fig:swat_test}
\end{figure}

\textbf{SWaT}~\cite{Mathur2016swat, 7469060} is a six-stage water treatment plant developed by the Singapore University of Technology and Design (SUTD) represented in Figure~\ref{fig:swat_test}. One PLC (plus one for backup) controls each stage, and the overall testbed leverages a distributed control approach. Furthermore, through a Human-Machine Interface (HMI), an operator can manually control all the system components.
Communication between PLCs and sensors/actuators are based on Ethernet ring topology, while PLCs communicate with each other through a separate network based on an Ethernet star topology. The protocols implemented in the systems are EtherNet/IP and Common Industrial Protocol (CIP). 
In the paper, the authors implemented various attacks to manipulate plant operations. The different attacks leverage different assumptions on the attack model. In particular, the attacks are categorized as single-stage attacks, targeting a specific stage process and multi-stage attacks, which combine the compromising of various stages. Furthermore, each attack may target a single system point or multiple system points. The attacks include different scenarios (e.g., an attacker with access to the local plant communication network or an attacker who is on-site and has physical access to the device) and different types of attacks (e.g., MitM, eavesdrop, or packets modification).
The testbed is accessible only for collaborations or by renting it. Recently, a python-based software simulation of the testbed was developed and released with open-source code~\cite{chen2018learning, SWATSimWeb}.
Also, datasets based on different data collection are openly available upon request. These datasets contain both network and physical packets in normal behavior and with the system under attacks~\cite{Goh2017swatdata}. We present the dataset in Section~\ref{sec:dataset}.

\textbf{Teixeira et al.}~\cite{Teixeira2018} implemented an ICS testbed to model a simple water storage tank's control system. The storage tank is equipped with two-level sensors to control the water level. When it reaches the maximum level, the upper sensor sends a signal to the PLC, which turns off the water pump used to fill the tank. At the same time, another pump is activated to draw water from the tank. When the water reaches the lower sensors, a signal is sent to the PLC, which will reverse the two pumps' state to fill up the tank again.
The SCADA system gets data from the PLC using the Modbus protocol and displays them to the system operator through the HMI interface.
To complete the study, the authors tested some attacks such as scanning, device identification, and not authorized read of actuators. By recording SCADA network traffic for 25 hours, a dataset has been released~\cite{TeixeiraWeb} and will be presented in Section~\ref{sec:dataset}.
In 2019, minor improvements of the testbed had been presented~\cite{Zolanvari2019}, such as embedding a turbidity sensor and a turbidity alarm to add analog input to the system.

\textbf{Turbo-Gas Power Plant} (T-GPP) testbed~\cite{Fovino2010} is an experimental platform presented by Fovino et al. at the Joint Research Centre of Ispra (Italy) to perform security research on a SCADA system. It is a physical testbed that replicates a power plant's dynamics process and its control systems providing additional mechanisms for running and analyzing the system.   
The testbed is composed of seven different functional elements: 1) Field Network, used to link PLCs with the SCADA servers, actuators, and sensors; 2) Process Network, that interconnects the different physical subsystems; 3) Intranet, the internal private network connecting PCs and server of the company; 4) Demilitarized Zone, used to separate IT area from OT components; 5) External Network, such as the Internet; 6) Observer Network, a network of meshed sensors to gather a massive quantity of raw data useful for the analysis; and 7) Horizontal Services Network, used for the management of the laboratory.
The paper profoundly analyzes such systems' vulnerabilities, highlighting those related to the protocols implemented (i.e., Modbus/TCP and DNP3), and describes various attacks deployed on the testbed: DoS, worm, and malware infection on the process network, phishing attack, and local DNS poisoning. Finally, the authors propose different countermeasures to the attacks.


\textbf{WADI}~\cite{Ahmed2017wadi, WADIWeb} is a scaled version of a water distribution testbed build by the Singapore University of Technology and Design (SUTD) to perform security researches. It consists of five stages controlled by three PLC and two RTU, which can supply 10 US gallons/min of water. The communication happens using Modbus/TCP protocol at Layer 0, while at Level 1 network between PLCs uses TCP over Ethernet instead of RTUs that exploit High-Speed Packet Access (HSPA) using GPRS modem to generate a precise real-world scenario.
The authors also implemented different attacks against the testbed by manipulating data from sensors to cut off the consumer tank's water supply.
The system is physically connected to SWaT, and it can be used to generate a more accurate scenario and study the cascade effects of a cyber attack on connected ICS.
Furthermore, WADI is available to organizations for joint research programs and usage, but a dataset generated upon request is available. We will analyze the dataset in Section~\ref{sec:dataset}.

In 2014, \textbf{Yang et al.}~\cite{6737311} proposed a physical SCADA power grid testbed specific designed to test their detection approach. At the control network level, the testbed is composed of an HMI, a database to log events and data, a host used to perform the attacks, and different networking components (e.g., protocol gateway, switch, firewall, router). Instead, the physical network is composed of various IED simulated, connected to a real photovoltaic system.
The connections between the Gateway and the IEC devices are based on the IEC 60870-5 series protocol, and then the Gateway translates the IEC 60870-5 to allow the communication with the HMI station.
The IDS proposed by the authors was installed between the HMI and the Protocol Gateway. It monitors all the incoming connections to the substation and the LAN network through a port mirroring.

\textbf{Zhang et al.}~\cite{Zhang2019} presented a security research on a physical process ICS testbed which simulates a two-loop nuclear power system. The primary loop includes a 9kW heater representing the reactor core, controlled by the SCADA master through an open-loop controller. It also contains a variable speed coolant pump, upper and lower delay tanks, and other instrumentation such as a flow meter and temperature detectors. The secondary loop is composed of valves, a magnetic flow meter, and two temperature detectors. The SCADA system consists of an engineering workstation as the SCADA master and a National Instruments chassis used to read data and control signal output modules as SCADA slave. The system is completed with data storage and an attacker machine with Kali Linux. LabVIEW was installed on the engineering workstation to record sensor data and send control commands to actuators.
In the same paper, the authors proposed some attacks to the testbed (e.g., MitM, DoS). Furthermore, they implemented some intrusion detection mechanisms based on Random Forest (RF), k-Nearest Neighbors (KNN), and Auto-Associative Kernel Regression (AAKR).

\subsection{Virtual}\label{subsec:testbed_virtual}


\textbf{Davis et al.}~\cite{Davis2006} is a power grid simulated testbed based on a client-server paradigm. 
The client mimicked a control room's graphical interface containing SCADA data and used it to control power elements. Each client can switch between different servers to monitor several systems from the same machine. The most common operating systems support client software.
On the other hand, the server is based on the PowerWorld~\cite{PowerWorldWeb} simulator and can model a complex power grid. The server sends the process data to the client via a custom TCP/IP protocol, converted to Modbus/TCP using an integrated protocol converter. Furthermore, the simulator can connect and interact with real hardware devices, but it is not mandatory.
The network is emulated using RINSE~\cite{Rinse} which allows clients to launch different commands to simulate attacks (e.g., DoS attacks), defense techniques (e.g., filtering), diagnostic tools, device controls, and simulator data.
The authors present various attacks, such as DDoS and network overload, comparing the results with and without security measures.
To the best of the authors' knowledge, the testbed is not available online.

In~\cite{krotofil2015rocking} the authors present \textbf{Damn Vulnerable Chemical Process (DVCP)}, an open-source framework developed for cyber-physical security experimentation based on two models of chemical processes. In particular, the framework includes \textbf{DVCP-TE} and \textbf{DVCP-VAC}, two simulated ICS testbed based respectively on Tennesse-Estman~\cite{lyman1995plant} and Vacuum-assisted closure (VAC)~\cite{argenta1997vacuum} chemical processes simulated with Matlab. The authors use these simulation models in hybrid scenarios with the simulated process and real industrial hardware (i.e., SIMATIC S7-1200+KTP400 Starter Kit). Furthermore, the Modbus and PROFINET protocols were implemented to enable communication between the simulated process, the PLC, and the HMI. However, for this implementation, the authors did not share any code or further implementation information.

\textbf{Genge et al.}~\cite{Genge2012} proposed a framework based on Emulab~\cite{EmulabWeb} for the emulation of the components and to Simulink~\cite{SimulinkWeb} for the physical processes simulation.
The architecture comprises three layers: the cyber layer containing the regular emulated ICT components used in SCADA systems, the physical layer providing the simulation of physical processes, and the link-layer to connect the cyber and physical layers through the use of a shared memory region.
The paper provides a qualitative comparison with other works, comparing the testbed with other related projects. Results show high performances in all the functionalities considered (e.g., repeatability, safe execution), except for the physical layer fidelity, where physical testbeds perform better.
Furthermore, estimating the cost to build and maintain a physical testbed is compared with the predicted expense related to the presented framework, showing considerable savings through the years.
A peculiarity of this framework is the possibility to attack the different components using specific malware. For instance, as a case study, the authors present Stuxnet~\cite{Falliere2011stuxnet} on a boiling water power plant, showing its effectiveness. Another attack example targets a chemical process by deleting and delaying some packets and changing the process parameters to reach their shut-down safety limits.
There are many supported protocols such as Modbus, Profinet, and DNP3. The testbed, implemented in C\#, is not available online to the best of the authors' knowledge.

\textbf{Giani et al.}~\cite{Giani2008} developed a virtual SCADA testbed for security-related researches purposes. However, this work represents a preliminary study presenting the testbed at a high-level, but without a practical implementation description.
At the center of the architecture, there is the SCADA master station containing the SCADA server and the HMI. 
The SCADA master station containing the SCADA server and the HMI is placed in the architecture center. SCADA master servers run the server-side applications that communicate with the RTUs using different strategies: dial-up modems, private leased line, wireless or radio channel, and LAN/WAN links.
The most used protocols for these communications are Modbus and DNP3. The SCADA server is also connected to the corporate network, connected in turn to the Internet, exposing the system to vulnerabilities, such as unauthorized remote access. 
The authors planned to employ a single simulation-based instantiation to build all the testbed elements in the same machine using software like Simulink~\cite{SimulinkWeb}. 
Other implementation strategies for the system architecture are possible, like the federated simulation-based, where a different machine simulates each element. 
Moreover, emulation-based and implementation-based instantiations that use actual commercial SCADA devices along with simulation and emulation of software modules, network, and physical processes are depicted but not implemented. 
Finally, the authors depict various possible attacks (e.g., DoS, integrity attacks, phishing attacks) and suggestions about security mechanisms.
To the best of our knowledge, the testbed is not publicly available online.

\begin{figure}[t]
    \centering
    \includegraphics[width=\columnwidth, trim={0 0 0 1cm}, clip]{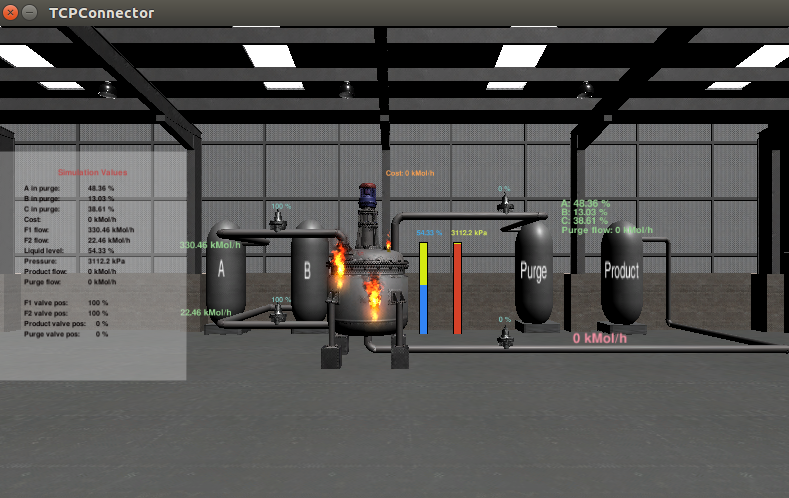}
    \caption{Example of GRFICS simulator rendering.}
    \label{fig:grfics}
\end{figure}

\textbf{GRFICS}~\cite{formby2018lowering} is a graphical and open-source~\cite{GRFICSWeb} ICS simulation tool based on the Tennessee Eastman process (Figure~\ref{fig:grfics}). Currently, the testbed is designed for educational purposes and allows only the use of pre-defined functions.
The ICS devices are simulated. In particular, the OpenPLC~\cite{alves2014openplc} is used for the PLCs, and the HMI Virtual Machine simulated an HMI using AdvancedHMI~\cite{AdvancedHMIWeb} software. The testbed allows running many pre-defined attacks such as MitM, Command Injection, False Data Injection, Reprogramming of PLCs (i.e., Stuxnet), Loading Malicious Binary Payload (i.e., TRITON), and Common IT attacks (i.e., password cracking, buffer overflow). Once the attacks are launched, the interface allows monitoring the testbed attacks' consequences, log the process information, and how much cost is wasted through the purge. Finally, the testbed allows the installation of the Snort detector~\cite{snort} and to customize it with new rules. The communications on the testbed are based on Modbus protocols.

Maynard et al.~\cite{Maynard2018open} proposed \textbf{Maynard SCADA}, an open-source, scalable framework for deploying a replication of a SCADA network.
The testbed is composed of a collection of scripts used to build and configure virtual machines that, by default, are emulated using Oracle VirtualBox~\cite{VirtualBoxWeb}. The resulting network can also support and integrate the connection with physical devices. Maynard SCADA supports IEC 60870-5-104 (IEC104) and OPC Unified Architecture (OCP-UA) to support additional industrial protocols such as Modbus or IEC 61850. The framework implements two types of profiles: an operation profile, which defines the deployment of nodes, simulators, and configuration of the network; and a configuration profile to configure nodes to represent specific industrial devices(e.g., HMI, RTU).
Such profiles can be developed by the community, adding new use cases and simplifying the testbed's deployment. 
The framework does not consider the physical process simulations, but it can be easily integrated using third-parties software (e.g., Simulink~\cite{SimulinkWeb}). 
Furthermore, the paper~\cite{Maynard2018open} shows a common metering application using seven virtualized nodes with detailed instructions to replicate it. The instructions also include an accurate description system's requirements and a comparison between some other testbeds in the same document. 
The framework is entirely open-source, and it is accessible on GitHub~\cite{MaynardWeb}, where are also available some datasets.

\textbf{MiniCPS}~\cite{Antonioli2015mini} by Antonioli and Tippenhauer is a toolkit used to create an extensible and reproducible research environment for network communications, control systems, and physical layer interactions in CPS. MiniCPS is an extension of Mininet~\cite{kaur2014mininet}, a widespread network simulator built around the Software-Defined Networking paradigm that exploits lightweight system virtualization using Linux containers. Connections between simulated devices are emulated using virtual Ethernet links with an easy drag and drop interface. These connections can be configured through Linux Traffic Control to emulate link performance such as delay, loss rate, and bandwidth. 
MiniCPS extends the classic Mininet by implementing ICS components such as PLCs and allowing the connection with real physical devices. 
The testbed is not focused on the physical process simulation that can be implemented using third-parties process simulation engines (e.g., Simulink~\cite{SimulinkWeb}).
The reproducibility is a significant advantage of this testbed: it is possible to write Python scripts that generate a complete ICS environment easily exportable and shareable.
On the top of the emulated Ethernet network, the testbed includes different industrial protocols, in particular, using the \textit{CPPPO} Python library, MiniCPS implement, for example, EtherNet/IP and Modbus/TCP. 
The paper~\cite{Antonioli2015mini} accurately describes all the design decisions and the consequent strengths and drawbacks of the testbed. 
The authors present an attack scenario of a MitM attack on a replicated model of the SWaT testbed~\cite{Mathur2016swat}, also providing different countermeasures based on a custom SDN controller. The testbed and its documentation are open-source and available on Github~\cite{MiniWeb}.

Morris et al.~\cite{Morris2015} presented a virtual gas pipeline system (called \textbf{Gas Pipeline testbed}) that is a simulation of a testbed previously built. 
The testbed consists of four components running in different virtual machines: a virtual physical process, a Python-based PLC simulation, a network simulation, and an HMI. The various components communicate through Modbus/TCP over a virtual network and may be connected to real devices.
The virtual system allows modeling a pump, a valve, a pipeline, a fluid, and a fluid flow. The models are based on a previous physical testbed~\cite{Morris2011testbeds}, allowing to compare measures from the two testbeds. The virtual testbed mimics the physical device's behavior but with some difference in pressure change frequency. Also, the startup process is similar but not identical. The authors present a command injection attack to the virtual testbed, but the resulting behavior is not compared with the physical testbed.
To the best of our knowledge, this virtual simulator is not publicly available online.

\textbf{Reavers \& Morris}~\cite{Reaves2012virtual} develop an open and complete platform for creating virtual testbeds.
The resulting system is highly scalable, and it is possible to install plenty of different virtual devices. The testbed's main components are process simulators, data loggers, and configuration files used to configure virtual devices and connections among them. All the simulations are implemented with Python without adopting off-the-shelf network simulation tools.
The process simulator includes four components: 1) a simulator module, 2) a communication interface, 3) an update queue, and 4) configuration files. The simulator communicates directly with the virtual test devices via a ``backchannel'' to transmit measurements and inputs. 
The virtual devices supported are RTU, MTU, IED, PLC, repeaters, and Programmable Automation Controllers (PAC), which can run as standalone processes or inside virtual machines. It is possible to connect physical devices such as wireless radios and HMI.
There are two main protocols for the communications: Modbus/TCP, which can be logged using standard applications such as Wireshark or tcpdump; and Modbus/RTU, which instead need a PortLogger, a class of proxy that reads the communication and then resends it to the channel.
In the paper~\cite{Reaves2012virtual}, the authors present two testbed applications. The first represents a gas pipeline, while the second models a water storage tank control system. To verify the simulated data's consistency, the authors implement these two testbeds as physical ones. 
Furthermore, to obtain more relevant results, the authors decided to compare the virtual and physical testbeds' behaviors during different attacks (e.g., data injection and DoS).
Results show that the attacks are effective in both scenarios, with some slight variations on time needed for the attack to succeed.
The paper also compares both the virtual and physical systems' normal behavior discovering many similarities, but with some detectable differences. The study's conclusion states that the virtual testbed is good for proof-of-concept, but a physical testbed is needed in some cases. 
To the best of our knowledge, the virtual platform is not publicly available online, but some datasets are available~\cite{MorrisWeb}.
Starting from this work, Thornton and Morris in 2015~\cite{Thornton2015improv}, deployed a similar platform that permits the usage of Simulink~\cite{SimulinkWeb} instead of Python to simulate the physical processes.

\textbf{RICS-el} testbed~\cite{Almgren2019} is a virtual testbed representing a power system built on top of the Cyber Range And Training Environment (CRATE) infrastructure at the Swedish Defence Research Agency (FOI)~\cite{CRATEWeb}. All the hosts of the testbed are run on virtual machines using VirtualBox~\cite{VirtualBoxWeb}. 
Researches and vendor experts designed the OT segment. It is divided into the OT DMZ, the OT LAN, the substation communication WAN, and the power grid simulator, including all the RTUs. In the OT DMZ, there are the FTP server, the HMI, and the historian. The WAN is used to enable communication between the 15 hosts. Three of these hosts are RTUs that communicate with the front-end through the IEC 60870-5-104 (IEC104) protocol. The power grid simulator is the key component of the architecture: it can generate realistic traffic and event in the whole RICS-el environment. In detail, this testbed simulates a backbone high voltage 400kW grid with twenty substations and some medium voltage transmission. Finally, to add more realist to the environment, the system is connected through another DMZ to an office IT segment. It contains a LAN to interconnect 17 office workstations, nine sales workstations, and some other servers.  
Ongoing work is focusing on adding realistic traffic to each segment by emulating users and different scenarios.

\textbf{SCADASim}~\cite{Queiroz2011} is a simulator for SCADA systems created on top of OMNET++~\cite{OmnetWeb}. The testbed is developed to satisfy specific requirements: 1) it allows plug-n-play to create simulations to allow system experts to set up the software; 2) allows connectivity to multiple external hardware or software that can be used to expand the simulator; and 3) supports multiple industry-standard protocols such as Modbus/TCP, DNP3, and the integration of proprietary protocols.
The simulator contains modules for the emulation of ICS devices (e.g., RTU, PLC, MTU, HMI) and components to implement different attacks (DoS, MitM, spoofing, eavesdropping).
SCADASim architecture includes three components: 1) a real-time scheduler; 2) a communication port implementing protocols for communication to the external environment; and 3) a simulation object that models external components within the simulation environment.
For the evaluation, the authors present two simulations: a smart meter and a wind power plant. A DoS attack and a spoofing attack are also deployed on the systems and analyzed in the paper.
SCADASim is an open-source project available on Github~\cite{SCADASimWeb}.

\textbf{SCADAVT}~\cite{Almalawi2013} is a framework to build a virtual SCADA model-based testbed designed for research in the security field purposes. 
The framework is developed on top of the CORE emulator~\cite{CoreWeb} by integrating the Modbus/TCP communication protocol between master, slave, and HMI server. 
Simulations of I/O modules are also integrated into the CORE emulator, which acts as a server, receives input data from the external environment, and sends output data when requested using a simple custom TCP-based protocol. 
The physical process is modeled using an EPANET server~\cite{EpanetWeb} which provides a graphical interface to reproduce water distribution systems.
Two attack scenarios are also presented: a DoS and manipulation of command messages.
The framework, which supports real devices' connection, is described in detail, but the source code is not publicly available.

\textbf{TASSCS} (Testbed for Analyzing Security of SCADA Control Systems)~\cite{Mallouhi2011} was developed by the University of Arizona mainly to test a novel technique to protect SCADA systems from attacks, which the authors called Autonomic Software Protection System (ASPS)~\cite{hariri2007autonomic}.
The testbed architecture is composed of different components. The Control HQ is the central command and control for all the resources and services offered. It contains the HMI, the control server, the data storage, and the engineering LAN. Through a WAN, the Control HQ is connected to a large scale electric grid modeled using the PowerWorld simulation tool~\cite{PowerWorldWeb}. Finally, a device is used to monitor all the ingress and egress communications that pass through the WAN to feed the ASPS. This last device acts as an active anomaly detector: it can identify attacks and stop them.
To enable communication, Modbus/TCP is used. The Modbus Server is simulated using Modbus RSim (the official website is not anymore available) and connected to an Opnet-based network simulator~\cite{OpnetWeb}.
The authors present various attacks, including spoofing, MitM, DoS, and data injection. Two of these attacks scenario are also implemented: a DoS attack and a compromised HMI scenario to force a complete network blackout. The protection system under testing was able to detect the two launched attacks.

\textbf{Virtual Tennessee-Eastman Testbed (VTET)}~\cite{Xie2018vtet} is a simple virtual testbed that simulates a chemical ICS with Matlab. 
The architecture is based on four components: a physical PLC, a PC used for network communication, and two other PCs simulating the physical process and a PLC. 
The process is the Tennessee-Eastman (TE)~\cite{lyman1995plant}, a nonlinear and continuous process widely used in the chemical field.
VTET can work in two different modes. In the full-virtualization mode, the physical PLC is disconnected, and the testbed is completely virtual and simulates the controller using NetToPLCSim~\cite{NetToPLCSimWeb} and PLCSim~\cite{PLCSimWeb}, the official simulator of Siemens PLC. Instead, the semi-virtualization mode allows replacing the simulated PLC with the real one. 
VTET supports three standard ICS protocols for network communication: Open Platform Communications (OPC), Modbus, and S7Comm.
The authors present and test five attacks, mainly using MitM and jamming techniques to disturb or disrupt the physical process. 
Unfortunately, the testbed is not available online to our knowledge, but the description is quite complete on the paper.


\subsection{Hybrid}\label{subsec:testbed_hybrid} 

\begin{figure}[t]
    \centering
    \includegraphics[width=\columnwidth]{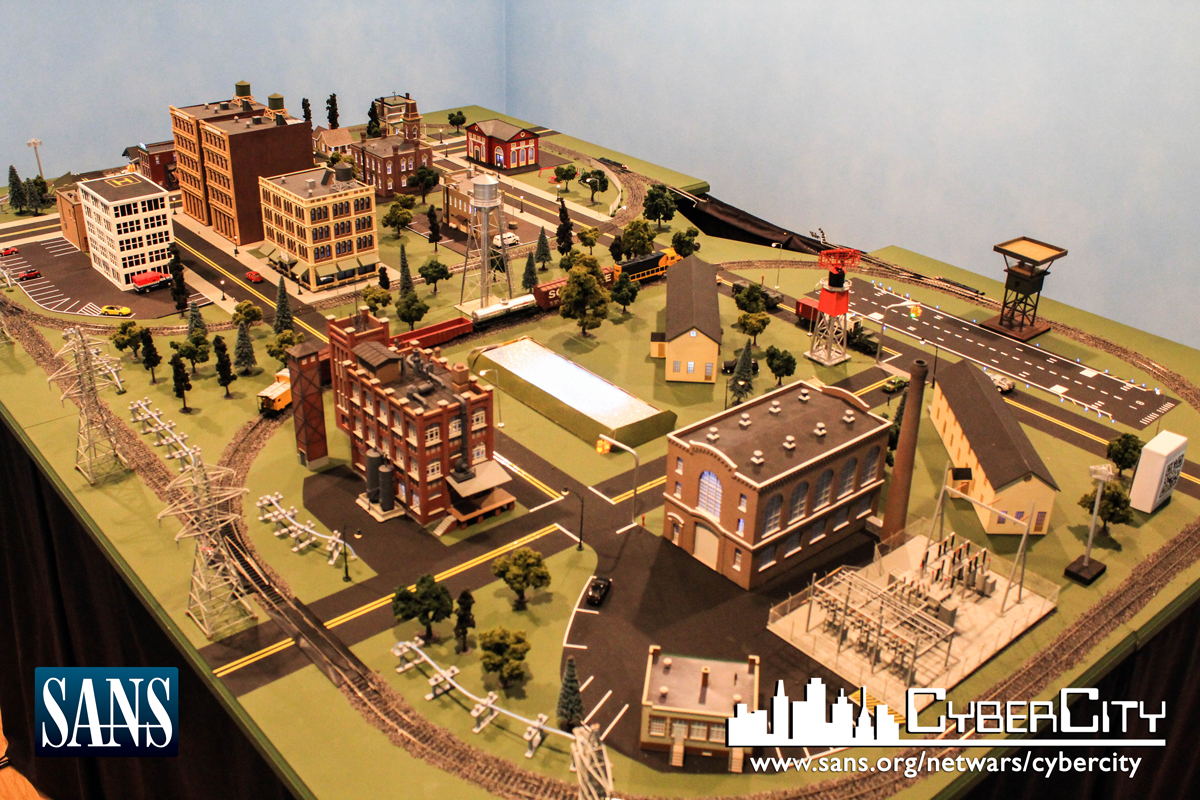}
    \caption{A figure representing CyberCity testbed.}
    \label{fig:cybercity}
\end{figure}

\textbf{CyberCity Testbed}~\cite{BorgesHink2016, CyberCityWP} is a physical representation of an entire city (Figure~\ref{fig:cybercity}) developed by the SANS Institute to test security measures on the ICS field. It includes a bank simulation, a hospital, a power plant, a train station, a water town, and many other available infrastructures. Furthermore, 15k ``people" who have e-mail accounts, work passwords, and bank deposits are generated to create a complete environment. A tabletop scale model of the town was built to visually show the effects of attacks on the electric train, the water tower, and the traffic light. 
Although the lack of official documentation, Borges Hink et al. in~\cite{BorgesHink2016} recover some details about the components of the testbed by studying a dataset generated from CyberCity, which is available online~\cite{CyberCityWeb}.
They discover a wide variety of components ranging from web server emulated using VMWare to physical Siemens PLCs, Cisco routers, and NetDuino+ controllers. The protocols used by the ICS components are mainly Modbus/TCP, EtherNet/IP, and NetBIOS.
Nowadays, the testbed is mainly used to teach cybersecurity on ICS as part of the SANS Institute courses and federal agencies to perform security research.

\textbf{Experimentation Platform for Internet Contingencies (EPIC)} by Siaterlis et al.~\cite{Siaterlis2013epic} is an innovative hybrid testbed to simulate CPSs based on Emulab~\cite{EmulabWeb}. It is developed by the Joint Research Center at the Institute for the Protection and Security of the Citizen in Ispra, Italy.
The testbed architecture comprises two control servers, a pool of physical resources used as experimental nodes (e.g., PCs, routers), and a set of switches employed to interconnect the nodes.
Every configuration step uses a web interface where a user can create a customized network.
The EPIC setup phases require a detailed description of the required topology using a formal language (i.e., an extension of Network Simulator (NS) language). 
The experiment is then instantiated by using Emulab~\cite{EmulabWeb} which can automatically configure network switches to recreate the desired virtual topology by connecting nodes using multiple VLANs. Finally, experiment-specific software can be launched through events defined in the setup script or manually by logging in to each station. 
Physical processes are simulated using Simulink Coder~\cite{SimulinkCoderWeb} and managed by a software simulation unit.
For the communications, EPIC provides tools to generate latencies for the simulation of different network types and integrate realistic background traffic datasets. It also supports industrial protocols such as Modbus through proxy units that translate calls between the simulation unit and other SCADA devices.
Furthermore, after a theoretical comparison between fidelity, repeatability, and measurement accuracy between EPIC and other popular testbeds, these characteristics are analyzed on EPIC with a deep testing phase.
The software part is open-source and freely available online ~\cite{EpicWeb} with complete documentation.

\textbf{EPS-ICS}~\cite{Gao2013eps} is a framework to implement a hybrid testbed, principally developed by the Technical Assessment Research Lab (CNITSEC) in Beijing, China. The testbed implements a multi-level design approach where Level 3, the corporate network, and Level 2, the supervisory control LAN, are emulated.
Instead, Level 1 devices, including Distributed Control Systems (DCS) controllers, PLCs, and RTUs, are real physical devices.
Finally, a mathematical model is used to simulate the physical process at Level 0, and it is implemented with Simulink~\cite{SimulinkWeb}.
This approach allows replicating the interactions between the ICS components. The communication interface between network testbed and physical devices is implemented through layer three switches with an IP routing. However, the industrial protocols used are not mentioned in the relative paper.

\textbf{Gillen et al.}~\cite{Gillen2020} presented a hybrid replication of the cooling system for Oak Ridge National Laboratory's 200-petaflop Summit supercomputer, currently declared the fastest open-science computer in the world~\cite{superc1, superc2}. Summit consists of over 4600 nodes and has a peak power draw of 13MW. The cooling system cycles through over 4000 gallons of water each minute. 
The developed replica is based on the same controller, an Allen Bradley Control- Logix PLC with 34 I/O modules distributed over six chassis connected with an Ethernet/IP ring-topology backbone. Furthermore, the HMI, the historian, the industrial switches, and the power supplies are perfect physical replicas. An engineering workstation is connected to the system to configure the different components.
On the other side, the over 500 sensors and actuators employed in the cooling system are instead emulated by using over 40 Raspberry PIs and 200 daughter boards. All the sensors and actuators communicate with the PLC using hard-wire electrical signals or an Ethernet-based signal line. For this last case, raw traffic from the production environment has been recorded. 
A software-based model of the protocol, traffic rate, and handshakes of the real cooling system was computed and employed by the Raspberry PIs to emulate the entire communication. 
The authors collected 30 days of data from the real Summit cooling system historian to correctly emulate sensors and actuators. Then they used emulation scripts to generate data from the devices. 
Each sensor and each actuator is connected to an independent display used to verify the correct measures, despite the HMI values. This is useful in the case of attacks targeting data visualization (e.g., replay attack). 
To validate the testbed, the authors compared its behavior with the real Summit supercomputer cooling system. Considering the alerts, logs, and historian data, all the data are replicated accurately. 
Instead, concerning the network traffic consistency, results show an hour-to-hour average variance under 0.01\% for the majority of the properties. Therefore, the fidelity is adequately accurate to simulate the original system properly.

Hui et al.~\cite{Hui2019hybrid} introduce \textbf{Hui Nuclear}, a hybrid testbed modeling a nuclear reactor built at the Center for Secure Information Technologies (CSIT) at the Queen's Univerity of Belfast. 
The testbed's scope is to generate a realistic network interaction and a simple way to collect network data to be used in the CPS security field.
The testbed implements four main sub-process controlled by four PLCs. The main reactor sub-process, the heat exchanger sub-process, and the heat exchanger sub-process controlled by physical Siemens PLCs. Instead, the generator sub-process is monitored by a Schneider PLC.  
The inter-communications between sub-processes are enabled by physical interactions or IP network communications through S7Comm protocol, Profinet, and a custom protocol based on TCP.  
For practical and safety reasons, the heating process and the turbine are simulated by two Raspberry PIs.
The network architecture exhaustive and contains all the Purdue model areas, together with firewalls, IDS, and logging services.

\textbf{HYDRA}~\cite{Bernieri2016hydra} is a low-cost and open-source physical emulator for critical infrastructures developed at the Università Roma Tre in Italy. 
It can be used for investigating fault diagnosis, cybersecurity strategies, and testing control algorithms. 
The testbed is designed to emulate a simple water distribution system's behavior. It employs seven tanks at the physical level deployed vertically. Each tank can be easily unconnected or moved to another position giving the testbed high modularity and flexibility.
The communications between sensors and actuators implement the Modbus protocol on a Local Area Network (LAN) to PLCs and RTUs simulated using Arduino Nano and Galileo.
The authors also present an attack scenario of a data modification attack. The code and all the testbed technical details are open-source and available on Github~\cite{HYDRAWeb}.

\textbf{Kim et al.}~\cite{Kim2019} proposed a platform to perform cybersecurity exercise for national critical infrastructure protection. The testbed was designed to replicate a realistic ICS environment that matches the characteristics of the Cyber Conflict Excercise (CCE). CCE is an annual national real-time attack-defense battlefield competition organized in South Korea and Locked Shields (LS). It is the world's largest international technical live-fire cyber defense exercise. The platform can scale and provide dozens of identical ICS setups to satisfy an increasing number of participants. With respect to standard testbeds, this project required a visualization layer representing the physical facilities and the damage caused by the attackers. To make it possible, a diorama city was considered the most cost-effective and modular approach. It contains symbolic structures representing the critical infrastructures, surrounded by residential and commercial buildings, and tri-color LED lights to introduce a physical representation of attacks' effects. 
The paper~\cite{Kim2019} describes an implementation of the proposed platform, which includes six different critical infrastructures: a power grid, a nuclear plant, a water purification plant, railroad control, airport control, and traffic light control. The system contains two PLCs of different vendors that control some typical actuators (e.g., mechanical relay, magnetic switch, motor). Furthermore, a platform with 255 LED lights was built to illustrate the state of the critical infrastructures. The control network layer is hosted by remote cloud servers and contains HMI, an engineering workstation, a historian DB, a patch management system, and office computers. The protocol adopted depends on the selected PLCs.

In~\cite{koutsandria2015real}, \textbf{Koutsandria et al.} presented a hybrid testbed for testing a real-time Network IDS. To simulate the ICS environment, the authors employ a combination of simulated and real devices. The testbed is based on Matlab Simulink to simulate the physical and control networks. In particular, the authors model the physical system with Simulink by simulating IED and field devices controlled by a PLC via Modbus in a master/slave communication model. In this setting, the authors describe the implementation of the master devices both with a SIMATIC S7-1200 PLC and a simulated PLC.
The network communication and information exchanged by the different device are collocated through a network tap implemented with a central hub and a Raspberry PIs~\cite{RaspberryWeb} running a packet dissector. The authors gave particular attention also to the data management and visualization part. All the traffic collected is saved in a historian server and managed with OSIsoft~\cite{OSIsoftWeb} PI System. Then the historian information is continuously analyzed and monitored by an IDS based on rules and behavior analysis. 
To validate this architecture and its capabilities, the authors also present three attacks (i.e., two network communication alterations and a physical behavior violation) scenario showing the effectiveness of the detection rules.

\textbf{KYPO4INDUSTRY}~\cite{Celeda2020krypto} is a training facility for students based on open-source hardware and software, built at Masaryk University in the Czech Republic.
This testbed consists of a laboratory room designed to help computer science students to learn cybersecurity in a simulated industrial environment.
The laboratory is divided into different tables to split the students into groups and give everyone the possibility to have hands-on experience on the entire system.
Tables can be moved and rearranged around the room to generate a flexible environment for every possible activity, ranging from team assignments to student presentations. 
A control panel exposes the I/O modules on each table, and the touchscreen is used to interact with PLCs (simulated using Raspberry Pi~\cite{RaspberryWeb}), linear motor, and communication gateway. The software stack includes the Linux OS, Docker ecosystem, and on-premise OpenStack cloud environment to achieve an automated orchestration. Thank the open-source hardware and software used in the system, and different industrial protocols can be implemented, such as Modbus or DNP3. 
Finally, the paper introduces the university's course syllabus that employs the facility, showing the arguments addressed on each of the 13 weeks of the course.

\textbf{LegoSCADA}~\cite{Rubio2017lego, LegoSCADAWeb} is a cost-effective hybrid testbed developed at the Universite Paris-Saclay in France. 
The testbed's conceptual architecture is based on three block elements: the controller, the system, and the sensors. 
The controller reads data from the sensors, computes new information, and transmits new commands to the actuators. Many RTU and PLCs can be connected to the controller based on the system that we want to represent. The protocols supported are Modbus and DNP3. 
To test the architecture, the authors have developed a test scenario based on Lego Mindstorms EV3 brick~\cite{rollins2014beginning} which emulates a PLC on a car, a Raspberry Pi~\cite{RaspberryWeb} to emulate an RTU connected to the vehicle, and a personal computer as a controller. 
The controller is always correcting the car speed and polling the distance between the car and an obstacle. Furthermore, a single RTU and a single controller can control more PLCs, and, therefore, more cars can be connected to the testbed.
MitM attacks are deployed on the developed testbed, in particular replay attacks and injection attacks. Moreover, a watermark authentication technique has been tested to stop the attacks with interesting results.

\textbf{LICSTER}~\cite{sauer2019licster, web:licster} is an open-source and open-hardware testbed presented at the Hochschule Augsburg in Germany. Its main target is to give students and researchers an affordable system to perform security research with an expense of about 500 euros. 
The system is composed of an OpenPLC~\cite{alves2014openplc}, an HMI built using a web server and a SCADA system. Each of these components is loaded on three dedicated Raspberry PIs.  
The physical process implemented is a representation of an industrial process provided by Fischertechnik~\cite{fischertechnikWeb}. A conveyor belt is used to move a plastic cylinder to a punching machine, which is then activated. In the end, the cylinder is taken back to the original position. The process can be easily substituted with others.    
Modbus/TCP is the protocol used to enable communication between components.  
Different attack scenarios on LICSTER are presented and tested. The authors cover widely used threats to levels 0, 1, and 2, such as passive/active sniffing, Dos, MitM, and manipulation over the network. For each attack, an evaluation is presented containing useful information (e.g., impact, skill level, detection difficulty). 
Scripts and instruction on the implementation are available on the Github repository~\cite{web:licster}.

\textbf{Microgrid}~\cite{Guo2013} is a flexible and adaptable testbed developed by The Ohio State University, composed of a hybrid setup of physical hardware and real-time simulations. The testbed contains Power Hardware-In-the-Loop (PHIL) able to emulate power hardware not installed in the testbed, along with a real-time SCADA system with an OPNET~\cite{OpnetWeb} based real-time System-In-the-Loop (SITL) communication network simulation system. 
PHIL can emulate several components like stationary battery unit, charging station, renewable energy resources, 9-bus or 14-bus systems. It is also possible to connect physical components to PHIL. The paper presents an implementation of a 5kVA charging system of a simulated electric vehicle, a photovoltaic system, local energy storage, and different power electronic circuits.  
The three main components of the simulated SCADA environment are the data acquisition, the real-time virtual communication network, and the real-time control center with the HMI. 
The authors introduce a case study implementation by connecting local energy storage and a second power grid PHIL simulation. Furthermore, the authors validate the case study with experimental results and analysis. 
The testbed is designed to study topics related to smart grid and provide hands-on experience to students.

\textbf{MSICST} (Multiple-Scenario Industrial Control System Testbed)~\cite{xu2019msicst} is a hybrid representation of four different ICS scenarios: a thermal power plant, a rail transit, a smart grid, and intelligent manufacturing. Physical processes are always simulated while the control systems are built using commercial hardware and software. Furthermore, in some scenarios, a combination of software simulation and actual physical equipment is used to build a more realistic scenario. MSICST also contains an attacker model and a monitoring network. 
The thermal plant comprises four PLCs of different manufacturers used to manage the three simulated physical systems: combustion system, steam-water system, and electrical system. A sand table is synchronized with the simulation to visualize what is occurring using Light Emitting Diode (LED), fans, and smoke generators. 
The rail transit scenario includes three stations, two trains, and a circular rail transit line. All the components are realized in a sand table as a scaled-down version of a real system. To achieve automatic control of trains and station components, the authors use two Siemens PLCs. 
Regarding the smart grid, the testbed mainly focuses on the power consumption part. It contains two smart meters, a concentrator, and a  station device, which can, for instance, display the power consumption of an area.
Finally, the intelligent manufacturing scenario is based on a Computer Numerical Control (CNC) that contains a controller, memory, and HMI. Moreover, a Distributed Numerical Control (DNC) system was developed to improve the manufacturing industry's intelligence level.
A DMZ containing the data historian and the HMI server is generated to separate the OT area from the enterprise zone. This latter simulates an office by using a PC with Windows 7.  
The protocol used for communication in the OT area are mainly Modbus/TCP, S7Comm, and EtherNet/IP, depending on the PLC used.
Some vulnerability discovery experiments have been done on MSICST, ranging from discovering vulnerabilities on a specific type of PLC to some attacks to known vulnerabilities of S7Comm and Modbus provided by the lack of encryption and identity authentication. Some security measures are presented as well, like a whitelist-based host protection software and a new IDS solution that combines traditional IT system IDS with behavior-based ICS-specific IDS.

\textbf{NIST} (National Institute of Standards and Technology) developed a cybersecurity testbed for ICS presented in detail in~\cite{Testbed2015nist}.
The testbed is designed to emulate three real-world industrial systems without replicating the entire plant or assembling a complete system. 
The first system is a Tennessee Eastman (TE) problem~\cite{lyman1995plant}, a widely used process in the chemical manufacturing field. The TE process is simulated using an open-source code~\cite{TESimWeb}, and it is connected to physical devices such as switches, PLCs, HMI, and terminals through different protocols such as OPC, Ethernet/IP, and DeviceNet. 
The second is an entirely physical cooperative robotic assembly system for smart manufacturing. It contains a PLC, controllers, buttons for emergency stops, HMI, and two robots. These devices are interconnected through Ethernet, EtherCAT, Serial, Modbus, and Analog/Digital signals based on the components' needs.
Finally, the third simulates a pipeline network with a Wide Area Network (WAN) SCADA infrastructure and an intelligent transportation system, including public infrastructure components, cooperative real-time embedded components, and wireless components. However, this last testbed was only introduced in the paper and was not implemented at the publication time (i.e., 2015).
The testbed is available upon request to academia, government, and industry to analyze new technologies.  
Based on the research on these testbeds, NIST published a long and complete guide to ICS security in 2015~\cite{Stouffer2015nist}.

\textbf{PNNL}~\cite{Edgar2011} by Edgar et al. at Pacific Northwest National Laboratory is a remotely-configurable and community-accessible hybrid testbed to support research on cyber-physical equipment. 
This testbed combines physical, simulated, and virtual components giving considerable implementation flexibility. In fact, the testbed allows simulating from small systems like traffic lights to extremely complex scenarios such as power grids. 
The testbed is composed of many different back-end functionalities that can be employed in user management. Each user can remotely deploy its own system configuration and manage the operation and data gathering process.
Furthermore, users can control different areas of the architecture, including the environment (used to simulate the physical process), devices (e.g., PLCs, RTUs), network communication (representing the backbone communication), simulation, and device integration.
The testbed is accessible following the indications provided in the PNNL website~\cite{PNNLWeb} and using Arion~\cite{ArionWeb} as modeling software.

\textbf{SNL Testbed}~\cite{Urias2012} is a complex hybrid testbed built by Sandia National Labs in Albuquerque, USA. It contains simulated components (i.e., represented using a model in OPNET~\cite{OpnetWeb}), emulated nodes (i.e., using real software running on an emulated machine), and physical (i.e., real software running in real hardware) devices. 
The reference paper also includes an accurate explanation concerning the connection between the various components. 
The testbed is presented as a case study used to model a complex scenario, containing: the corporate network (connected to the Internet), a DMZ, a control system network (containing HMI, the SCADA Server, Engineering Workstation, and Front End Processor), and the field layer (containing sensors, RTUs, and IEDs).
The protocols implemented are Modbus/TCP, DNP3, and IEC 60870.
Finally, the authors present a security assessment of the testbed considering different threats and attacks such as reconnaissance, resistance to standard penetration tools (e.g., Metasploit~\cite{maynor2011metasploit}), and MitM.

\textbf{VPST} (Virtual Power System Testbed)~\cite{Bergman2009} of the University of Illinois is designed to be integrated with other testbeds across the country to explore SCADA protocols and equipment's performance and security.
Thanks to its easy integration with real devices and testbeds, VPST has the advantage of having actual HIL and a faithful communication system. The architecture is divided into three main subsystems: the first handle electrical simulation using PowerWorld~\cite{PowerWorldWeb}, the second simulates the communication systems using RINSE~\cite{Rinse}, and the third includes all the actual devices.
Furthermore, a framework for the Inter-Testbed Connection (ITC) is integrated with VPST. This framework is based on low bandwidth and reliable control plane and a high bandwidth data plane. ITC requires secure connectivity, which is achieved using OpenVPN and IPSec. Moreover, the implementation of performance, reproducibility, and resource allocation properties are addressed in the paper. Fidelity is another essential property achieved by implementing real industrial protocols such as DNP3 or Modbus, leaving the possibility of testing new versions and protocols (e.g., DNP3SA that provides Secure Authentication).
The paper presents some example use cases: attack robustness analysis, incremental deployment analysis, and Human-in-the-loop event analysis. However, thanks to its flexibility, the testbed is suited for many different types of research.

\section{ICS Datasets}\label{sec:dataset} 

In this section, we provide a description of the ICS dataset available in the literature, highlighting the key design point and the most interesting and performant IDS applied to them. In Section~\ref{subsec:dataset_classification} we outline the classification method that we use in the following sections, while in Section~\ref{subsec:data_challenges_req} we introduce the main requirements and challenges in developing a dataset. Furthermore, we summarized in 
In Section~\ref{subsec:metrics} we briefly recall the common evaluation metric for IDS. Then, in Section~\ref{subsec:dataset_physical} we present the datasets offering only physical level data, while in Section~\ref{subsec:dataset_network} we describe network level datasets. Finally, in Section~\ref{subsec:dataset_phynet} we highlight datasets containing both the information.

\subsection{Datasets Classification}\label{subsec:dataset_classification}

Datasets are a collection of data recorded from a testbed or synthetically produces, which can be used to train and test an IDS. Unlike datasets concerning IT systems, which are composed only of network traffic, to characterize an ICS, a dataset must contain both network traffic, representing the communications between the various devices, and the physical processes' measurements. 

Datasets are generally shared as csv, arff, or pcap format files, depending on the typology of data collected. An interesting solution introduced by Morris et al.~\cite{Morris2014} consists of providing also some datasets containing only a subset of the data. They can be used, for instance, to quickly look at the data without downloading huge files or training a preliminary algorithm during the early stages of development. 

There are many ways to categorize datasets. For example, Choi et al. in~\cite{Choi2019datasetSurv} groups datasets based on attack path.
In this survey, we decided to divide datasets based on the typology of the collected data.
The capturing can contain data at \emph{physical level} i.e., field data such as measures from sensors, actuator, and other physical level devices, or \emph{network level} data, i.e., packet or flow sent in the channel under control.
However, datasets can contain both the typology of data, and so they are considered both \emph{physical and network level}.
Sometimes, it is possible to find other types of data, like device logs, to better understand the ICS's behavior.
To perform our study and provide reliable statistics, we downloaded every dataset and analyzed it reporting the main interesting properties. 

Table~\ref{tab:dataset} summarizes the main features and statistics of the presented datasets. We reported the following features.
\begin{itemize}
    \item\textbf{Name} of the dataset (or of the authors if a name is not provided);
    \item\textbf{Sector} indicates the field of the source ICS; 
    \item\textbf{Data} type provided. Can be 
    \begin{itemize}
        \item\emph{Logs} if logging information of the system during the process are available;
        \item\emph{Network} if network traffic data are provided;
        \item\emph{Physical} if measurements of sensors and actuator states are available;
    \end{itemize}
    \item\textbf{Time} provide an approximation of the duration of the recording;
    \item\textbf{Entries} indicates an approximate number of entries contained in the dataset. In case of datasets containing different versions, the most used or the most recent is considered;
    \item\textbf{Reference} includes a reference to a description of the dataset;
    \item\textbf{Resource} indicates a webpage in which the dataset is downloadable or information about how to retrieve it are available;
    \item\textbf{Attacks} specified the categories of attacks contained in the dataset, if any. Can be Reconnaissance, Replay, MitM, DoS, Injection
    , or Others which contains less used categories. More information about the attacks are presented in Section~\ref{subsec:attacks}.
    \item\textbf{\%} indicates the percentage of data under attack on the total entries, if any;
    \item\textbf{Format} indicates the format of the files containing the capture. Can be:
    \begin{itemize}
        \item\texttt{pcap} is a widely used format containing network packets;
        \item\texttt{csv} is an extension for files containing Comma Separated Values;
        \item\texttt{log} contains textual logging of events;
        \item\texttt{xslx} is a format for spreadsheet files;
        \item\texttt{arff} is a format used to save data for databases in a textual format. It is generally used with Weka~\cite{wekaWeb};
        \item\texttt{inp} contains data of emulations. In this context, it is generally used with Epanet~\cite{EpanetWeb}.
    \end{itemize}
    \item\textbf{IDS} contains a reference to the best IDS available in literature applied on the testbed at the best of authors knowledge; 
    \item\textbf{F1-score}, \textbf{Accuracy}, and \textbf{Precision} represent the evaluation metrics of the IDS specified, according to Section~\ref{subsec:metrics}.
\end{itemize}

The detection algorithms selected are implemented on the whole dataset and not on a fraction of it. Furthermore, the selection does not take into consideration the rank of the publication venue of the paper. 
For some datasets and IDSs were not possible to obtain all the information since the related paper does not provide exhaustive information. Thus, the degree of depth of analysis may not be the same for all work. 

\subsection{Datasets Challenges \& Requirements}\label{subsec:data_challenges_req}
There are several challenges in generating a valuable dataset. Therefore it is fundamental to create it by following a suitable methodology and keeping in mind the design requirements.
Gomez et al.~\cite{Gomez2019electra} described a framework useful to generate reliable anomaly ICS datasets to be employed in anomaly detection tasks. Firstly, it is important to select a priori, one or more attacks that will be implemented. To do so, researchers must know the main protocols used in the field of interest, discover the related threat, and design attacks according to the related vulnerabilities. 
Then, attacks can be deployed, carefully choosing the nodes affected, each attack's duration, and its starting time. Finally, it is possible to capture network packets and/or data from sensors and actuators: it is essential to define the data capture duration, the sampling frequency and smartly choose the collecting point. Generally, the latter should be a central node of the system. 
The last step is the final dataset generation. To generate the dataset to release, it is important to carefully choose the features useful to describe the system under consideration. The behavior of the system can be represented at packet-level, flow-level, or physical-level data.
The \textbf{deployment of attacks} in datasets is probably the most challenging phase. In fact, if not accurately performed, the attacks generated can lead to an inaccurate system representation or bias in the detection methodology. There are principally two ways to generate attacks. The first one, and the most accurate one, is to attack the testbed in real-time, recording the corresponding network traffic or the ICS's physical state. 
Another strategy is to insert synthetic malicious data, a posterior, in a dataset with regular operation. However, this strategy could lead to inaccuracies and may not accurately represent the real system behavior response. In fact, if we want to inject packets on a dataset with normal operations, we must consider all the complex cascade relations of the systems. By breaking these relations, we would leave a trace that an IDS can exploit to detect anomalies, creating detection bias. Since this property is not present in real systems, the IDS will miss most of the attacks in physical environments, reducing the detection generalization in other systems. It is one of the main problems of Lemay et al. dataset~\cite{Lemay2016}, which use tools such as Metasploit~\cite{maynor2011metasploit} to inject the malicious traffic.
Another critical concern causing the lack of available datasets from real environments (i.e., ICS of companies) is related to the collected \textbf{data's privacy}. 
In fact, companies may be reluctant to share their internal configurations, intellectual property, or proprietary protocols. Moreover, giving the public access to an industrial site data may allow malicious users to identify vulnerabilities and exploit them to attack the company.
As a result, many datasets are generally generated from scale-down testbeds and the few real ICS environments.
Since many intrusion detection techniques are supervised, a complete dataset must provide \textbf{labels} indicating normal or abnormal data.
Furthermore, labels are essential as ground truth for the evaluation of detection performances during the test phase.
However, the labeling process is not always straightforward. For example, some attacks can move the system in abnormal behavior after a long time the malicious packets have been sent. In this scenario, the data labeled as malicious should start when the actual attack starts or when the system's behavior starts to be compromised?
An analysis of this problem can be found in~\cite{8805036} and~\cite{Lemay2016}. 
In both cases, the solution could raise a problem in the ground truth.
Therefore, there is no right or wrong answer to this question. It depends on the context and the attack type, but it must be specified in the dataset's documentation to allow researchers to act accordingly. 

\subsection{Evaluation Metrics}\label{subsec:metrics}

In this section, we briefly recall the metrics used to evaluate the performances of the detection algorithms. According to the literature, the most common metrics are Accuracy and F1-Score.
They are defined as follows.

\begin{itemize}

    \item \textbf{Accuracy}: represents the fraction of correct predictions of the model under consideration. In the binary classification case, the accuracy is defined in terms of positives and negatives samples classified as follows:
    \begin{equation}
        Accuracy= \frac{TP+TN}{TP+TN+FP+FN},
    \end{equation}
    where TP = True Positives, TN = True Negatives, FP = False Positives, and FN = False Negatives.
    
    \item \textbf{F1-Score}: is a metric used to evaluate a classification, defined as the harmonic mean between \textit{precision} and \textit{recall} as follows:
    \begin{equation}
        F1-Score=2 \cdot \frac{precision \cdot recall}{precision+recall},
    \end{equation}
    
    where the \textit{true negative rate}, or precision is: 
    \begin{equation}
        precision=\frac{TP}{TP+FP},
    \end{equation}
    
    while the \textit{positive and negative predictive values}, or recall, is:
    \begin{equation}
        recall=\frac{TP}{TP+FN}.
    \end{equation}
    
\end{itemize}



{
\clearpage
\onecolumn
\renewcommand{\arraystretch}{1.4}

\begin{landscape}
\begin{center}
\tablecaption{Summary of datasets presented in the literature. The \textbf{Data} type indicated as L: Logs, N: Network; P: Physical. \textbf{Times} of recording are estimations and measure units are h: hours, d: days, m: months. \textbf{Entries} numbers are estimations, too. We denote the \textbf{Attacks} launched during the recording as RC: Reconnaissance; RP: Replay; M: MitM; I: Injection; D: DoS; O: Others. The \textbf{\%} column indicate the percentage of data under attack with respect to the whole dataset. File \textbf{Formats} are indicated as P: pcap; C: csv; L: log; A: arff; I: inp; and X: xlsx. *: the version of WADI dataset considered is the one dated November 2019; the one of SWaT dataset is instead A1 dated 2015, the most used one as the best of the authors' knowledge.}

\tablefirsthead{%
\hline
\textbf{Name}                & \cellcolor[HTML]{EFEFEF}\textbf{Sector} & \cellcolor[HTML]{EFEFEF}\textbf{Data} & \cellcolor[HTML]{EFEFEF}\textbf{Time} & \cellcolor[HTML]{EFEFEF}\textbf{Entries} & \cellcolor[HTML]{EFEFEF}\textbf{Ref.} & \cellcolor[HTML]{EFEFEF}\textbf{Res.} & \cellcolor[HTML]{EFEFEF}\textbf{Attacks} & \cellcolor[HTML]{EFEFEF}\textbf{\%} & \cellcolor[HTML]{EFEFEF}\textbf{Formats} & \cellcolor[HTML]{EFEFEF}\textbf{IDS} & \cellcolor[HTML]{EFEFEF}\textbf{F1} & \cellcolor[HTML]{EFEFEF}\textbf{Acc.} & \cellcolor[HTML]{EFEFEF}\textbf{Prec.} \\ \hline
}

\tablehead{%
    \multicolumn{14}{c}%
    {{\bfseries \tablename~\thetable{} -- continued from previous page}} \\
    \hline
    \textbf{Name}                & \cellcolor[HTML]{EFEFEF}\textbf{Sector} & \cellcolor[HTML]{EFEFEF}\textbf{Data} & \cellcolor[HTML]{EFEFEF}\textbf{Time} & \cellcolor[HTML]{EFEFEF}\textbf{Entries} & \cellcolor[HTML]{EFEFEF}\textbf{Ref.} & \cellcolor[HTML]{EFEFEF}\textbf{Res.} & \cellcolor[HTML]{EFEFEF}\textbf{Attacks} & \cellcolor[HTML]{EFEFEF}\textbf{\%} & \cellcolor[HTML]{EFEFEF}\textbf{Formats} & \cellcolor[HTML]{EFEFEF}\textbf{IDS} & \cellcolor[HTML]{EFEFEF}\textbf{F1} & \cellcolor[HTML]{EFEFEF}\textbf{Acc.} & \cellcolor[HTML]{EFEFEF}\textbf{Prec.} \\ \hline
}

\tabletail{%
    \hline
}
\begin{supertabular}{|>{\columncolor[HTML]{EFEFEF}}l |l|l|l|l|l|l|l|l|l|l|l|l|l|}
\textbf{D5: Energy M.S.D.} & Energy Manag.                           & L                                     & 30d                                   & 6M                                       & -                                     & \cite{MorrisWeb}                      & -                                        & -                                   & C                                       & -                                    & -                                   & -                                     & -                                      \\ \hline
\textbf{QUT\_DNP3}           & Power Grid                              & N, L                                  & 40d                                   & 31M                                      & \cite{Rodofile2013}                   & \cite{RodofileDNP3Web}                & RC, RP, M, I, O                          & $\sim$0.01                          & P, L                                    & -                                    & -                                   & -                                     & -                                      \\ \hline
\textbf{QUT\_S7Comm}         & Mining Refinery                         & N, L                                  & 17.5h                                 & 2M                                       & \cite{Rodofile2017qut}                & \cite{RodofileS7Web}                  & M                                        & $\sim$10                            & C, L, P                                 & -                                    & -                                   & -                                     & -                                      \\ \hline
\textbf{4SICS}               & Generic ICS                             & N                                     & 46h                                   & 3M                                       & -                                     & \cite{4SICSWeb}                       & unk                                      & unk                                 & P                                       & \cite{Yu2018}                        & $\sim$1                             & $\sim$1                               & -                                      \\ \hline
\textbf{CyberCity Dataset}   & City                                    & N                                     & 16d                                   & 170K                                     & \cite{BorgesHink2016}                 & \cite{CyberCityWeb}                   & I, M, D, RC, O                           & 16.58                               & P                                       & -                                    & -                                   & -                                     & -                                      \\ \hline
\textbf{D2: Gas Pipeline}    & Gas Pipeline                            & N                                     & -                                     & 400K                                     & \cite{Beaver2013Morris2}              & \cite{MorrisWeb}                      & I                                        & 0.97                                & C                                       & \cite{Beaver2013Morris2}             & 0.75                                & -                                     & 0.75                                   \\ \hline
\textbf{D3b: Water S. T.}    & Water Storage                           & N                                     & -                                     & 230K                                     & \cite{Morris2014}                     & \cite{MorrisWeb}                      & RC, I, D                                 & 27                                  & A                                       & \cite{Demertzis2017}                 & 0.981                               & 0.981                                 & 0.981                                  \\ \hline
\textbf{D4: New Gas P.}      & Gas Pipeline                            & N                                     & -                                     & 270K                                     & \cite{Morris2015}                     & \cite{MorrisWeb}                      & M, I, O                                  & 21.86                               & A                                       & \cite{Demertzis2017}                 & 0.988                               & 0.988                                 & 0.988                                  \\ \hline
\textbf{Electra Modbus}      & Power System                            & N                                     & $>$12h                                & 16M                                      & \cite{Gomez2019electra}               & \cite{ElectraWeb}                     & RC, I, RP                                & 5.2                                 & C                                       & \cite{Gomez2019electra}              & 0.987                               & -                                     & 0.988                                  \\ \hline
\textbf{Electra S7Comm}      & Power System                            & N                                     & $>$12h                                & 387M                                     & \cite{Gomez2019electra}               & \cite{ElectraWeb}                     & RC, I, RP                                & 1.42                                & C                                       & \cite{Gomez2019electra}              & 0.996                               & -                                     & 1.000                                  \\ \hline
\textbf{HVAC\_Traces}        & HVAC                                    & N                                     & 7d                                    & 40M                                      & \cite{Ndonda2019hvac}                 & \cite{HVACWeb}                        & -                                        & -                                   & P                                       & -                                    & -                                   & -                                     & -                                      \\ \hline
\textbf{Lemay Covert}        & Breakers                                & N                                     & 6.55h                                 & 1.6M                                     & \cite{Lemay2016}                      & \cite{LemayWeb}                       & Covert Channel                           & 100                                 & P, C                                    & -                                    & -                                   & -                                     & -                                      \\ \hline
\textbf{Lemay SCADA}         & Breakers                                & N                                     & $\sim$6h                              & 900K                                     & \cite{Lemay2016}                      & \cite{LemayWeb}                       & RC, I, O                                 & 3.29                                & P, C                                    & \cite{Anton2018}                     & 1.000                               & 1.000                                 & -                                      \\ \hline
\textbf{Modbus SCADA \#1}    & Liquid Pump                             & N                                     & $\sim$24d                             & 41M                                      & \cite{tjcruz2018dataset}              & \cite{tjcruzWeb}                      & M, D                                     & 4.81                                & P                                       & \cite{Radoglou2020}                  & 0.775                               & 0.812                                 & 0.964                                  \\ \hline
\textbf{S4x15 ICS}           & Generic ICS                             & N                                     & $<$1h                                 & 310K                                     & -                                     & \cite{S4x15Web}                       & unk                                      & unk                                 & P                                       & \cite{Yu2018}                        & $\sim$1                             & $\sim$1                               & -                                      \\ \hline
\textbf{WUSTL-IIOT-2018}     & Water Control                           & N                                     & 25h                                   & 7M                                       & \cite{Teixeira2018}                   & \cite{TeixeiraWeb}                    & O                                        & 6.07                                & C                                       & \cite{Teixeira2018}                  & 1.000                               & 1.000                                 & 1.000                                  \\ \hline
\textbf{D1: Power System}    & Power System                            & P, L                                  & -                                     & 78K                                      & -                                     & \cite{MorrisWeb}                      & I, O                                     & 71.02                               & C, A                                    & \cite{BorgesHink2014}                & 0.955                               & 0.950                                 & 0.980                                  \\ \hline
\textbf{EPIC Dataset}        & Generic ICS                             & P, N                                  & 4h                                    & 5K                                       & \cite{Adepu2018epic}                  & \cite{iTrustDatasets}                 & -                                        & -                                   & P, C                                    & -                                    & -                                   &                                       &                                        \\ \hline
\textbf{QUT\_S7 (Myers)}     & Generic ICS                             & P, N                                  & 8.5h                                  & 15M                                      & \cite{Myers2018}                      & \cite{MyersWeb}                       & I, M                                     & $<$0.001                            & P, L, X                                 & \cite{Myers2018}                     & 0.744                               & -                                     & 0.727                                  \\ \hline
\textbf{SWaT Dataset$^*$}        & Water Treatment                         & P, N                                  & 11d                                   & 950K                                     & \cite{Goh2017swatdata}                & \cite{iTrustDatasets}                 & I, O                                     & 5.76                                & P, C, X                                 & \cite{Abdelaty2020}                  & 0.889                               & -                                     & 0.919                                  \\ \hline
\textbf{BATADAL}             & Water Distribution                      & P                                     & 22m                                   & 13K                                      & \cite{Taormina2018batadal}            & \cite{BatadalWeb}                     & RP, M, O                                 & 1.69                                & C, I                                    & \cite{Housh2018}                     & 0.970                               & 0.989                                 & 0.987                                  \\ \hline
\textbf{HAI Dataset}         & Power Plant                             & P                                     & 10d                                   & 1M                                       & \cite{Choi2020}                       & \cite{HAIWeb}                         & RP, M                                    & 1.83                                & C                                       & \cite{Kim2020}                       & 0.780                               & -                                     & 0.950                                  \\ \hline
\textbf{WADI Dataset$^*$}        & Water Distribution                      & P                                     & 16d                                   & 950K                                     & \cite{Ahmed2017wadi}                  & \cite{iTrustDatasets}                 & M                                        & 1.04                                & C                                       & \cite{Abdelaty2020}                  & 0.804                               & -                                     & 0.908                                  \\ \hline
\end{supertabular}
\label{tab:dataset}
\end{center}
\end{landscape}
}
\clearpage
\twocolumn

\subsection{Physical Level}\label{subsec:dataset_physical} 

\textbf{BATADAL (BATtle of the Attack Detection ALgorithms)}~\cite{Taormina2018batadal, iTrust, BatadalWeb} was a design challenge aimed at the creation of an attack detection algorithm. Every participant was provided with three datasets containing observations of a simulated C-Town network~\cite{taormina2019epanetCPA}, a real-world, medium-sized water distribution system operated through PLCs and SCADA systems, which allows modeling the hydraulic response of a water distribution network under attack. The dataset, provided in csv format, contains SCADA reading for 43 system's variables and is designed for different purposes: two are thought for training (1 year in normal behavior and 6 months with some partially labeled attacks) and one for testing (4 months with unlabeled attacks). The 14 cyberattacks conducted on the system include malicious activation of actuators, change of actuators settings, replay, and MitM attacks. In the paper~\cite{Taormina2018batadal}, the authors present the dataset and the evaluation criteria for the competition (time-to-detection and classification performance). Furthermore, it briefly explains the strategies employed by each participant. The dataset is free and available in csv format. There is available also an inp file that can be used with EPANET2 to simulate the system. A new version of the dataset is also available at~\cite{Batadal_new} contains sensor readings without concealment and is discussed in~\cite{erba2020concealment}.

The challenges' participants developed different algorithms for intrusion detection ranging from Random Forest to Recurrent Neural Network. 
Housh and Ohar~\cite{Housh2018} achieved the best result by proposing a model-based fault detection approach that employs a simulator to generate benign data and then compare them to the available SCADA readings to detect anomalous behaviors. 
This approach is composed of three main phases: 1) available SCADA data are used in a Mixed-Integer Linear Program to estimate the water demand in each node; 2) EPANET simulator is used to generate reference values, which are used to produce simulation errors when compared to actual readings; and 3) a multi-level classification approach is implemented to classify the obtained simulation errors into events and normal conditions. The result shows a Precision of 0.987, and Accuracy of 0.989, and an F1-Score of 0.970.
In~\cite{Kravchik2019} Kravchik and Shabtai present a detection approach base on under-complete Autoencoder in the frequency domain, which could reach an F1-Score of 0.937, which is high, considering the simplicity and non-specificity of the used algorithm. The presented paper was applied to the first version of BATADAL. Finally, we must consider that BATADAL is synthetically generated. Therefore this dataset does not suffer from significant noisy problems, making anomaly detection easier.


Morris et al. presented different ICS datasets, which are available online~\cite{MorrisWeb}. Each dataset's name is labeled with a number from 1 to 5 and the involved industrial sector.

\textbf{Dataset 1: Power System Datasets} are a collection of three datasets provided by Morris et al.~\cite{MorrisWeb, Pan2015morris1} containing the same data but with various labels. 
One dataset has binary labels (i.e., Normal and Attack). The second dataset has three-class labels (i.e., Attack, Natural, and No Events), which identity as ``Natural" events single line-to-ground (SLG) faults and line maintenance and as ``No Events'' the normal operations.
Finally, the third dataset includes 41 different labels containing more information about the attacks and various events. In particular, one label is reserved for ``No Events" (i.e., Normal Operation), eighth labels contain different classes of the ``intensity'' of the ``Natural" samples previously mentioned. The remaining 32 labels identify different attacks such as Data Injection, Command Injection, and Relay Setting Change. All the details about the labeling process are available in the readme file at~\cite{MorrisWeb}.
Physical measures and logs from the control panel, relays, and Snort captures are collected from a physical testbed containing two power generators, four IEDs that can switch four breakers, all connected through switches and routers. 
Data are provided as a csv file (for the first two datasets) and ARFF format (for the third dataset).

Different interesting IDSs are implemented on these datasets~\cite{Pan2015morris1, BorgesHink2014, Pan2015d, Pan2015c}.   
In~\cite{BorgesHink2014} different machine learning-based anomaly detection algorithms are tested against the three datasets. The most performant method was JRipper algorithm~\cite{JRipper} together with Adaboost~\cite{Adaboost} to improve the performance. Results show an F1-score, recall, and precision almost always greater than 0.8, with a peak of F1-Score of 0.955 in the three-class dataset. Also, Accuracy was always greater than 0.85.
Even if the authors did not include any numerical results based on common metrics, it is worth mention another approach presented in~\cite{Pan2015morris1}. The authors presented a specification-based intrusion detection framework, which is tested in the discussed dataset.
They implemented a Bayesian network to model different threat scenarios. The authors' purpose was to build a network with a unique path for each threat scenario. In other words, each scenario must be described as a sequence of system states, actions, and events that uniquely identify it. 
For each threat identified, the system collected related measurable variables and events. Then, each scenario is divided into actions that cause the system state transition. Finally, the Bayesian network is built on an independent path of states, computed for each threat.
An IDS was implemented starting from the Bayesian network obtained, which reads states and logs to track the system states. The obtained IDS can classify ten different scenarios containing both faults and cyber-attacks by monitoring the state transitions, with different precision based on the relay location.

\textbf{Dataset 5: Energy Management System Data}~\cite{MorrisWeb} is a large anonymized log collected by an Energy Management System (EMS) in a utility in the United States of America. 


The dataset's csv contains the timestamp and ID of each event, the SCADA category (i.e., information of the type of event), each device type, the event message, the priority code, the name of the substation, and the area of responsibility (i.e., the controlling authority). Data are collected in a period of 30 days.
Since the dataset contains only normal operation, no attacks are provided. For this reason, to the best of the authors' knowledge, there are not IDS implemented on this dataset.

\textbf{HAI Dataset} (HIL-based Augmented ICS)~\cite{Shin2020, Choi2020, HAIWeb} is a collection of physical data from three physical control systems (a GE's turbine, Emerson's boiler, and a FESTO's water treatment systems) combined through the dSPACE HIL


\noindent simulator~\cite{Shin2019}. 
Data were sampled every second in 59 points representing the variables measured or controlled by the control system.
Basing on the GitHub repository~\cite{HAIWeb} (which currently differs from~\cite{Shin2020}), the data collected contains seven days of normal system behavior, a day with 20 different attack scenarios on each control loop, and two days with 14 attacks on multiple control loops, for a total of 10 days of capturing. Totally, there are around 1 million samples, 1.83\% of them are labeled as under MitM attacks, in particular relay and modification attacks. However, all the attacks are deeply explained in~\cite{Shin2020}.
Data are provided in a csv format with a document that accurately depicts the testbed architecture and the dataset's data.

Due to the novelty of the dataset, released in 2020, there is a lack of IDS implemented on this dataset. However, in~\cite{Kim2020} the authors present an anomaly detection strategy based on clustered deep one-class classification (CD-OCC). It is an unsupervised approach that combines clustering algorithms with deep learning (DL) models. In particular, K-means was applied for clustering on the training set. Then, different types of neural networks (e.g., DNN, CNN, RNN) were implemented to predict the clusters and return softmax values classified with the iForest algorithm.
Currently, on the HAI Dataset, the higher precision is achieved using DNN as cluster predictor (0.95) while the overall higher scores are obtained with CNN as cluster predictor (F1-score: 0.78; Precision: 0.78). 
To complete the research, the same algorithms are tested on another popular dataset, SWaT~\cite{Goh2017swatdata}, showing the best results with the same algorithms (i.e., CNN and DNN).

\textbf{WADI}~\cite{Ahmed2017wadi, iTrustWADIWeb} is a dataset with data collected from WADI, a water distribution testbed, created as an extension of the SWaT testbed~\cite{Mathur2016swat}. The system comprises three subsystems: a primary grid, a secondary grid, and a return water grid. It is also able to simulate water consumption following time-varying demand patterns. 
The dataset collects 16 days of continuous operation: 14 under regular operation and two days within an attack scenario (a total of 15 attacks). The adversary aimed to cut off the water supply to the consumer tanks. In the attacker model, the adversary has remote access to the SCADA system. The data recorded represent the state of all the 123 sensors and actuators connected using Modbus/TCP protocol. The dataset is free upon request~\cite{iTrustDatasets}, and it is provided as csv files.

There are many IDSs designed and tested on WADI Dataset in literature. 
MAD-GAN~\cite{Li2019} is an unsupervised multivariate anomaly detection method based on Generative Adversarial Networks (GANs). This method uses a generative model to create a fake time serie and a discriminator to distinguish between normal and abnormal data. 
A peculiarity of this work is that, instead of considering each data stream independently, the framework considers the entire variable set concurrently to capture the latent interactions among variables. To do so, the authors implement a sliding-window approach to divide the multivariate time series into sub-sequences.
On WADI, MAD-GAN obtains a precision of 0.53 and an F1-Score of 0.62. 
Better results were achieved by Kravchik and Shabtai~\cite{Kravchik2019} which obtain a Precision of 0.83 and an F1-Score of 0.75. They employ an Autoencoder with sequences of length 7 in the time domain. With respect to SWaT~\cite{Goh2017swatdata} and BATADAL~\cite{Taormina2018batadal}, the authors also mention that it was impossible to apply the AutoEncoder on the frequency domain because most of the features do not have a clear dominant frequency. 
However, the best results on WADI was obtained by DAICS~\cite{Abdelaty2020}, a deep learning solution for anomaly detection in ICSs. The authors propose a 2-branch feature extraction framework. The wide branch, containing only one fully connected layer, is used to memorize the normal state of sensors and actuators. Instead, the deep branch comprises two fully connected layers between two convolution layers and provides the generalization degree required to handle events not covered in the training set.
Moreover, DAICS introduces the \textit{few-time-steps algorithm} which can be used to efficiently reconfigure DAICS in a production environment when operators encounter false alarms.
DAICS can achieve a Precision of 0.919 and an F1-Score of 0.804 on WADI.

\subsection{Network Level}\label{subsec:dataset_network} 

\textbf{CyberCity Dataset}~\cite{BorgesHink2016, CyberCityWP, CyberCityWeb} is a dataset collected by the SANS Institute from their own ICS CyberCity testbed. CyberCity testbed is a complete simulation of an entire city containing a bank, a hospital, a power plant, and many other generally available components in a small town. There is also a tabletop scale model of the city, which shows an electric train's behavior, a water tower, and a miniature traffic light. A pcap file is freely downloadable online~\cite{CyberCityWeb} containing over 170k network packets recorded as a dataset for the Holiday Hack cybersecurity challenge in 2013. The data are unlabeled, but in~\cite{BorgesHink2016} the authors estimate that about 16\% of the data is under attack. Various attacks are included, such as scanning, information disclosure, command injection, MitM, and DoS. The ICS components use Modbus/TCP, EtherNet/IP, and NetBIOS as communication protocols. 
For each attack presented, some preventative measures are proposed and evaluated. Some examples are awareness training, system patching, IDS, or anti-virus, but it is remarked that neither one is 100\% effective.
It is worth noting that, at the best of the authors' knowledge, there is no precise and official documentation of the dataset provided by the SANS Institute.

\textbf{Dataset 2: Gas Pipeline Datasets}~\cite{MorrisWeb, Beaver2013Morris2} contains a collection of labeled Modbus/RTU telemetry streams from a gas pipeline system in Mississipi State University's Critical Infrastructure Protection Center~\cite{Morris2011testbeds}. Each stream is composed of some selected features, including, for instance, an identification bit to discriminate between command and responses, states of components,  length of data, and physical measurements. 
The authors include different command injection and data injection attacks, alongside some data in normal behavior.
The dataset contains about 397k samples, divided into csv files with a name indicating the particular attack. The dataset also includes a feature to identify the samples that are effectively part of an attack, with information about the attacker's action in the particular moment. The total percentage of samples with abnormal behavior is 0.97\%.
Unfortunately, the dataset does not include each sample's timestamps, making it impossible to analyze timing information.
The dataset was used to test different machine learning algorithms as a discriminator of malicious RTU transactions to detect the deployed attacks~\cite{Beaver2013Morris2}. K-Nearest Neighbors and Random Forest are the two algorithms that provided better results across all the attacks, with a Recall/Precision of 0.75 or higher for five of the seven attacks. More in detail, the most problematic attacks were burst values (i.e., sending multiple successive pressure values, faster than the data display rate, to the operator interface) and setpoint value injection (i.e., the attacker sends false pressure values equal to the setpoint).
Yüksel et al.~\cite{Yuksel2016} formally describes the user-understandable framework with effective anomaly detection techniques for ICSs. The test implemented using Modbus/RTU employs the Dataset 2: Gas Pipeline by dividing the attacks into scanning, illegal values, timing, and illicit command. 
The results are highly variable depending on the trade-off between the detection rate and the false positive rate. However, by fine-tuning the algorithm, it was possible to achieve a detection rate of 0.9991 and a false positive rate of 0.001.

\textbf{Dataset 3: Gas Pipeline and Water Storage Tank} by Morris et al.~\cite{Morris2014, MorrisWeb} are two different datasets from physical testbeds containing both physical data field and network traffic. The first comprises data deriving from a gas pipeline, while the second contains data from a water storage tank. Both the datasets come from testbeds at the Mississipi State University's Critical Infrastructure Protection Center~\cite{Morris2011testbeds} and are shared as ARFF files. 
A bump-in-the-wire approach was used to capture data logs and inject attacks in Modbus communication in both cases. The implemented attacks are reconnaissance, response and command injection, and DoS. They cover around 27\% of the total data.
The authors also provide two short datasets created using 10\% of the complete datasets, suited for rapid tests during the preliminary IDS development phases.
As explained in~\cite{MorrisWeb, Morris2015}, the gas pipeline dataset contains unintended patterns that cause some algorithms to identify attacks and non-attacks in unrealistic ways easily. Therefore, we do not report this work in the corresponding dataset tables. Instead, we consider the second version of this dataset, called Dataset 4: New Gas Pipeline.

\textbf{Dataset 4: New Gas Pipeline}~\cite{Morris2015, MorrisWeb} is a new version of the Dataset 3: Gas Pipeline dataset. This version was proposed to fix dataset problems causing machine learning algorithms models that do not match real system behavior and lead to overly optimistic classification accuracy. 
In this version, the authors implement 35 attacks and precisely document them in the paper and the dataset. The dataset includes different labels for each attack, which cover 21.86\% of the capture. Like the previous version, the protocol used is Modbus and data are available as an ARFF dataset containing both physical data and information about the network packets.

D3 and D4 datasets are widely used in the study of IDS for ICS. 
Feng et al.~\cite{Feng2017} presented a multi-level anomaly detector using package signatures and LSTM networks. The detection architecture provided is composed of two-level. First, a packet-level anomaly detector based on a Bloom Filter is applied; second, the first-level not-anomaly data are used as input to a stacked LSTM neural network model time-series level anomaly detection. The anomaly detector was tested on Dataset 4: New Gas Pipeline using two LSTM layers of 256 nodes, each achieving a Precision of 0.94, Accuracy of 0.92, and an F1-Score of 0.85. The most problematic attack to be detected was the injection of malicious state commands for which a Gaussian Mixture Model performed better.
Demertizis et al.~\cite{Demertzis2017} proposed the Spiking One-Class Anomaly Detection Framework (SOCCADF), which employs the advanced evolving Spiking Neural Netowork (eSNN). eSNN is a modular connectionist-based system that evolves its structure and functionality in a continuous, self-organized, on-line, adaptive, and interactive way using incoming information. The framework is supervised and was tested on both the Dataset 3: Water Storage Tank  (Precision 0.981; Accuracy 0.981; F1-Score 0.981) and the Dataset 4: New Gas Pipeline (Precision 0.988; Accuracy 0.988; F1-Score 0.988).
The same authors adopted eSNN on GRYPHON~\cite{Demertzis2020}, which simplifies the validation mechanisms to work in a semi-supervised way, getting as input only data in standard behavior (i.e., labeled as normal packets). This approach was able to get a Precision of 0.980, an Accuracy of 0.980, and an F1-Score of 0.980 on the Dataset 3: Water Storage Tank, while a Precision of 0.975, an Accuracy of 0.977, and an F1-Score of 0.970 on the Dataset 4: New Gas Pipeline. 
Another interesting work is the metaheuristic approach by Mansouri et al.~\cite{Mansouri2018}. In this work, the authors provide an anomaly detector based on neural networks with a pre-processing step able to act with a different algorithm based on the packet's delay to have as little impact as possible on the real-time communications. When computational speed is required, computationally efficient Evolutionary System~\cite{jagerskupper2006es} optimization is used. Insted, a more accurate but computationally expensive Grey Wolf optimizer~\cite{mirjalili2014grey} is used if with higher latency scenarios. A neural network is then used to detect malicious data with an accuracy up to 98\% on the Dataset 4 New Gas Pipeline. 

\textbf{Electra}~\cite{Gomez2019electra} 
dataset was obtained from a real scenario of an electric traction station used in the railway industry. 
Electra is composed of 5 PLCs, a SCADA system, a switch, and a firewall. All the communications between the components implement Modbus and S7comm over TCP/IP with a master-slave model. There are two different datasets, one for each communication protocol. The implemented and labeled attacks are false data injection, replay attack, and reconnaissance attacks in both cases. The attacks were deployed with a new device attached to the network with a MitM configuration. In both \textbf{Electra Modbus} and \textbf{Electra S7comm} datasets, the capture lasts about 12 hours in which the 94\% and the 98\% of the data are in normal condition, respectively. The data amount is enormous, containing 387M entries for S7Comm (36.8GB) and 16M for Modbus (1.5GB). The two datasets are freely available on the web~\cite{ElectraWeb} in csv format. 

Together with the datasets' presentation, Gómez et al. provided an implementation of the main algorithms used for anomaly detection. The authors try both supervised and semi-supervised algorithms. On Electra Modbus, a simple supervised Random Forest with 200 estimators was sufficient to achieve a Precision of 0.988 and an F1-Score of 0.987, while a single layer supervised Neural Network with 128 neurons was able to reach a Precision of 0.9999 and an F1-Score of 0.996 on the S7Comm version. On the other hand, the semi-supervised OCSVM performed properly on both the dataset, reaching 0.996 of Precision in Electra S7Comm.
In the successive year, the same authors proposed SafeMan~\cite{PeralesGomez2020}, a framework to manage both cybersecurity and safety in the manufacturing industry.
It is composed of a set of applications and services used to monitor and analyze the industrial process in real-time. SafeMan is based on Edge Computing (EC) to achieve low latency and fast deployment of applications and services. Furthermore, EC allows performing the necessary computing tasks close to the manufacturing activity or the network edge. The framework contains several components to assist the deployment, and the risk assessment, together with the cyber threats detection application proposed in~\cite{Gomez2019electra}.  
A different and innovative approach was introduced by Li et al.~\cite{Li2020} who design an anomaly detection method based on cross-domain knowledge transferring. The authors employ the TrAdaBoost algorithm to train a neural network using not only a part of the data of the Electra Datasets but also employing data from different domains, both from other ICS (e.g., SWaT Dataset~\cite{Goh2017swatdata}) or other CPS fields (e.g., KDDCup99 Dataset~\cite{Tavallaee2009}). Then, they compared the error rate with respect to a standard SVM and a standard LSTM, showing better results, especially when employing a small fraction ($<10\%$) of the Electra Dataset in the training phase.

\textbf{HVAC\_Traces} by Ndonda and Sadre~\cite{Ndonda2019hvac, HVACWeb} is a dataset recorded on a Heating, Ventilation, and Air Conditioning (HVAC) system powered by Honeywell and used to provide thermal comfort and acceptable indoor air quality on a university campus. The Building Management System (BMS) is fully automated, and it is suited to monitor from 15 to 20 structures, each containing different PLCs and RTUs. Operators can access the system through the HMI. The protocols implemented are proprietary (e.g., DCE/RPC, NetBIOS, S7Comm) and use TCP/IP at the transport layer. The data capture was produced using \emph{tcpdump}, at two routers via port mirroring. 
To obtain an accurate timestamp on each packet in two separate recording points, the authors synchronized the clocks using Network Time Protocol (NTP)~\cite{mills1991internet}. However, it was not sufficient to ensure good accurate timing. To overcome this problem, the authors introduced a correction factor calculated using ad-hoc ICMP messages sent periodically on the network.  
The anonymized dataset is publicly accessible in pcap files, where each file contains one hour of traffic. In total, there are about 7 days of collected data in normal conditions, without any attacks.
Since the dataset does not contain attacks and is a novel collection, there are no IDS tested to the best of our knowledge.

Lemay et al.~\cite{Lemay2016} present a dataset of a SCADA network, also called \textbf{Lemay SCADA}, virtually implemented with SCADA Sandbox. The simulations contain different MTUs and controllers connected with the Modbus/TCP protocol. The attacks are generated with an infected machine that launches various exploits to infect other devices. Then,  the compromised machines launch different attacks by leveraging Metasploit~\cite{maynor2011metasploit} (e.g., Malware Injection, Reconnaissance). The authors give particular attention to the labeling process and to maintain normal intra-packet time properties. The captured data are divided into various collections with an explicit name indicating the types of implemented attacks.
The authors also implemented a cover-channel attacks dataset presented as \textbf{Lemay Covert}. In these attacks, the least significant bit of the Modbus packets are used to carry information. To the best of our knowledge, this is the only available dataset containing side-channel attacks. Unfortunately, none of the attacks were designed considering Modbus protocol vulnerabilities. Instead, they are implemented with Off-the-Shelf Tools (i.e., Metasploit~\cite{maynor2011metasploit}).
Data collection lasts about 6.25h, and the samples labeled as attacks are about 0.15\% of the total for the first attack, while the covert channel packets are present in the whole capture. The datasets are shared in both pcap and csv format.

Schneider and Böttinger~\cite{Schneider2018} proposed an unsupervised anomaly detection framework. They employ deep autoencoders with pipelining parallel processing strategies to speed up the training. While the proposed framework performs well on the SWaT dataset~\cite{Goh2017swatdata}, it shows very different results depending on the attack type when applied on the Lemay SCADA dataset. In particular, to correctly detect an attack, the framework requires a minimum duration of it. For attacks lasting longer than the minimum threshold, the Precision reaches 100\%.
Anton et al.~\cite{Anton2018} implemented different standard classification machine learning algorithms on Lemay SCADA. The authors extract 14 basic features from the packets and nine additional features derived from timing and frequency information. Algorithms are tested on three different batches of packets resulted from merging different Lemay SCADA datasets. Both Random Forest and SVM result in an F1-Score and an Accuracy greater than 0.999 with all the batch, while k-means clustering report the lowest results.
In a follow-up work of Anton et al.~\cite{Anton2019} data are considered as time series. Each second of network traffic was aggregated into a single data point. Three different algorithms were implemented to detect anomalies inside the three batches of captures defined in~\cite{Anton2018}. The first algorithm implemented was Matrix Profiles, and it performs well on data with periodic characteristics, requiring only one hyperparameter. Second, the Seasonal ARIMA-process performed well on periodical data and is more resistant to noise but requires a more complicated tuning of the three hyperparameters. Finally, the authors implemented LSTM, which requires a high training effort compared with the other two light-weight approaches. They tested the algorithms on a subset of the Lemay SCADA datasets containing seven attacks divided into three categories (fake command, executable upload, file moving). The attacks are almost all correctly classified with every algorithm. With LSTM, the accuracy is always greater than 0.90, while the F1-Score is really variable based on the threshold selection methodology.

\textbf{Modbus SCADA \#1}~\cite{tjcruz2018dataset, tjcruzWeb} by Cruz et al. is a dataset containing data recorded from a small physical testbed simulating a liquid pump. The testbed comprises an HMI, an Adruino-based RTU, a PLC, a Variable-Frequency Drive (VFD), and a 3-phase motor. The protocols used are Modbus/TCP and Modbus/RTU.
Data are divided into subfolders based on the attack deployed. Moreover, each pcap file is named with an intuitive strategy that includes the duration of both the capture and the attack.
The attacks implemented range from MitM to different flooding types: ping flooding, TCP SYN flooding, and Modbus Query flooding. All these flooding attacks are aimed at the generation of DoS.
The data recording lasts for 24 days, containing 4.81\% of data flagged as under attack.

This dataset was used by Radoglou et al.~\cite{Radoglou2020} to test an IDS. Firstly, the authors present an expansion of Smod~\cite{luswata2018smod}, a penetration testing tool for Modbus/TCP, to enable the generation of  DoS, MitM, and replay attacks. Then, they deployed an IDS to detect DoS attacks and a server for machine learning offloading computation. Among the various algorithms tested, AdaBoost~\cite{Adaboost} and Random Forest achieve the best results with a Precision of 0.96, an Accuracy of 0.81 and an F1-Score of 0.77. 

\textbf{QUT\_DNP3}~\cite{Rodofile2013, RodofileDNP3Web} is a dataset presented in the Ph.D. dissertation of the author. The dataset contains data collected from a small section of a transmission substation SCADA network.
The testbed involves GOOSE and DNP3 protocols, enabling the communication between the Master, the Slave, the IED, and the attacker machine. All the communications pass through an industrial switch. 
The attacks are categorized into six categories: Injection, Flooding, Masquerading, Replay, MitM, and all attacks. Each category also contains Reconnaissance packets. The attacks are launched by an attacker machine, which also generates a log providing information about each attack sequence's start and end.
Each dataset file has a different duration based on the attack frequency during the capture creation since the authors implement a random time between two attacks.
Moreover, the dataset is divided into two categories based on the attack frequency: frequent attacks (i.e., approximately an attack every half an hour) and infrequent attacks (i.e., approximately an attack every random time between one and four hours).
For each frequency category, the authors provide two datasets, respectively, for training and testing. 
Furthermore, a control dataset with only legitimate communications (i.e., without any attacks) is available, and it covers 24 hours of recording.
In total, the dataset contains 40 days of recording.
It is worth mention that the labeling process was performed with particular care since it was the main topic of the thesis work.
The dataset is available on Github~\cite{RodofileDNP3Web}.
However, to the best of our knowledge, there are no IDS tested on this dataset.

\textbf{QUT\_S7Comm}~\cite{Rodofile2017qut, RodofileS7Web} developed by Rodofile et al. is an open-source dataset collected in a three time-based sub-processes testbed of a mining refinery plant. The plant testbed is composed of one Siemens PLC acting as Master and three PLCs actings as slaves, all connected with a switch to an HMI and communicating using S7Comm protocol.
The attack dataset comprises 9 hours of data and 64 attacks from 13 different possible typologies. Data are provided with pcap files and four process logs: a master log, a conveyor log, a tank log, and a reactor log. The labels of the attack samples are contained in separated csv files. 
The control dataset comprises 8.5 hours of network traffic and process log data, with 32 different processes. 
This dataset's peculiarity is the particular division of the network traffic in separate files based on the node capture perspective: a file collected from the attacker's point of view, one from the HMI, and one from the master PLC.
This particular composition could be initially complex to use, but on the other hand, it provides higher flexibility with respect to datasets with the entire capture.
The dataset is available on Github~\cite{RodofileS7Web}.
To the best of our knowledge, there are no IDSs implemented on this dataset.

\textbf{4SICS}~\cite{4SICSWeb} is a pcap dataset collected by Netresec from an ICS lab at the Industrial Cyber Security Conference. At this conference, there was an ICS testbed composed of heterogeneous devices such as PLCs, RTUs, servers, and industrial network equipment (e.g., switch, firewalls). It was available for hands-on testing by the conference attendees and, since the testbed was left almost uncontrolled, the data recorded are not labeled. Furthermore, it is impossible to know what the users have done and, eventually, what kinds of attacks are present. The dataset includes a wide variety of ICS protocol traffic such as S7Comm, Modbus/TCP, EtherNet/IP, and DNP3. 

\textbf{S4x15} ICS Village CTF Dataset~\cite{S4x15Web} provided by Digital Bond, contains network traffic collected during a capture-the-flag (CTF) competition in the ICS Village. The system was composed of different interconnected PLCs, and the dataset contains, without labeling, the attacks launched by the players to the system. The dataset contains pcap files with Modbus/TCP and BACnet packets.
 
Basing on this dataset, Yu et al.~\cite{Yu2018} proposed an anomaly detection method based on TCP and UDP payload inspection. 
The detector's architecture comprises an offline module for the expected behavior model and an online module containing the actual anomaly detector and a packet signature generator. In the proposed work~\cite{Yu2018}, the authors use 4SICS~\cite{4SICSWeb} dataset to model the normal traffic behavior for Modbus/TCP protocol, while the normal traffic behavior of BACnet protocol is based S4x15 dataset. Instead, malicious packets are retrieved from Quickdraw-Snort~\cite{QuickdrawWeb}, a collection of Snort rules for ICS environments, which also provides some testing packets. Results show Accuracy and Recall close to 100\% and a very low false alarm rate.

\textbf{WUSTL-IIOT-2018}~\cite{Teixeira2018, TeixeiraWeb} is a dataset recorded from a testbed simulating a water tank control system. Network traffic was monitored for 25 hours, collecting 25 features. Then, the authors performed a data cleaning process to delete corrupted or missing values and outliers. In this phase, about 10k observations were erased, leaving the final version with about 7037k entries. Furthermore, only the six more relevant features are available in the provided csv file, together with a column indicating if the observations are related to an attack. Various attacks have been launched during the capture: port scanning using Nmap~\cite{nmap}, address scan attacks, device identification, and unauthorized access to actuators status by using known exploits of the Mobus protocol. The final dataset contains 6.07\% of data under attack.
On the same paper~\cite{Teixeira2018}, the authors developed IDSs employing standard Machine Learning algorithms. Best results were achieved by Decision Tree and KNN with an accuracy of up to 100\% considering the offline evaluation. Instead, regarding the online phase, Decision Tree and Random Forest obtain the best results with an accuracy of 0.999. Furthermore, these last two models performed well in terms of False Alarm Rate (i.e., percentage of the normal flows misclassified as abnormal flows) and Un-Detection Rate (i.e., the fraction of the abnormal flows misclassified as normal flows), which are close to 1.


\subsection{Physical and Network levels}\label{subsec:dataset_phynet}

\textbf{Electric Power and Intelligent Control (EPIC)} is a collection of data from 8 scenarios collected with the EPIC testbed~\cite{Adepu2018epic, iTrustEPICWeb}. Each collection scenario lasts for 30 minutes under normal operations, resulting in more than 5000 readings of sensors and actuators, together with the corresponding Modbus network traffic. 
Blaq\_0~\cite{iTrustDatasets} is a dataset obtained from the same testbed under different attacks. Blaq\_0 contains network-level data collected from a three-day Hackaton 2018 where different teams attack the EPIC testbed. 
Both datasets are free upon request, but the second one is not widely used. The EPIC dataset contains both pcap and csv files.
Both the datasets are free upon request on the iTrust website~\cite{iTrustDatasets}.

\textbf{QUT\_S7 (Myers)}~\cite{Myers2018} is a dataset generated from a small scale ICS testbed. It is composed of a bi-directional conveyor belt system, a water pump system, and a ``reactor" pressure vessel system. All these devices are connected to a power meter, and an HMI is used to collect logs. The protocol used is S7 Communication, the standard protocol for Siemens PLCs.
The dataset contains device logs with information about each component's state and pcap files with network traffic. Data are divided into two folders containing control data (i.e., data in a normal behavior) and attack data. 21 cyberattacks were launched, consisting of two major types: Injection Attacks and Flooding Attacks. Furthermore, the authors also provide an xlsx file containing pre-processed data.
The dataset is freely available for the download~\cite{MyersWeb}. 
In the same paper, Myers et al.~\cite{Myers2018} proposed a novel process mining based anomaly detection technique. The detector idea is to collect logs in order to produce a record of each device's status. From this data, the authors compute a model containing the expected behavior of the ICS. The model is designed to manage the entire process instance from start to finish with only acceptable events. Finally, process mining is used to perform conformance checking activity, calculating the fit of a given event log by replaying it on the model.
Results show that only 16 attacks out of 21 were correctly identified with several false positives (i.e., Precision: 0.727; F1-Score: 0.744; Recall: 0.762). The authors motivate that for most false positives, the starting condition was altered by previous attacks. It is a common problem that can be mitigated with a correct generation of the dataset, as will be discussed in Section~\ref{sec:good}, especially taking care of the labeling part.

\textbf{SWaT}~\cite{Goh2017swatdata, iTrustSWATWeb} is the most popular dataset in the ICS field. It contains monitoring data from a fully operational scaled-down water treatment plant. 
The testbed contains two separate networks: a level 1 star network that allows the communication between the SCADA system and the six PLCs and a level 0 ring network that transmits sensor and actuator data to the corresponding PLC.
The protocols employed for communications are CIP and EtherNet/IP. 
The dataset received various updates and improvements over the years. More precisely, up to today, there are seven different data collections (the last one is dated June 2020).
The first version (December 2015, described in~\cite{Goh2017swatdata, iTrustDatasets}) is the largest and most used. This version includes both network traffic and recordings from all the 51 sensors and actuators for eleven days. Of these eleven days, seven days are under normal operation, and four days contain 36 different attacks, classified into four types based on the attack number of stage and devices affected. 
The first version of SWaT contains about 944k physical samples (5.76\% are labeled as attack samples). The singular network packets are instead unlabelled, with only a flag indicating the presence (or not) of malicious data in each packet's batch.
In 2017, a new version was released, which collected about 136 hours of network traffic together with measurements of sensors and actuators provided in csv form. No attacks were performed during the recording phase.
In 2019 about 4 hours of recording were saved as a dataset. About one hour contains 6 different attacks like spoofing and tampering with some switch. In this case, physical data are provided as xlsx files. 
In the same year, another version was released containing both network and physical data of about 3 hours, during which two malware attacks were launched. The first try to exfiltrate historian data, while the second disrupt sensors reading and process. 
The most recent version is dated 2020 and contains 4 runs, each one lasted 2 or 4 hours. Physical data without any attacks are provided in xlsx form.
In 2017 during the SUTD Security Showdown, a competition where researchers could access and attack the SWaT testbed, all the network flows are saved in a dataset called S317~\cite{S317Web}. The dataset records three days of competition and contains historical data and the description of the attack scenarios.
Both datasets are free upon request, but the second one is not widely used. Data are provided in different csv, xlsx, and pcap formats.
As reported in~\cite{turrin2020statistical}, the various SWAT releases are very different from the operational point of view, also implementing different actuators control logic. It makes it difficult to transfer the detection framework among the dataset releases. Furthermore, even in the same SWaT version, the systems behave very differently. It is probably due to the testbed recovery time after an attack. To overcome this problem, in more recent versions, the authors restart the testbed after each attack.
All the datasets are free upon request at the iTrust datasets webpage~\cite{iTrustDatasets}.

The SWaT dataset is probably the most used dataset on which researchers try their IDSs. In almost all the papers using the SWaT testbeds, there is no explicit mention of the dataset version used. However, from the data description, we can infer that the first version is used in almost all the cases. 
Several innovative detection methodologies have been tested on this dataset, ranging from sensor noise fingerprint~\cite{ahmed2018noise}, a graphical-based detection approach~\cite{lin2018tabor} and a framework to generate invariants with association rules mining~\cite{feng2019systematic}.
Kravchik and Shabtai~\cite{Kravchik2019} employ the SWaT physical data to test different detection approaches such as Dynamic PCA, 1D-CNN, and AutoEncoder, both in the time and frequency domains. 
Thanks to the linearity of many relations between sensors and actuators in SWaT, PCA performed very well, especially using a sliding-window approach, which results in 0.92 of Precision and 0.879 F1-Score. Also AutoEncoder in the frequency domain reaches similar scores (i.e., Precision: 0.924; F1-Score: 0.873).
Similarly to WADI, excellent results on SWaT physical data were achieved by Abdelaty et al.~\cite{Abdelaty2020}. This paper introduces DAICS, 2-branch neural networks with automatic tuning mechanisms to update the system model. DAICS scores 0.9185 of Precision and 0.889 of F1-Score.
It is worth noting that, despite a not high Precision (i.e., 0.70) and F1-Score (i.e., 0.81), MAD-GAN~\cite{Li2019} on SWaT achieved a Recall of 0.954, the higher one to the best of our knowledge. 
Instead, by relying on SWaT network traffic, Schneider and Böttinger~\cite{Schneider2018} implement an autoencoder-based unsupervised anomaly detection framework leveraging pipelining parallel processing strategies to speed up the training.
To overcome the problem of unlabelled network packets and generate the ground truth, the authors developed a semi-automatic label estimation mechanism to detect the packets with a higher probability of being anomalous and then use a manual investigation to label them.
Results reveal a Precision of 0.998 and an F1-Score of about 0.988 in scenarios like TCP session reset attack and SYN Flood attack, while other types of attacks as duplicate acknowledgments or TCP retransmissions are essentially not detected.

\section{Lesson Learned: Good Practices}\label{sec:good}

\begin{figure}
    \centering
    \includegraphics[width=\columnwidth]{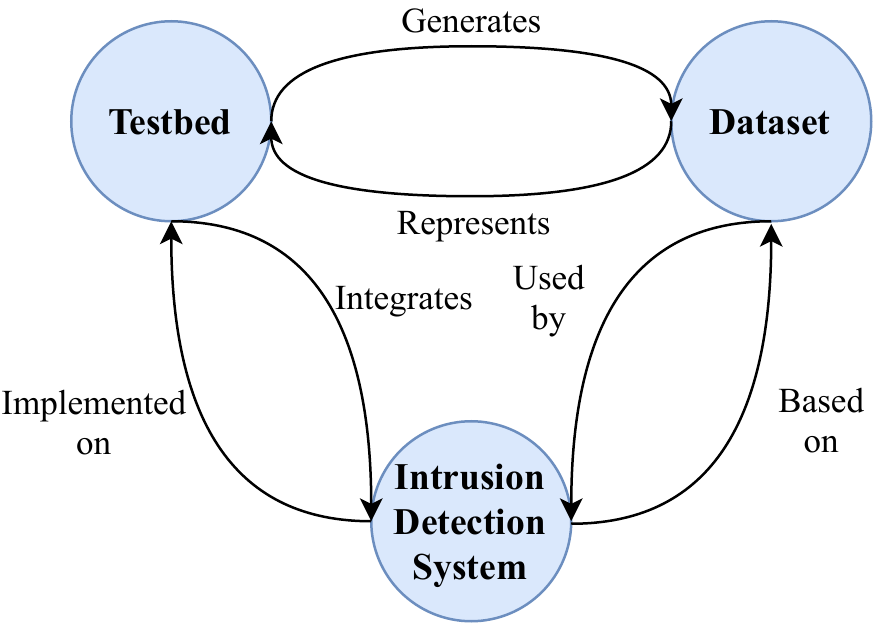}
    \caption{Relations between Testbed, Dataset, and IDS.}
    \label{fig:relations}
\end{figure}

Basing on the knowledge acquired on surveying the different works and analyzing the common mistakes and solutions implemented, in this section, we summarize concepts and good practices to consider when selecting a testing system. In particular, we summarize the good practices in creating an effective testbed in Section~\ref{subsec:good_testbed} and to develop a dataset in Section~\ref{subsec:good_dataset}. Furthermore, we also provide additional insight to help the standardization of the IDS results in Section~\ref{subsec:good_ids}. During the designing phase of each of the three resources, the designer must consider the final use of such resources and the other two resources' requirements in future integration. Figure~\ref{fig:relations} graphically represents the relation between the three resources. More precisely, a testbed should allow an efficient data collection to produce a well representative dataset and integrate IDSs to validate the case studies in a real scenario. A dataset must be designed to be an exhaustive and precise representation of a testbed and easily allow data analysis tasks and the implementation of an IDSs. Finally, an IDS, which represents the higher-level products with respect to the other two resources, should generalize on different datasets to avoid construction biases. Moreover, the design of a dataset should consider an easy integration into a real-world scenario such as a testbed. 


\subsection{Good Practices: Testbed}\label{subsec:good_testbed}

An effective testbed development passes through various steps and challenges, each composed of a notable complexity level. These challenges should be considered during the design phase.

\paragbf{Scope Identification}
During the design phase, the designer must consider the final application. The applications of a testbed can be~\cite{tippenhauer2019design}: i) Discovery, to study and obtain knowledge about a particular ICS field or system functioning; ii) Demonstration, to validate or experiment the research findings; and iii) Education, to use the testbed to educate students, researchers, and stakeholders.
Every scope implies different requirements to deal with and different funding. For instance, if a testbed is specifically designed for IDS development, the authors must consider developing an attack chain and data collection accurately. On the contrary, the Educational testbeds do not have this requirement. Instead, they should be composed of an easily understandable and representative process. In this case, water systems are an excellent choice due to their immediate visualization.
Once the scope is identified, the designer can give the system's specific layer adequate importance to satisfy the scope.

\paragbf{Fidelity}
If the testbed is used for Discovery or Education, data's perfect fidelity is generally not needed. In these cases, Virtual or Hybrid testbeds are the preferable platforms to be used due to their flexibility and cheapness. 
Contrarily, in the case of validation tests, Physical testbeds are the best solutions since the smallest variation of measures is fundamental for the research. A complete work that can help researchers to identify the correct design criteria is~\cite{ani2020design}.
The US National Institute of Standard and Technology (NIST) has recommended that a SCADA testbed for security assessment should consider four general areas~\cite{Stouffer2015nist}: the control center, the communication architecture, the field devices, and the physical process itself.

\paragbf{Expensiveness}
Expensiveness in the construction and the maintenance of Physical testbeds are the first barriers a research group will encounter when deciding to build one.
Suppose a research group can deal with this limitation. In that case, it is useful to share with the community datasets collected from the Physical testbed and the related documentation, as iTrust laboratory of SUTD~\cite{iTrust} is doing with SWaT and WADI.
Furthermore, provide a simple way for other researchers to access the testbed can be an added value not only for the community, which can take advance of it but also for the owner who can have a more critical view of the system.

\paragbf{Reproducibility and Comparability}
Robust and innovative researches need to be reproducible and peer validated. Basing on this, testing on physical testbeds is not recommended since it creates difficulties in reproducibility. An intelligent solution is to create one or more datasets capturing network traffic and physical measures from the testbed and share them with the community. In this way, the reviewer can easily verify the study, while the community will benefit from newly available datasets.
On the other hand, if a virtual testbed is used, it is not required to provide a dataset to support the research. Instead, it is possible to provide the software with the entire simulation. However, it is fundamental to precisely indicate the architecture and the state of the ICS at the beginning of the experiment to avoid reproducibility errors due to a different scenario. 

\paragbf{Missing Representation} We identified that the most common scenario represented with a testbed is relative to water management (e.g., Water Distribution, Water Treatment). This is probably because Water Systems are the most easier scenario to implement in terms of equipment and maintenance costs. Another very represented scenario is related to the Electric Plant (e.g., Power Grid, Power Power). However, IDS rarely investigate this scenario. We believe this is due to the difficulty in developing detection systems that deal with high-dynamic environments such as Electricity. Also, the majority of the related dataset does not include attacks scenario.
For future organizations that want to approach a testbed development, we identified a low contribution in scenarios such as large-scale manufacturing, Nuclear Plants, Transportation Systems, health-care infrastructure, IIoT, or HVAC.

\subsection{Good Practices: Dataset}\label{subsec:good_dataset}
A well-designed dataset should exhaustively represent a testbed's behavior and allow easy implementation of research findings. To do this, the design process of an effective dataset must consider the following points.

\paragbf{Labeling}
When designing a dataset, the labeling process must be precisely described in the documentation to allow researchers to process the data accordingly. 
Packets that are part of attacks must be carefully labeled to provide ground truth to researchers. 
Furthermore, in a valuable dataset, labels must also contain information about the attack type (e.g., Injection, Replay, DoS) and the attack phase. 
This last element is essential due to the recovery time of many ICSs: after an attack occurs, the system may need some time to stabilize itself. This behavior can be wrongly considered part of the attack by an inaccurate labeling process. Hence, a good strategy is to flag this kind of packet as \emph{recovery}, leaving the decision on how to consider them to researchers. The work~\cite{8805036} explained that the authors of the SWaT dataset decided to label a process data sample as ``Attack'' when the attack was launched, instead of when the system behavior started to change. This approach can lead to a ground truth problem if not correctly documented or managed.
Furthermore, it is important to consider the label generation methodology accurately. A manual approach to flag each entry of an attack as malicious is costly, and if the data amount is large may be impracticable. On the other hand, fully automatic strategies are possible, and they work quite well if attacks are at the same time automatically generated. However, automatic labeling cannot provide high accuracy in case of complex attacks on highly distributed systems. Semi-supervised approaches provide a trade-off that efficiently spends an expert's work supported by a visualization platform such as RiskID~\cite{Torres2019}.

\paragbf{Documentation} 
Many of the datasets surveyed lack in documentation. To allow correct and easy use of the dataset, the designer should include detailed information with the dataset's characteristics or a description of the source testbed design. Exhaustive documentation should include the system's control logic, a description of the implemented attacks, and the configuration settings.
In the SWaT case, as reported in~\cite{turrin2020statistical}, the recent versions of the dataset implement different control logic. However, the authors never mentioned such modifications.

\paragbf{Attacks Similarity} A complete dataset should also include attacks. Similarly, in a testbed, a researcher needs to be able to deploy attacks easily. However, the designer must approach this phase with caution.
Attacks should be as similar as possible to real cases. If a dataset is collected from a testbed, it is sufficient to launch the attacks following an adversary approach and various system information. The authors should also include a clear and complete analysis of the attacker model. 
On the other hand, it is important and challenging to capture traffic while the attack is occurring in order to generate the datasets. 
Adding synthetic packets in the resulting capture, if not accurately managed, could disrupt the fidelity of the dataset, making it unrealistic. 
Furthermore, if data are captured in different monitoring points inside the ICS, it is required a synchronization mechanism to provide consistent data.

\paragbf{Domain Shift} A common problem in a dataset is the so-called Domain Shift, i.e., the difference between the training entries and the testing data~\cite{Abdelaty2020, turrin2020statistical}. To support researchers and IDS development, datasets should be released with complete documentation explaining the system's initial state. Another problem observed in~\cite{turrin2020statistical} is related to the testbed remains unstable for a long time. More precisely, after the end of an attack, a system's behavior may need time to recover, remaining unstable. In this case, its behavior will be identified as anomalous by detectors even if flagged as Normal. To deal with this problem, when designing a dataset, the authors should consider adding another label to classify the dataset, e.g., ``System Unstable''. If the authors can directly interact with the testbed, another solution could be to restart the system after an attack.
An imbalanced dataset is a collection of data that contains a significantly low number of samples from one class with respect to the other~\cite{Zolanvari2019imbalanced}. It is a critical issue that can influence the performances of Machine Learning based classifiers. 
Some datasets provided by researchers contain a low percentage of data classified as under attack, as reported in Table~\ref{tab:dataset}.
This happens because attacks generally last for seconds or minutes, while the ICS is expected to run for much longer. 
There are different techniques to get better results from an imbalanced dataset~\cite{Ramyachitra2014}. One solution is to act at the data level by re-balance the data in a pre-processing phase using different sampling strategies (e.g., down-sampling). Another novel solution is Data Augmentation, recently introduced to improve Anomaly detection performances in~\cite{8594955}. This technique leverages generative models, such as GAN, to generate synthetic samples.
In~\cite{Zolanvari2019imbalanced} the authors performed an experiment to understand the impact of an imbalanced dataset in the ICS security field. Different datasets were obtained from an extensive network traffic capture collected from a water control system testbed. To do this, the authors associated to a fixed number of attack samples a variable number of normal samples to create five different datasets with different imbalance ratios (i.e., the percentage of data under attack over the whole dataset).
Ratios span from 0.1\% to 10.0\%. Results show a high Recall variance with better results on higher ratios (Recall$>$0.99 for ratio$\geq$0.70\%; Recall$<$0.12 for ratio$\leq$0.30). At the same time, the Undetected Rate (UR) (i.e., the fraction of the attack samples classified as normal) shows near-zero values for high ratio and large misclassification on low ratio datasets (UR$<$0.01 for ratio$\geq$0.70\%; UR$>$0.88 for ratio$\leq$0.30).
Basing on these results, the authors conclude that it is advisable to generate datasets with at least 1\% of data under attack to reduce imbalance problems when testing IDSs.

\subsection{Good practices: Intrusion Detection System}\label{subsec:good_ids}

Nowadays, the majority of IDS are based on Machine Learning and Deep Learning techniques. To build a model using such techniques is required a notable amount of well-organized dataset (e.g., balanced, labeled). Since these techniques are not always straightforward to understand and implement, researchers should implement a clear and well-approved pipeline. The following aspects can help the development of an effective IDSs.

\paragbf{Results Baseline}
While reading the different papers concerning the implementation of IDSs, we noticed the absence of an evaluation baseline in many cases. Defining a good baseline could help researchers evaluate if their proposed research is effective and improve the current state of the art. Furthermore, various works base their results on a subset of an available dataset. If not used for a specif reason (e.g., isolate and test a specific attack), this approach could cause a problem in understanding an IDS's effectiveness. For this reason, we believe that our Table~\ref{tab:dataset} can support the future evaluation baseline. 
We also identified issues in the evaluation metrics. Many research not always use the same metrics, making it difficult to compare different approaches. We suggest using as many common metrics as possible, such as F1-Score, Accuracy, Precision, and Recall.
Baseline problems are also mentioned in other fields, such as Review Helpfulness predictions~\cite{diaz2018modeling}, where the authors proposed to used the same features on the different models proposed by researchers. In this way, it is easier to compare the efficiency of a prediction model.
In the ICS field, IDSs model features can be very different and based on diverse approaches. However, comparing a proposed model with the best IDSs of the art state could help future researchers identify the right research directions. Furthermore, to avoid the effect of design bias of the dataset, an IDS should be validated on multiple datasets.

\paragbf{Data Verification}
Generally, researchers rarely investigate the causes of the weak performance of IDSs. Sometimes, the reasons may be due to a problem related to the distribution of the dataset. In~\cite{turrin2020statistical}, the authors showed that SWaT dataset behavior and data distribution between Training Set and Test Set significantly differ. However, even if the SWaT dataset currently represents the most used dataset to test IDSs, no previous works analyzed the statistical distribution through statistical tests.
Finally, to allow the community to validate the research approach and improve it, a good practice is to release the code repository source code (e.g., GitHub, Bitbucket).

\section{Conclusion}\label{sec:conclusion}

In recent years the interconnection between IT and OT networks has opened up modern ICSs to new risks and novel vulnerability surfaces. These vulnerabilities were highlighted in many works but also by the dangerous malware targeting industrial companies. Therefore, it is vital to develop new security mechanisms to protect such systems.

In this paper, we provide a comprehensive overview of the ICS field by presenting the architecture and the typical devices employed. We then proposed an analysis of the industrial protocols used in ICSs, highlighting security measures offered by the protocols, their expansions, and analysis of their diffusion in the real world. 
Furthermore, we surveyed and analyzed the different platforms to test new security mechanisms in the ICS field. To do this, we categorized the testbeds as Physical, Virtual, or Hybrid based on their functioning and explaining the various challenges and requirements to consider during the development or selection phases. Also, we presented the different ICS datasets dividing them based on the type of data provided and useful information that can help the reader choose the dataset (e.g., attack implemented, format, and various data information). To do this, we accessed every dataset and analyzed it separately. We also reported the IDS with the best performance present in the literature to offer a baseline to further works for each dataset.
Finally, we depicted different good practices and suggestions for researchers who want to use this kind of testing method and institutions that want to build testbeds or collect datasets.

We believe this survey can help address future research on this field and new researcher approaching the ICS area. In the future, we aim to continue collecting new testbeds and datasets on the website to create a collection of useful information the research community can exploit for researches and studies on this essential field. 

\section*{Acknowledgment}
This work was supported by a grant from the Cariparo Foundation and Yarix S.r.l. which we would like to thank.

\bibliographystyle{IEEEtran}
\bibliography{biblio}
\ifCLASSOPTIONcaptionsoff
  \newpage
\fi

\end{document}